\pdfoutput=1
\documentclass[prd,showpacs,letterpaper,10pt,aps,notitlepage,nofootinbib,superscriptaddress]{revtex4-2}

\usepackage[utf8]{inputenc}
\usepackage{amsfonts}
\usepackage{amsmath}
\usepackage{graphicx}
\usepackage{hyperref}
\usepackage{placeins}
\usepackage{mathrsfs}
\usepackage{bm}
\usepackage[utf8]{inputenc}
\usepackage{multirow}
\usepackage{slashed}
\usepackage[export]{adjustbox}
\hypersetup{colorlinks, linkcolor = [rgb]{0,0.0,0.75}, citecolor = [rgb]{0,0.0,0.75}, urlcolor = [rgb]{0,0.0,0.75}}

\newcommand{\nb}{\phantom{0}}
\newcommand{\wm}{\phantom{-}}
\newcommand{\bs}[1]{\ensuremath{{\boldsymbol{#1}}}}

\begin{document}

\title{\texorpdfstring{$\bm{\Lambda_c \to \Lambda^*(1520)}$}{Lambdac to Lambda*} form factors from lattice QCD and improved analysis of the \texorpdfstring{\\}{} \texorpdfstring{$\bm{\Lambda_b \to \Lambda^*(1520)}$}{Lambdab to Lambda*} and \texorpdfstring{$\bm{\Lambda_b \to \Lambda_c^*(2595,2625)}$}{Lambdab to Lambdac*} form factors}

\author{Stefan Meinel}
\affiliation{Department of Physics, University of Arizona, Tucson, AZ 85721, USA}

\author{Gumaro Rendon}
\affiliation{Physics Department, Brookhaven National Laboratory, Upton, NY 11973, USA}

\date{January 9, 2022}

\begin{abstract}
We present the first lattice-QCD calculation of the form factors governing the charm-baryon semileptonic decays $\Lambda_c \to \Lambda^*(1520)\ell^+\nu_\ell$. As in our previous calculation of the $\Lambda_b \to \Lambda^*(1520)$ form factors, we work in the $\Lambda^*(1520)$ rest frame, but here we use four different heavy-baryon momenta instead of just two. Because of the lower mass of the $\Lambda_c$, the moderately-sized momenta used here are sufficient to determine the form factors in the full kinematic range of the semileptonic decay. We also update the analysis of our lattice results for the $\Lambda_b \to \Lambda^*(1520)$ and $\Lambda_b \to \Lambda_c^*(2595,2625)$ form factors by imposing exact relations among the different form factors at zero recoil that follow from rotational symmetry. Imposing these relations ensures the correct behavior of the angular observables near the endpoint.
\end{abstract}

\maketitle

\FloatBarrier
\section{Introduction}
\FloatBarrier

 Studying weak decays of charm or bottom quarks bound inside $\Lambda_c$ or $\Lambda_b$ baryons has proven very fruitful in recent years. Two examples are the determination of $|V_{ub}/V_{cb}|$ from the ratio of $\Lambda_b \to p \mu^-\bar{\nu}_\mu$ and $\Lambda_b \to \Lambda_c \mu^-\bar{\nu}_\mu$ decay rates \cite{Aaij:2015bfa} and the analysis of $b\to s \mu^+\mu^-$ Wilson coefficients using the full angular distribution of $\Lambda_b \to \Lambda(\to p \pi^-)\mu^+\mu^-$ decays \cite{Blake:2019guk}. For semileptonic $J^P=\frac12^+\to J^P=\frac12^+$ transitions of heavy baryons, lattice-QCD calculations of the relevant form factors are already available for $\Lambda_b \to \Lambda_c$ \cite{Bowler:1997ej,Gottlieb:2003yb,Detmold:2015aaa,Datta:2017aue}, $\Lambda_b \to p$ \cite{Detmold:2013nia,Detmold:2015aaa}, $\Lambda_b \to \Lambda$ \cite{Detmold:2012vy,Detmold:2016pkz}, $\Lambda_c \to \Lambda$ \cite{Meinel:2016dqj}, $\Lambda_c \to n$ \cite{Meinel:2017ggx}, and $\Xi_c \to \Xi$ \cite{Zhang:2021oja}. Recently, we have also performed first lattice-QCD calculations of $\Lambda_b$ transition form factors to $J^P=\frac32^-$ and $J^P=\frac12^-$  baryons in the final state: $\Lambda_b \to \Lambda^*(1520)$ \cite{Meinel:2020owd} and $\Lambda_b \to \Lambda_c^*(2625)$, $\Lambda_b \to \Lambda_c^*(2595)$ \cite{Meinel:2021rbm}. These transitions provide further opportunities to test the Standard Model at the LHC \cite{Legger:2006cq,Hiller:2007ur,Boer:2018vpx,Descotes-Genon:2019dbw,Das:2020cpv,Amhis:2020phx,Albrecht:2020azd,Bernlochner:2021vlv}, and can also lead to new insights into the structure of the negative-parity baryons in the final states and heavy-quark effective theory \cite{Roberts:1992xm,Leibovich:1997az,Mannel:2015osa,Ikeno:2015xea,Boer:2018vpx,Cohen:2019zev,Nieves:2019kdh,Nieves:2019nol,Das:2020cpv,Bordone:2021bop,Papucci:2021pmj}.  In the following, we present first lattice-QCD results for the \emph{charm}-to-strange $\Lambda_c \to \Lambda^*(1520)$ form factors, while also improving our analysis of the $\Lambda_b \to \Lambda^*(1520)$ and $\Lambda_b \to \Lambda_c^*(2625)$, $\Lambda_b \to \Lambda_c^*(2595)$ form factors.
 
 Presently, the most precise measurements of absolute branching fractions of $\Lambda_c$ semileptonic decays come from the BESIII experiment, using $e^+e^-\to\Lambda_c\overline{\Lambda_c}$ production at threshold. Results are available for the exclusive semileptonic branching fractions to the lightest $\Lambda$ baryon \cite{Ablikim:2015prg, Ablikim:2016vqd}, and also for the inclusive semipositronic branching fraction \cite{Ablikim:2018woi}.
With future larger $e^+e^-\to\Lambda_c\overline{\Lambda_c}$ data sets from BESIII or other high-intensity $e^+e^-$ machines \cite{Ablikim:2019hff,Bondar:2013cja,Luo:2019xqt,Kou:2018nap}, and perhaps also with the LHCb experiment, it may be possible to observe the $p K^-$ invariant-mass distribution of $\Lambda_c\to p K^- \ell^+\nu_\ell$. This distribution is expected to be sensitive to several $\Lambda^*$ resonances \cite{Zyla:2020zbs}. Measurements, or lattice-QCD calculations, of the $\Lambda_c\to \Lambda^* \ell^+\nu_\ell$ branching fractions can provide new insights into the internal structure of these resonances \cite{Ikeno:2015xea}. Due to its narrow width, the $\Lambda^*(1520)$ with $J^P=\frac32^-$ is the most accessible for lattice QCD and likely also for the experiments. Lattice-QCD results for the $\Lambda_c\to\Lambda^*(1520)$ form factors may also further constrain $1/m_Q$ and $1/m_Q^2$ contributions in heavy-quark effective theory fits \cite{Bordone:2021bop} when combined with our previous results for $\Lambda_b\to\Lambda^*(1520)$ \cite{Meinel:2020owd}.
 
To avoid mixing with unwanted lighter states on the lattice, we found it necessary to set the spatial momentum of the $\Lambda^*(1520)$ baryon to zero, and determine the $q^2$-dependence of the form factors by varying the spatial momentum $\mathbf{p}$ of the initial-state $\Lambda_Q$ baryon instead \cite{Meinel:2020owd}. The four-momentum transfer squared is then equal to $q^2=m_{\Lambda_Q}^2-2E_{\Lambda_Q}m_{\Lambda^*}+m_{\Lambda^*}^2$ where $E_{\Lambda_Q}=\sqrt{m_{\Lambda_Q}^2+\mathbf{p}^2}$. In the case $Q=b$, the large mass of the $\Lambda_b$ has the effect that very large values of $\mathbf{p}$ are needed to appreciably move $q^2$ away from $q^2_{\rm max}=(m_{\Lambda_Q}-m_{\Lambda^*})^2$. In Ref.~\cite{Meinel:2020owd}, we performed the calculation for the two values $\mathbf{p}=(0,0,2)\frac{2\pi}{L}$ and $\mathbf{p}=(0,0,3)\frac{2\pi}{L}$, where $\frac{2\pi}{L}\approx0.47$ GeV for the spatial lattice size $L\approx 2.7$ fm, corresponding to $q^2/q^2_{\rm max}\approx 0.986$ and $q^2/q^2_{\rm max}\approx 0.969$, respectively. The situation is much more favorable for $Q=c$,  because $q^2_{\rm max}$ is much smaller and because the energy $E_{\Lambda_c}$ increases more rapidly with $\mathbf{p}$. Here we use the four different values $\mathbf{p}=(0,0,1)\frac{2\pi}{L}$, $\mathbf{p}=(0,1,1)\frac{2\pi}{L}$, $\mathbf{p}=(1,1,1)\frac{2\pi}{L}$, and $\mathbf{p}=(0,0,2)\frac{2\pi}{L}$, and these values are in fact sufficient to determine the shapes of the form factors in the full kinematic range relevant for the semileptonic decays $\Lambda_c\to\Lambda^*(1520)\ell^+\nu_\ell$, using only small extrapolations/interpolations. Consequently we are able to make Standard-Model predictions also for the fully integrated decay rates. These predictions and their implications are presented in an accompanying Letter \cite{Meinel:2021grq}.
 
We use helicity-based definitions of the $\frac12^+\to \frac12^-$ and $\frac12^+\to \frac32^-$ form factors \cite{Meinel:2020owd,Meinel:2021rbm}. It is known that helicity amplitudes, and hence helicity form factors, satisfy certain exact relations at the kinematic endpoint $q^2=q^2_{\rm max}$ that follow from rotational symmetry \cite{Zwicky:2013eda,Hiller:2013cza}. For the $\frac12^+\to \frac12^+$ form factors, such relations were found by relating the helicity-based and non-helicity-based (``Weinberg'') form factors in Refs.~\cite{Detmold:2015aaa,Detmold:2016pkz} and were already incorporated in the parametrizations used to fit the lattice results. When fitting our lattice QCD results for $\Lambda_b \to \Lambda^*(1520)$ and $\Lambda_b \to \Lambda_c^*(2595,2625)$ in Refs.~\cite{Meinel:2020owd,Meinel:2021rbm}, we did not impose any endpoint relations. Since then, we have found such relations (presented in Sec.~\ref{sec:endpointrelations}) also for the $\frac12^+\to \frac12^-$ and $\frac12^+\to \frac32^-$ cases by matching the helicity and non-helicity form factors, and they were proven rigorously in Ref.~\cite{Hiller:2021zth}. Our analysis of the $\Lambda_c\to\Lambda^*(1520)$ form factors (Sec.~\ref{sec:LcLstar}) imposes these endpoint relations at $q^2_{\rm max}$, as well as further exact relations at $q^2=0$. Given that the values of angular observables near $q^2_{\rm max}$ may be affected significantly by any small deviations from these relations, here we also provide updated fits of the lattice-QCD results for $\Lambda_b \to \Lambda^*(1520)$ (Sec.~\ref{sec:LbLstar}) and $\Lambda_b \to \Lambda_c^*(2595,2625)$ (Sec.~\ref{sec:LbLcstar}) in which we impose the constraints at $q^2_{\rm max}$. We also present the correspondingly updated Standard-Model predictions for $\Lambda_b \to \Lambda^*(1520)(\to pK^-)\mu^+\mu^-$ and $\Lambda_b \to \Lambda_c^*(2595,2625)\ell^-\bar{\nu}_\ell$.

\FloatBarrier
\section{\label{sec:endpointrelations}Endpoint relations for the helicity form factors}
\FloatBarrier

\FloatBarrier
\subsection{\texorpdfstring{$\bm{\frac12^+\to\frac12^-}$}{1/2+ to 1/2-}}
\FloatBarrier

Our definitions of the $\frac12^+\to\frac12^-$ helicity form factors $f_0^{(\frac12^-)}$, $f_+^{(\frac12^-)}$, $f_\perp^{(\frac12^-)}$, $g_0^{(\frac12^-)}$, $g_+^{(\frac12^-)}$, $g_\perp^{(\frac12^-)}$, $h_+^{(\frac12^-)}$, $h_\perp^{(\frac12^-)}$, $\widetilde{h}_+^{(\frac12^-)}$, $\widetilde{h}_\perp^{(\frac12^-)}$ can be found in Ref.~\cite{Meinel:2021rbm}. Suitable non-helicity-based definitions are given in Refs.~\cite{Leibovich:1997az,Pervin:2005ve,Mott:2011cx,Gutsche:2017wag,Gutsche:2018nks}, and can also be obtained from the ``Weinberg'' form factors discussed for $\frac12^+\to\frac12^+$ in Refs.~\cite{Detmold:2015aaa,Detmold:2016pkz} by inserting an extra $\gamma_5$. The relations between our helicity form factors and the form factors used in Refs.~\cite{Leibovich:1997az,Pervin:2005ve,Mott:2011cx,Gutsche:2017wag,Gutsche:2018nks} are given in the appendix of Ref.~\cite{Meinel:2021rbm}. Assuming that the non-helicity form factors remain finite for $q^2\to q^2_{\rm max}=(m_{\Lambda_Q}- m_{\Lambda_q^*})^2$, the vanishing of the variable $s_-$ that appears in the helicity form factors [$s_\pm = (m_{\Lambda_Q}\pm m_{\Lambda_q^*})^2 - q^2$] leads to the relations
\begin{eqnarray}
 f_\perp^{(\frac12^-)}(q^2_{\rm max})&=&f_+^{(\frac12^-)}(q^2_{\rm max}), \label{eq:J12qsqrmaxconstraintfirst} \\
 h_\perp^{(\frac12^-)}(q^2_{\rm max})&=&h_+^{(\frac12^-)}(q^2_{\rm max}), \label{eq:J12qsqrmaxconstraintlast}
\end{eqnarray}
which are also proven directly in Ref.~\cite{Hiller:2021zth}. These relations are analogous to those satisfied by the $\frac12^+\to\frac12^+$ form factors $g_\perp$, $g_+$, $\widetilde{h}_\perp$, $\widetilde{h}_+$  \cite{Detmold:2015aaa,Detmold:2016pkz}, with the role of the vector and axial-vector currents flipped. Even though we did not impose these relations in our fits of the lattice-QCD results for $\Lambda_b \to \Lambda_c^*(2595)$ in Ref.~\cite{Meinel:2021rbm}, it can be seen in Figs.~4 and 5 of Ref.~\cite{Meinel:2021rbm} that our numerical results are consistent with them within the uncertainties. Similarly, at $q^2=0$, the vector and axial-vector helicity form factors satisfy
\begin{eqnarray}
 f_0^{(\frac12^-)}(0)&=&f_+^{(\frac12^-)}(0), \\
 g_0^{(\frac12^-)}(0)&=&g_+^{(\frac12^-)}(0),
\end{eqnarray}
which is identical to the $\frac12^+\to\frac12^+$ case \cite{Feldmann:2011xf,Detmold:2016pkz}. An additional relation for the tensor form factors at $q^2=0$ follows from the identity $\sigma^{\mu\nu}\gamma_5 = \frac{i}{2} \epsilon^{\mu\nu\alpha\beta} \sigma_{\alpha\beta}$:
\begin{eqnarray}
 \widetilde{h}_\perp^{(\frac12^-)}(0)&=&h_\perp^{(\frac12^-)}(0).
\end{eqnarray}
This relation is in fact also satisfied by the $\frac12^+\to\frac12^+$ form factors $\widetilde{h}_\perp$ and $h_\perp$, although this was not previously noted in Refs.~\cite{Feldmann:2011xf,Detmold:2016pkz}.

\FloatBarrier
\subsection{\label{sec:endpointrelationsJ32}\texorpdfstring{$\bm{\frac12^+\to\frac32^-}$}{1/2+ to 3/2-}}
\FloatBarrier

Our definitions of the $\frac12^+\to\frac32^-$ helicity form factors $f_0^{(\frac32^-)}$, $f_+^{(\frac32^-)}$, $f_\perp^{(\frac32^-)}$, $f_{\perp^\prime}^{(\frac32^-)}$, $g_0^{(\frac32^-)}$, $g_+^{(\frac32^-)}$, $g_\perp^{(\frac32^-)}$, $g_{\perp^\prime}^{(\frac32^-)}$, $h_+^{(\frac32^-)}$, $h_\perp^{(\frac32^-)}$, $h_{\perp^\prime}^{(\frac32^-)}$, $\widetilde{h}_+^{(\frac32^-)}$, $\widetilde{h}_\perp^{(\frac32^-)}$, $\widetilde{h}_{\perp^\prime}^{(\frac32^-)}$ can be found in Refs.~\cite{Meinel:2020owd,Meinel:2021rbm}. For the vector and axial-vector form factors, suitable non-helicity-based definitions are given in Refs.~\cite{Leibovich:1997az,Pervin:2005ve,Mott:2011cx,Gutsche:2017wag,Gutsche:2018nks}, and the relations between our helicity form factors and the form factors used in Refs.~\cite{Leibovich:1997az,Pervin:2005ve,Mott:2011cx,Gutsche:2017wag,Gutsche:2018nks} are given in the appendices of Refs.~\cite{Meinel:2020owd,Meinel:2021rbm}. For the tensor form factors, it was noticed in Ref.~\cite{Papucci:2021pmj} that the form factor basis given in Ref.~\cite{Mott:2011cx} is incomplete, and that there are 7 relevant terms in the decomposition of the tensor-current matrix elements instead of just 6. To find the endpoint relations, we therefore use the basis of Ref.~\cite{Papucci:2021pmj}, with the matching to our form factors given in the appendix of Ref.~\cite{Papucci:2021pmj}. We find that using the complete non-helicity basis with 7 tensor form factors is crucial to obtain the correct (non-vanishing) behavior of the helicity tensor form factors at $q^2_{\rm max}$.
Again assuming that the non-helicity form factors remain finite for $q^2\to q^2_{\rm max}=(m_{\Lambda_Q}- m_{\Lambda_q^*})^2$, the vanishing of $s_-$ (or, equivalently, $w-1$, where $w=v\cdot v^\prime$) at that point implies the following relations among the helicity form factors:
\begin{eqnarray}
 f_\perp^{(\frac32^-)}(q^2_{\rm max}) + f_{\perp^\prime}^{(\frac32^-)}(q^2_{\rm max}) &=& 0, \label{eq:J32qsqrmaxconstraintfirst} \\
 2(m_{\Lambda_Q}-m_{\Lambda_{q,3/2}^*})\: f_\perp^{(\frac32^-)}(q^2_{\rm max}) + (m_{\Lambda_Q}+m_{\Lambda_{q,3/2}^*})\: f_+^{(\frac32^-)}(q^2_{\rm max})  &=&0, \\
 g_\perp^{(\frac32^-)}(q^2_{\rm max})-g_{\perp^\prime}^{(\frac32^-)}(q^2_{\rm max})-g_+^{(\frac32^-)}(q^2_{\rm max}) &=& 0, \\
g_0^{(\frac32^-)}(q^2_{\rm max})  &=&0 , \\
 h_\perp^{(\frac32^-)}(q^2_{\rm max}) + h_{\perp^\prime}^{(\frac32^-)}(q^2_{\rm max}) &=& 0, \\
2(m_{\Lambda_Q} + m_{\Lambda_{q,3/2}^*})\: h_\perp^{(\frac32^-)}(q^2_{\rm max})  + (m_{\Lambda_Q}-m_{\Lambda_{q,3/2}^*})\: h_+^{(\frac32^-)}(q^2_{\rm max})  &=& 0, \\
 \widetilde{h}_\perp^{(\frac32^-)}(q^2_{\rm max}) - \widetilde{h}_{\perp^\prime}^{(\frac32^-)}(q^2_{\rm max}) - \widetilde{h}_+^{(\frac32^-)}(q^2_{\rm max}) &=& 0. \label{eq:J32qsqrmaxconstraintlast}
\end{eqnarray}
These relations are proven directly in Ref.~\cite{Hiller:2021zth}. We did not impose any of these relations in our fits of the lattice-QCD results for the  $\Lambda_b \to \Lambda^*(1520)$ and $\Lambda_b \to \Lambda_c^*(2625)$ form factors in Refs.~\cite{Meinel:2020owd,Meinel:2021rbm} but we again find that the numerical results are consistent with them within $2\sigma$ or better.

At $q^2=0$, the $\frac12^+\to\frac32^-$ helicity form factors satisfy the relations \cite{Descotes-Genon:2019dbw}
\begin{eqnarray}
  f_0^{(\frac32^-)}(0) &=& \frac{(m_{\Lambda_Q}+ m_{\Lambda_{q,3/2}^*})^2}{(m_{\Lambda_Q}- m_{\Lambda_{q,3/2}^*})^2} \:f_+^{(\frac32^-)}(0) , \label{eq:J32qsqr0constraintfirst} \\
  g_0^{(\frac32^-)}(0) &=& \frac{(m_{\Lambda_Q}- m_{\Lambda_{q,3/2}^*})^2}{(m_{\Lambda_Q}+ m_{\Lambda_{q,3/2}^*})^2}  \: g_+^{(\frac32^-)}(0) , \\
  \widetilde{h}_\perp^{(\frac32^-)}(0) &=& \frac{(m_{\Lambda_Q}+ m_{\Lambda_{q,3/2}^*})^2}{(m_{\Lambda_Q}- m_{\Lambda_{q,3/2}^*})^2} \: h_\perp^{(\frac32^-)}(0), \\
  \widetilde{h}_{\perp^\prime}^{(\frac32^-)}(0) &=& -\frac{(m_{\Lambda_Q}+m_{\Lambda_{q,3/2}^*})^2}{(m_{\Lambda_Q}-m_{\Lambda_{q,3/2}^*})^2} \: h_{\perp^\prime}^{(\frac32^-)}(0). \label{eq:J32qsqr0constraintlast}
\end{eqnarray}
As already mentioned, our lattice results for $\Lambda_b \to \Lambda^*(1520)$ and $\Lambda_b \to \Lambda_c^*(2595,2625)$  are limited to small kinematic regions near $q^2_{\rm max}$, so the relations at $q^2=0$ are not applicable. For  $\Lambda_c \to \Lambda^*(1520)$, however, our lattice results cover nearly the full kinematic range and we impose the endpoint relations at both $q^2=q^2_{\rm max}$ and $q^2=0$ in the following.

\FloatBarrier
\section{\label{sec:LcLstar}\texorpdfstring{$\bm{\Lambda_c \to \Lambda^*(1520)}$}{Lambdac to Lambda*(1520)} form factors}
\FloatBarrier

\FloatBarrier
\subsection{\label{sec:latticeparams}Lattice parameters and extraction of the form factors}
\FloatBarrier

Our $\Lambda_c \to \Lambda^*(1520)$ lattice calculation closely follows the one for $\Lambda_b \to \Lambda^*(1520)$ \cite{Meinel:2020owd}, and uses gauge-field configurations generated by the RBC and UKQCD Collaborations \cite{Aoki:2010dy, Blum:2014tka} with $2+1$ flavors of domain-wall fermions. For the $u$, $d$, $s$ valence quarks, we use the same Shamir domain-wall action with the same $N_5=16$ and $a M_5=1.8$ as used for the $u$, $d$, $s$ sea quarks \cite{Aoki:2010dy, Blum:2014tka}. The valence $u,d$ masses are set equal to the sea $u,d$ masses, while the valence $s$ masses are set equal to the physical values as determined with sub-MeV precision in Ref.~\cite{Blum:2014tka}. There, $m_\pi$, $m_K$, and $m_\Omega$ were used to determine the light and strange quark masses and the lattice spacing. The main parameters of the three ensembles and of the quark propagators computed thereon are given in Table \ref{tab:latticeparams}. For the valence charm quark, we use an anisotropic clover action with the mass $a m_Q^{(c)}$, anisotropy parameter $\nu^{(c)}$, and clover coefficients $c_E^{(c)}=c_B^{(c)}$ tuned nonperturbatively such that the $D_s$-meson rest mass, kinetic mass, and hyperfine splitting match the experimental values \cite{Zyla:2020zbs}. These observables calculated on each ensemble are found to agree with experiment within 0.4\%, 1.0\%, and 1.4\% (or better) precision, respectively. The $c\to s$ currents are renormalized using the mostly nonperturbative method described in Refs.~\cite{Hashimoto:1999yp, ElKhadra:2001rv}. That is, the renormalized currents are written as
\begin{equation}
 J_\Gamma=\rho_\Gamma\sqrt{Z_V^{(ss)} Z_V^{(cc)}} \left[ \bar{s}\: \Gamma\: c + a\, d_1^{(c)}\,\bar{s}\: \Gamma\: \bs{\gamma}_{\rm E}\cdot\bs{\nabla}  c \right], \label{eq:improvedcurrent}
\end{equation}
where $Z_V^{(ss)}$ and $Z_V^{(cc)}$ are the matching factors of the temporal components of the $s\to s$ and $c\to c$ vector currents, determined nonperturbatively using charge conservation, $\rho_\Gamma$ are the residual matching factors that are numerically close to 1 and are computed using one-loop lattice perturbation theory, and the term with coefficient $d_1^{(c)}$ removes $\mathcal{O}(a)$ discretization errors at tree level. The values of these parameters are given in Table \ref{tab:matching}. The residual matching factors for the vector and axial-vector currents were computed by C.~Lehner at one loop in mean-field-improved lattice perturbation theory, originally for Ref.~\cite{Meinel:2016dqj}. The perturbative calculation was performed for a slightly different tuning of the charm-action parameters, and we therefore assign a larger systematic uncertainty to the residual matching factors, as discussed in Sec.~\ref{sec:LcL1520extrap}. Here we also determine the $\Lambda_c \to \Lambda^*(1520)$ tensor form factors for completeness, even though they are not needed to describe $\Lambda_c \to \Lambda^*(1520)\ell^+\nu_\ell$ in the Standard-Model. One-loop results are not available for the tensor-current residual matching factors and we set them equal to $1\pm 0.05$ (this estimate should be viewed as corresponding to a renormalization scale $\mu=m_c$). As in Refs.~\cite{Meinel:2020owd,Meinel:2021rbm}, in our estimates of systematic uncertainties we will also account for the incomplete (tree-level only) $\mathcal{O}(a)$ improvement of the currents.

\begin{figure}
 \includegraphics[width=0.6\linewidth]{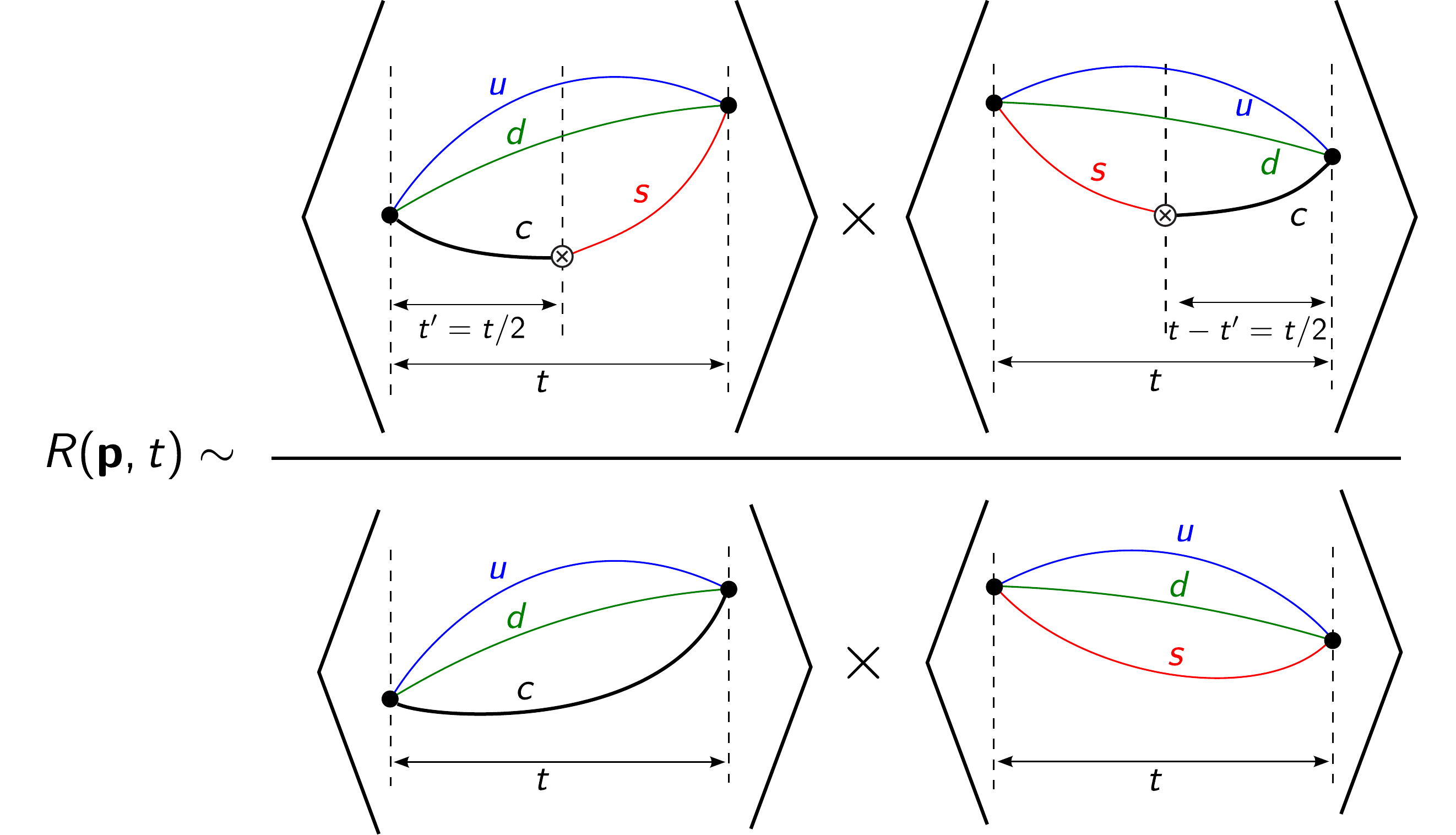}
 \caption{\label{fig:ratioillustration}Illustration of the ratio of products of three-point and two-point functions used in $R_{\perp^\prime}^{(\frac32^-)V}(\mathbf{p}, t)$, showing the relevant Euclidean time separations. The crossed circles denote the weak currents in the three-point functions. The detailed definitions are given in Ref.~\cite{Meinel:2020owd}. }
\end{figure}

\begin{figure}
\flushleft

\includegraphics[width=0.245\linewidth,valign=t]{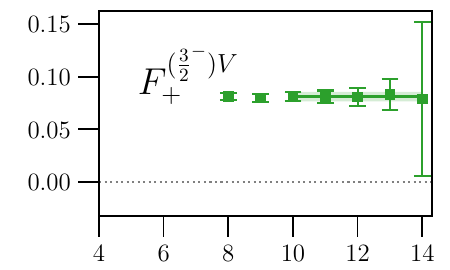} \includegraphics[width=0.245\linewidth,valign=t]{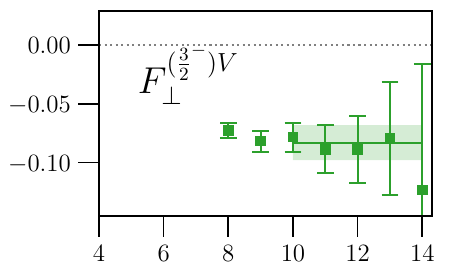} \includegraphics[width=0.245\linewidth,valign=t]{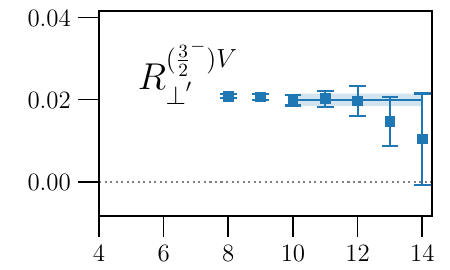} \includegraphics[width=0.245\linewidth,valign=t]{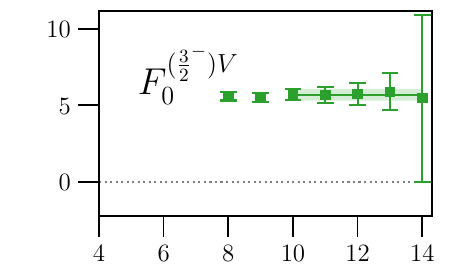}

\includegraphics[width=0.245\linewidth,valign=t]{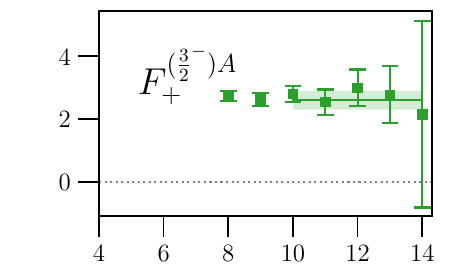} \includegraphics[width=0.245\linewidth,valign=t]{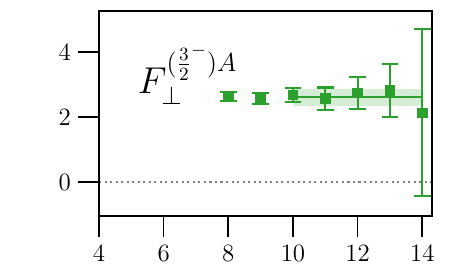} \includegraphics[width=0.245\linewidth,valign=t]{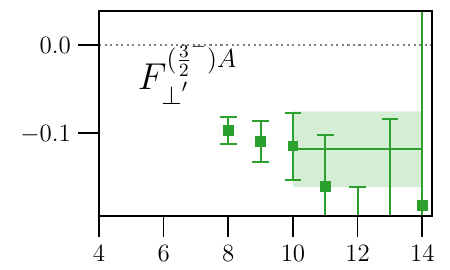} \includegraphics[width=0.245\linewidth,valign=t]{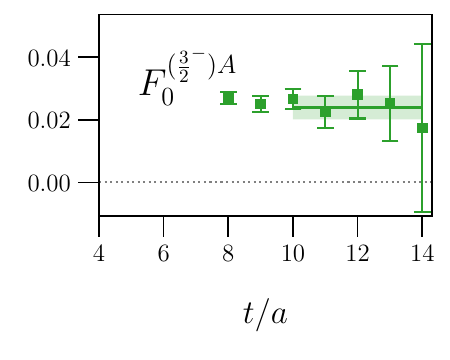}

\vspace{-6.0ex}

\includegraphics[width=0.245\linewidth,valign=t]{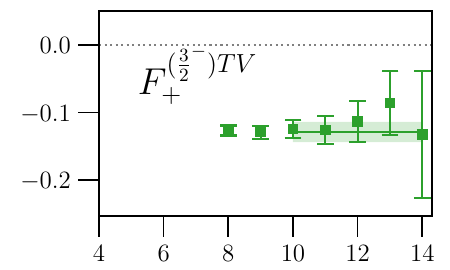} \includegraphics[width=0.245\linewidth,valign=t]{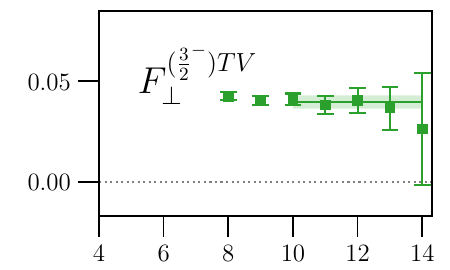} \includegraphics[width=0.245\linewidth,valign=t]{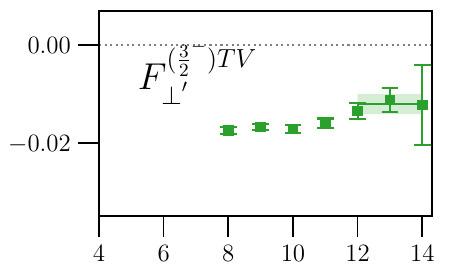}

\includegraphics[width=0.245\linewidth,valign=t]{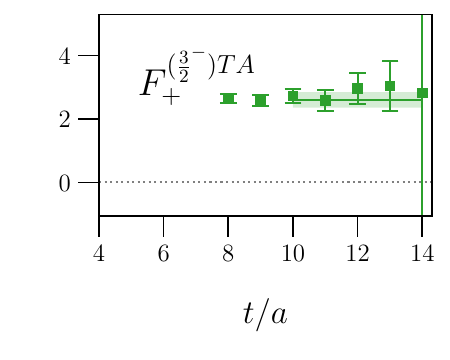} \includegraphics[width=0.245\linewidth,valign=t]{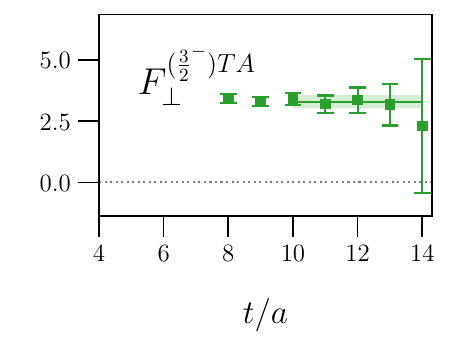} \includegraphics[width=0.245\linewidth,valign=t]{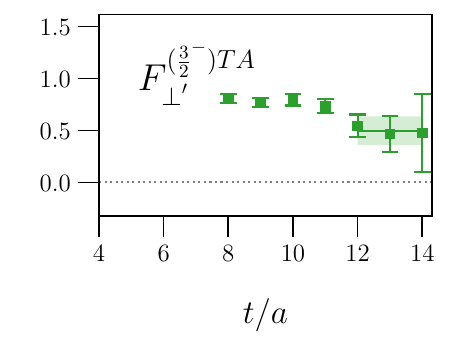}

\vspace{-1ex}
 \caption{\label{fig:ratios}Numerical results for the $\Lambda_c \to \Lambda^*(1520)$ quantities $F^{(\frac32^-)X}_\lambda(\mathbf{p},t)$, defined in Eq.~(\protect\ref{eq:FX}), as a function of the source-sink separation, for $\mathbf{p}=(0,0,1)\frac{2\pi}{L}$ and for the C005 ensemble. Also shown is $R_{\perp^\prime}^{(\frac32^-)V}(\mathbf{p}, t)$, which is used to extract the square of $f_{\perp^\prime}^{(\frac32^-)}$. The horizontal lines indicate the ranges and extracted values of constant fits.}
\end{figure}

We computed the $\Lambda_c \to \Lambda^*(1520)$ three-point functions for four different $\Lambda_c$ momenta,
\begin{equation}
 \frac{L}{2\pi}\mathbf{p}=(0,0,1),(0,1,1),(1,1,1),(0,0,2).
\end{equation}
In addition, the range of source-sink separations used here is shifted to slightly larger values compared to Ref.~\cite{Meinel:2020owd}, taking advantage of the improved signal-to-noise ratio in the charm case:
\begin{equation}
 t/a=8,9,...,14\:\:\text{ (C01 and C005 data sets), }\hspace{6ex}t/a=10,11,...,16\:\:\text{ (F004 data set)}.
\end{equation}
The extraction of the form factors from the three-point and two-point correlation functions was performed as in Ref.~\cite{Meinel:2020owd} by computing the quantities
\begin{align}
F^{(\frac32^-)X}_\lambda(\mathbf{p},t)=\frac{S^{(\frac32^-)X}_\lambda(\mathbf{p},t,t/2)}{S^{(\frac32^-)V}_{\perp^\prime}(\mathbf{p},t,t/2)}\sqrt{R_{\perp^\prime}^{(\frac32^-)V}(\mathbf{p})}, \label{eq:FX}
\end{align}
where $X\in\{V, A, TV, TA\}$ and  $\lambda \in \{0, +, \perp, \perp^\prime\}$. Above, $R_{\perp^\prime}^{(\frac32^-)V}(\mathbf{p})$ denotes the result of a constant fit to $R_{\perp^\prime}^{(\frac32^-)V}(\mathbf{p}, t)$ in the plateau region, where $R_{\perp^\prime}^V(\mathbf{p}, t)$ is a ratio of products of three-point and two-point functions that becomes equal to $f_{\perp^\prime}^2$ at large $t$, and is illustrated in Fig.~\ref{fig:ratioillustration}. The objects $S^{(\frac32^-)X}_\lambda(\mathbf{p},t,t/2)$ are linear projections of the three-point functions, each proportional to the helicity form factor corresponding to $(X,\lambda)$, normalized such that all unwanted factors cancel in Eq.~(\ref{eq:FX}) for large $t$. The individual form factors were then obtained from constant fits to Eq.~(\ref{eq:FX}) in the plateau regions. Example numerical results for $R_{\perp^\prime}^{(\frac32^-)V}(\mathbf{p}, t)$ and $F^{(\frac32^-)X}_\lambda(\mathbf{p},t)$ are shown in Fig.~\ref{fig:ratios}, and all fit results are listed in Table \ref{tab:FFvalues}.

\begin{table}
 \begin{tabular}{lccccccccccccc}
\hline\hline
Label & $N_s^3\times N_t$ & $\beta$  & $a$ [fm] & $2\pi/L$ [GeV] &  $am_{u,d}$ &  $m_\pi$ [GeV] & $am_{s}^{(\mathrm{sea})}$ 
& $am_{s}^{(\mathrm{val})}$ & $\wm a m_Q^{(c)}$ & $\nu^{(c)}$ & $c_{E,B}^{(c)}$  & $N_{\rm ex}$ & $N_{\rm sl}$ \\
\hline
C01  & $24^3\times64$ & $2.13$   & $0.1106(3)$ & $0.4673(13)$ & $0.01\nb$ & $0.4312(13)$ & $0.04$      & $0.0323$  & $\wm0.1541$ & $1.2004$ & $1.8407$  & 142 & 4544  \\
C005 & $24^3\times64$ & $2.13$   & $0.1106(3)$ & $0.4673(13)$ & $0.005$   & $0.3400(11)$ & $0.04$      & $0.0323$  & $\wm0.1541$ & $1.2004$ & $1.8407$  & 311 & 9952  \\
F004 & $32^3\times64$ & $2.25$   & $0.0828(3)$ & $0.4680(17)$ & $0.004$   & $0.3030(12)$ & $0.03$      & $0.0248$  & $-0.0517$   & $1.1021$ & $1.4483$  & 188 & 6016  \\
\hline\hline
\end{tabular}
\caption{\label{tab:latticeparams}Parameters of the three data sets used to determine the $\Lambda_c \to \Lambda^*(1520)$ form factors. The ensemble generation is described in Ref.~\cite{Aoki:2010dy} and the lattice spacings were determined in Ref.~\cite{Blum:2014tka}. Above, $L=N_s a$ is the spatial lattice size and we provide the values of the momentum unit, $2\pi/L$, for convenience. The parameters $a m_Q^{(c)}$, $\nu^{(c)}$, $c_E^{(c)}= c_B^{(c)}$ are the mass, anisotropy parameter, and chromoelectric/chromomagnetic clover coefficients in the anisotropic clover action used for the charm quark \cite{Meinel:2021rbm}. We use all-mode averaging \cite{Blum:2012uh,Shintani:2014vja} with 32 sloppy and 1 exact sample per gauge configuration; $N_{\rm ex}$ and $N_{\rm sl}$ are the total numbers of exact and sloppy samples, respectively.}
\end{table}

\begin{table}
\begin{center}
\small
\begin{tabular}{cllll}
\hline\hline
Parameter          & \hspace{2ex} &  Coarse lattice    & \hspace{2ex} &   Fine lattice       \\
\hline
 $Z_V^{(cc)}$                && $1.35761(16)$      &&  $1.160978(74)$   \\[0.2ex]
 $Z_V^{(ss)}$                && $0.71273(26)$      &&  $0.7440(18)$     \\[0.2ex]
 $\rho_{V^0}=\rho_{A^0}$     && $1.00274(49)$      &&  $1.001949(85)$   \\[0.2ex]
 $\rho_{V^j}=\rho_{A^j}$     && $0.99475(62)$      &&  $0.99675(68)$    \\[0.2ex]
 $d_1^{(c)}$                 && $0.0412$           &&  $0.0301$         \\[0.2ex]
\hline\hline
\end{tabular}
\caption{\label{tab:matching}Matching and $\mathcal{O}(a)$-improvement factors for the $c\to s$ currents. The parameters are defined and explained in Eq.~(\ref{eq:improvedcurrent}) and the text below.}
\end{center}
\end{table}

\begin{table}
\scriptsize
 \begin{tabular}{cccllllll}
  \hline\hline
  Form factor                & & $|\mathbf{p}|/(2\pi/L)$ & & \hspace{2ex}C01 & & \hspace{2ex}C005 & & \hspace{2ex}F004 \\
  \hline
  $f_+^{(\frac32^-)}$                             &&   $\sqrt{1}$   &&   $\wm  0.0796(51)$   &&   $\wm  0.0811(45)$   &&   $\wm  0.0782(40)$\\
                                    &&   $\sqrt{2}$   &&   $\wm  0.1123(80)$   &&   $\wm  0.1066(70)$   &&   $\wm  0.1063(61)$\\
                                    &&   $\sqrt{3}$   &&   $\wm   0.142(10)$   &&   $\wm  0.1279(89)$   &&   $\wm  0.1231(70)$\\
                                    &&   $\sqrt{4}$   &&   $\wm   0.149(13)$   &&   $\wm  0.1439(89)$   &&   $\wm  0.1417(73)$\\
  $f_0^{(\frac32^-)}$                             &&   $\sqrt{1}$   &&   $\wm    6.32(48)$   &&   $\wm    5.68(38)$   &&   $\wm    5.53(36)$\\
                                    &&   $\sqrt{2}$   &&   $\wm    6.11(45)$   &&   $\wm    5.18(35)$   &&   $\wm    5.30(33)$\\
                                    &&   $\sqrt{3}$   &&   $\wm    5.93(44)$   &&   $\wm    4.86(35)$   &&   $\wm    4.83(29)$\\
                                    &&   $\sqrt{4}$   &&   $\wm    5.18(48)$   &&   $\wm    4.53(29)$   &&   $\wm    4.63(27)$\\
  $f_{\perp}^{(\frac32^-)}$                       &&   $\sqrt{1}$   &&   $   -  0.079(15)$   &&   $   -  0.083(15)$   &&   $   -  0.105(15)$\\
                                    &&   $\sqrt{2}$   &&   $   -  0.006(16)$   &&   $   -  0.012(17)$   &&   $   -  0.026(17)$\\
                                    &&   $\sqrt{3}$   &&   $\wm   0.046(20)$   &&   $\wm   0.028(17)$   &&   $\wm   0.016(17)$\\
                                    &&   $\sqrt{4}$   &&   $\wm   0.092(19)$   &&   $\wm   0.068(18)$   &&   $\wm   0.047(17)$\\
  $f_{\perp^{\prime}}^{(\frac32^-)}$              &&   $\sqrt{1}$   &&   $\wm  0.1421(73)$   &&   $\wm  0.1411(54)$   &&   $\wm  0.1540(53)$\\
                                    &&   $\sqrt{2}$   &&   $\wm  0.1308(83)$   &&   $\wm  0.1278(63)$   &&   $\wm  0.1437(55)$\\
                                    &&   $\sqrt{3}$   &&   $\wm  0.1305(85)$   &&   $\wm  0.1143(70)$   &&   $\wm  0.1357(62)$\\
                                    &&   $\sqrt{4}$   &&   $\wm   0.116(11)$   &&   $\wm  0.1169(66)$   &&   $\wm  0.1313(58)$\\
  $g_+^{(\frac32^-)}$                             &&   $\sqrt{1}$   &&   $\wm    3.03(33)$   &&   $\wm    2.60(30)$   &&   $\wm    2.50(30)$\\
                                    &&   $\sqrt{2}$   &&   $\wm    3.13(29)$   &&   $\wm    2.70(26)$   &&   $\wm    2.53(26)$\\
                                    &&   $\sqrt{3}$   &&   $\wm    2.83(28)$   &&   $\wm    2.24(24)$   &&   $\wm    2.12(23)$\\
                                    &&   $\sqrt{4}$   &&   $\wm    2.42(28)$   &&   $\wm    2.09(21)$   &&   $\wm    2.06(21)$\\
  $g_0^{(\frac32^-)}$                             &&   $\sqrt{1}$   &&   $\wm  0.0265(39)$   &&   $\wm  0.0240(38)$   &&   $\wm  0.0228(38)$\\
                                    &&   $\sqrt{2}$   &&   $\wm  0.0547(52)$   &&   $\wm  0.0503(53)$   &&   $\wm  0.0470(51)$\\
                                    &&   $\sqrt{3}$   &&   $\wm  0.0718(68)$   &&   $\wm  0.0598(63)$   &&   $\wm  0.0572(58)$\\
                                    &&   $\sqrt{4}$   &&   $\wm  0.0786(80)$   &&   $\wm  0.0726(67)$   &&   $\wm  0.0732(62)$\\
  $g_{\perp}^{(\frac32^-)}$                       &&   $\sqrt{1}$   &&   $\wm    2.98(30)$   &&   $\wm    2.60(26)$   &&   $\wm    2.54(24)$\\
                                    &&   $\sqrt{2}$   &&   $\wm    3.03(25)$   &&   $\wm    2.51(21)$   &&   $\wm    2.47(20)$\\
                                    &&   $\sqrt{3}$   &&   $\wm    2.92(23)$   &&   $\wm    2.30(19)$   &&   $\wm    2.26(17)$\\
                                    &&   $\sqrt{4}$   &&   $\wm    2.47(23)$   &&   $\wm    2.15(16)$   &&   $\wm    2.16(14)$\\
  $g_{\perp^{\prime}}^{(\frac32^-)}$              &&   $\sqrt{1}$   &&   $   -  0.106(49)$   &&   $   -  0.118(43)$   &&   $   -  0.100(40)$\\
                                    &&   $\sqrt{2}$   &&   $   -  0.053(36)$   &&   $   -  0.059(33)$   &&   $   -  0.070(29)$\\
                                    &&   $\sqrt{3}$   &&   $   -  0.112(33)$   &&   $   -  0.053(29)$   &&   $   -  0.073(26)$\\
                                    &&   $\sqrt{4}$   &&   $   -  0.089(28)$   &&   $   -  0.096(25)$   &&   $   -  0.075(24)$\\
  $h_+^{(\frac32^-)}$                             &&   $\sqrt{1}$   &&   $   -  0.138(16)$   &&   $   -  0.129(15)$   &&   $   -  0.166(14)$\\
                                    &&   $\sqrt{2}$   &&   $   -  0.055(15)$   &&   $   -  0.045(17)$   &&   $   -  0.085(16)$\\
                                    &&   $\sqrt{3}$   &&   $   -  0.017(18)$   &&   $   -  0.011(17)$   &&   $   -  0.044(17)$\\
                                    &&   $\sqrt{4}$   &&   $\wm   0.021(18)$   &&   $\wm   0.015(18)$   &&   $   -  0.026(16)$\\
  $h_{\perp}^{(\frac32^-)}$                       &&   $\sqrt{1}$   &&   $\wm  0.0419(35)$   &&   $\wm  0.0396(33)$   &&   $\wm  0.0401(30)$\\
                                    &&   $\sqrt{2}$   &&   $\wm  0.0686(52)$   &&   $\wm  0.0637(48)$   &&   $\wm  0.0628(46)$\\
                                    &&   $\sqrt{3}$   &&   $\wm  0.0887(68)$   &&   $\wm  0.0772(60)$   &&   $\wm  0.0769(53)$\\
                                    &&   $\sqrt{4}$   &&   $\wm  0.0968(87)$   &&   $\wm  0.0875(62)$   &&   $\wm  0.0900(55)$\\
  $h_{\perp^{\prime}}^{(\frac32^-)}$              &&   $\sqrt{1}$   &&   $   - 0.0108(18)$   &&   $   - 0.0121(20)$   &&   $   - 0.0178(15)$\\
                                    &&   $\sqrt{2}$   &&   $   - 0.0102(25)$   &&   $   - 0.0129(27)$   &&   $   - 0.0202(22)$\\
                                    &&   $\sqrt{3}$   &&   $   - 0.0102(43)$   &&   $   - 0.0145(34)$   &&   $   - 0.0200(23)$\\
                                    &&   $\sqrt{4}$   &&   $   - 0.0078(33)$   &&   $   - 0.0112(34)$   &&   $   - 0.0224(28)$\\
  $\widetilde{h}_+^{(\frac32^-)}$                 &&   $\sqrt{1}$   &&   $\wm    2.97(29)$   &&   $\wm    2.60(25)$   &&   $\wm    2.60(24)$\\
                                    &&   $\sqrt{2}$   &&   $\wm    2.97(24)$   &&   $\wm    2.52(21)$   &&   $\wm    2.45(18)$\\
                                    &&   $\sqrt{3}$   &&   $\wm    2.85(22)$   &&   $\wm    2.25(19)$   &&   $\wm    2.12(16)$\\
                                    &&   $\sqrt{4}$   &&   $\wm    2.46(23)$   &&   $\wm    2.18(15)$   &&   $\wm    2.22(14)$\\
  $\widetilde{h}_{\perp}^{(\frac32^-)}$           &&   $\sqrt{1}$   &&   $\wm    3.88(33)$   &&   $\wm    3.30(27)$   &&   $\wm    3.32(25)$\\
                                    &&   $\sqrt{2}$   &&   $\wm    3.83(29)$   &&   $\wm    3.23(24)$   &&   $\wm    3.23(23)$\\
                                    &&   $\sqrt{3}$   &&   $\wm    3.61(28)$   &&   $\wm    2.92(23)$   &&   $\wm    2.95(21)$\\
                                    &&   $\sqrt{4}$   &&   $\wm    3.18(31)$   &&   $\wm    2.64(20)$   &&   $\wm    2.75(19)$\\
  $\widetilde{h}_{\perp^{\prime}}^{(\frac32^-)}$  &&   $\sqrt{1}$   &&   $\wm    0.56(14)$   &&   $\wm    0.50(14)$   &&   $\wm    0.90(11)$\\
                                    &&   $\sqrt{2}$   &&   $\wm    0.50(13)$   &&   $\wm    0.54(12)$   &&   $\wm    0.93(13)$\\
                                    &&   $\sqrt{3}$   &&   $\wm    0.49(19)$   &&   $\wm    0.57(13)$   &&   $\wm    0.84(11)$\\
                                    &&   $\sqrt{4}$   &&   $\wm    0.38(13)$   &&   $\wm    0.43(11)$   &&   $\wm    0.86(11)$\\
  \hline\hline
 \end{tabular}
 \caption{\label{tab:FFvalues} The values of the $\Lambda_c\to\Lambda^*(1520)$ form factors on the lattice extracted for each momentum and each data set.}
\end{table}

\FloatBarrier
\subsection{\label{sec:LcL1520extrap}Chiral and continuum extrapolations}
\FloatBarrier

The final step in determining the physical $\Lambda_c\to\Lambda^*(1520)$ form factors is to fit suitable parametrizations describing the dependence on the momentum transfer, the lattice spacing, and the light-quark mass (or, equivalently $m_\pi^2$) to the form factor data points shown in Table \ref{tab:FFvalues}. Because we will impose the constraints discussed in Sec.~\ref{sec:endpointrelationsJ32}, which relate different form factors, we perform global, fully correlated fits to all form factors: one ``nominal'' fit, and one ``higher-order'' fit that will be used to estimate systematic uncertainties.

As in Refs.~\cite{Meinel:2020owd,Meinel:2021rbm}, we fit the shapes of the form factors using power series in the dimensionless variable $(w-1)$, where
\begin{equation}
 w=v\cdot v^\prime=(m_{\Lambda_c}^2+m_{\Lambda^*}^2-q^2)/(2m_{\Lambda_c}m_{\Lambda^*})
\end{equation}
such that $w=1$ corresponds to $q^2=q^2_{\rm max}=(m_{\Lambda_c}-m_{\Lambda^*})^2$. This expansion is expected to converge for $|w-1|$ smaller than $|w_s-1|$, where $w_s$ denotes the position of the branch point or pole that is closest to $w=1$. Such singularities arise from on-shell intermediate states with four-momentum $q$ produced by the $\bar{s}\Gamma c$ weak current. The $D$-$K$ two-particle branch cut (in infinite volume) starts at $q^2=(m_D+m_K)^2$ which, for physical hadron masses, corresponds to $|w-1|\approx 0.72$. The exact isospin symmetry in our calculation forbids $D_s$-$\pi$ intermediate states. The three-particle $D_s$-$\pi$-$\pi$ branch cut starts at at $q^2=(m_{D_s}+2m_\pi)^2$ corresponding to $|w-1|\approx 0.64$. In addition, single-particle intermediate states result in poles at $q^2$ equal to the masses of these particles squared. The experimentally observed masses \cite{Zyla:2020zbs} of the lightest $\bar{c}s$ mesons with the $J^P$ quantum numbers occurring in the different form factors are given in Table \ref{tab:polemassesLcLstar}. The lowest-lying single-particle state is the pseudoscalar $D_s$ meson (which contributes a pole to the form factor $g_0^{(\frac32^-)}$), corresponding to $|w-1|\approx 0.47$. There are no closer singularities in any of the form factors. The region of interest for the semileptonic decay is $1\leq w \leq (m_{\Lambda_c}^2+m_{\Lambda^*}^2)/(2m_{\Lambda_c}m_{\Lambda^*})$, which corresponds to $|w-1|\leq 0.085$. Thus, the series is expected to converge in the entire region of interest (using the lattice hadron masses instead of the experimental masses changes the numerical values slightly but does not affect this conclusion).

Because we now have data for four different $\Lambda_c$ momenta, we are able to go beyond the first order in the expansion in $(w-1)$; we find that second order is sufficient. Furthermore, we choose to factor out the lowest-lying poles from the single-particle states. While this is not necessary for convergence, it may make the convergence slightly more rapid. The pole masses are set to the physical values listed in Table \ref{tab:polemassesLcLstar}. In the nominal fit, each form factor $f$ is parametrized as
\begin{equation}
 f(q^2) = \frac{1}{1-q^2/(m_{\rm pole}^f)^2} \sum_{n=0}^{2} a_n^f L_n^f \:(w-1)^n, \label{eq:LcLparam}
\end{equation}
where the factors
\begin{equation}
 L_n^f = \left[1+C_n^{f}\frac{m_{\pi}^2-m_{\pi,\rm phys}^2}{(4\pi f_{\pi})^2}+D_n^{f}a^2\Lambda^2\right]
\end{equation}
describe the dependence on the pion mass and lattice spacing (we set $f_{\pi}=132\,\text{MeV}$, $\Lambda=300\,\text{MeV}$). This functional form corresponds to the lowest nontrivial order in an expansion in the light-quark mass $m_{u,d}\propto m_\pi^2$ and the lattice spacing (as discussed at the beginning of Sec.~\ref{sec:latticeparams}, the strange and charm quark masses are already tuned accurately to their physical values on each ensemble, requiring no extrapolation).  The use of the chirally symmetric domain-wall action for the light and strange quarks, and of a nonperturbatively tuned relativistic-heavy-quark action \cite{Aoki:2012xaa} for the charm quark ensure the absence of $\mathcal{O}(a)$ discretization errors, except for the effects of the incomplete improvement of the $c\to s$ current. Systematic uncertainties from this incomplete improvement and from neglected higher-order or nonanalytic terms are estimated by varying the fit form, as discussed later in this section.

\begin{table}
\begin{tabular}{ccccc}
\hline\hline
 $f$     & \hspace{1ex} & $J^P$ & \hspace{1ex}  & $m_{\rm pole}^f$ [GeV]  \\
\hline
$f_+^{(\frac32^-)}$, $f_\perp^{(\frac32^-)}$, $f_{\perp^\prime}^{(\frac32^-)}$, $h_+^{(\frac32^-)}$, $h_\perp^{(\frac32^-)}$, $h_{\perp^\prime}^{(\frac32^-)}$                         && $1^-$   && $2.112$  \\
$f_0^{(\frac32^-)}$                                                      && $0^+$   && $2.318$  \\
$g_+^{(\frac32^-)}$, $g_\perp^{(\frac32^-)}$, $g_{\perp^\prime}^{(\frac32^-)}$, $\widetilde{h}_+^{(\frac32^-)}$, $\widetilde{h}_\perp^{(\frac32^-)}$, $\widetilde{h}_{\perp^\prime}^{(\frac32^-)}$ && $1^+$   && $2.460$  \\
$g_0^{(\frac32^-)}$                                                      && $0^-$   && $1.968$  \\
\hline\hline
\end{tabular}
\caption{\label{tab:polemassesLcLstar}Masses of the $D_s$ mesons \cite{Zyla:2020zbs} used in the pole factors of the $\Lambda_c\to\Lambda^*(1520)$ form-factor parametrizations.}
\end{table}

\begin{figure*}
 
\includegraphics[width=0.6\linewidth]{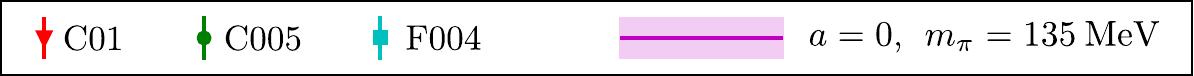}
 
 \vspace{2ex}

 \includegraphics[width=0.43\linewidth]{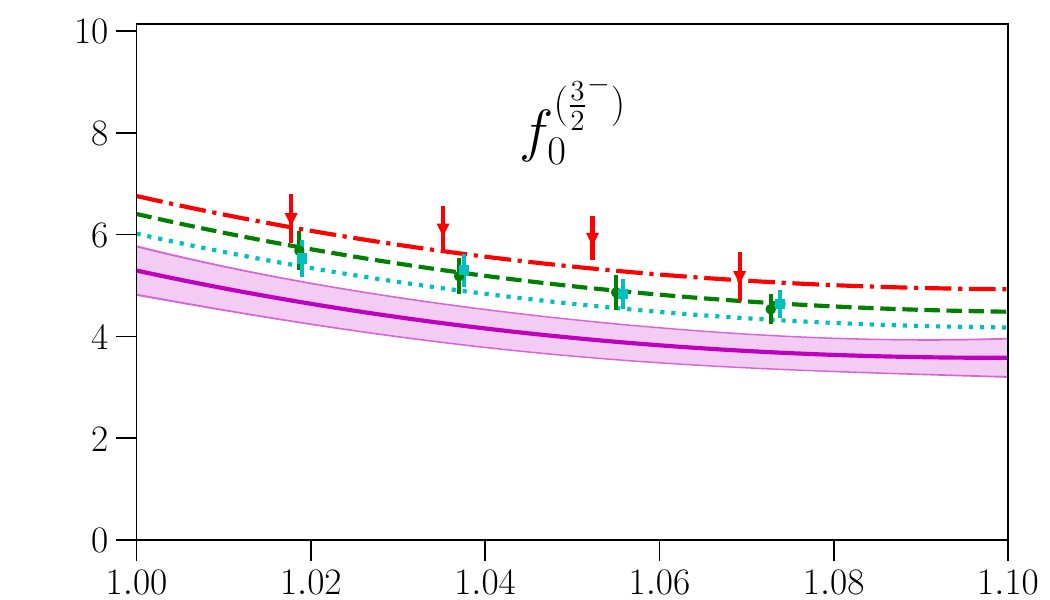}  \includegraphics[width=0.43\linewidth]{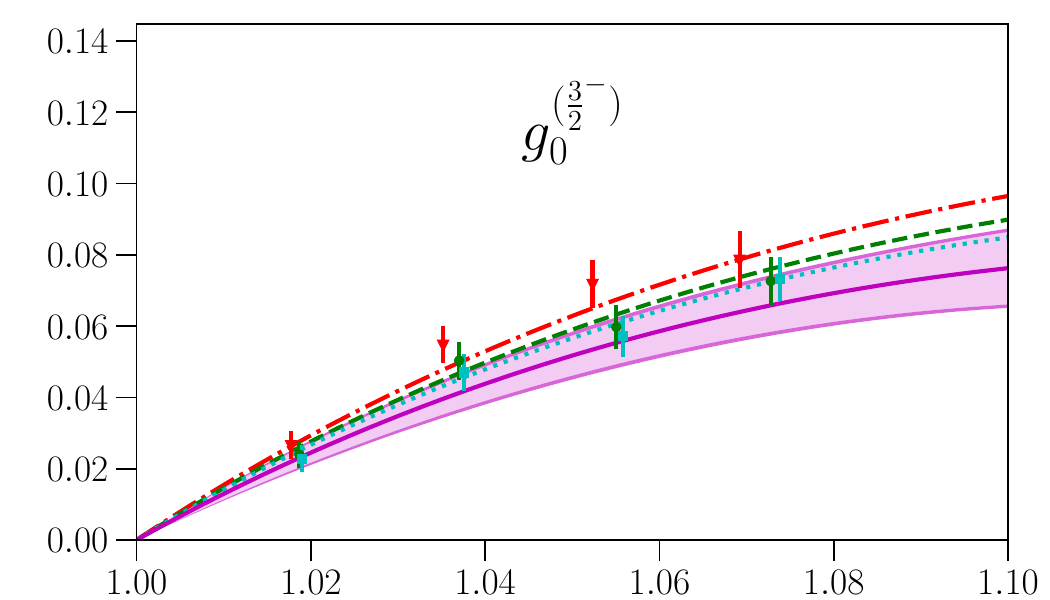} \\
 \includegraphics[width=0.43\linewidth]{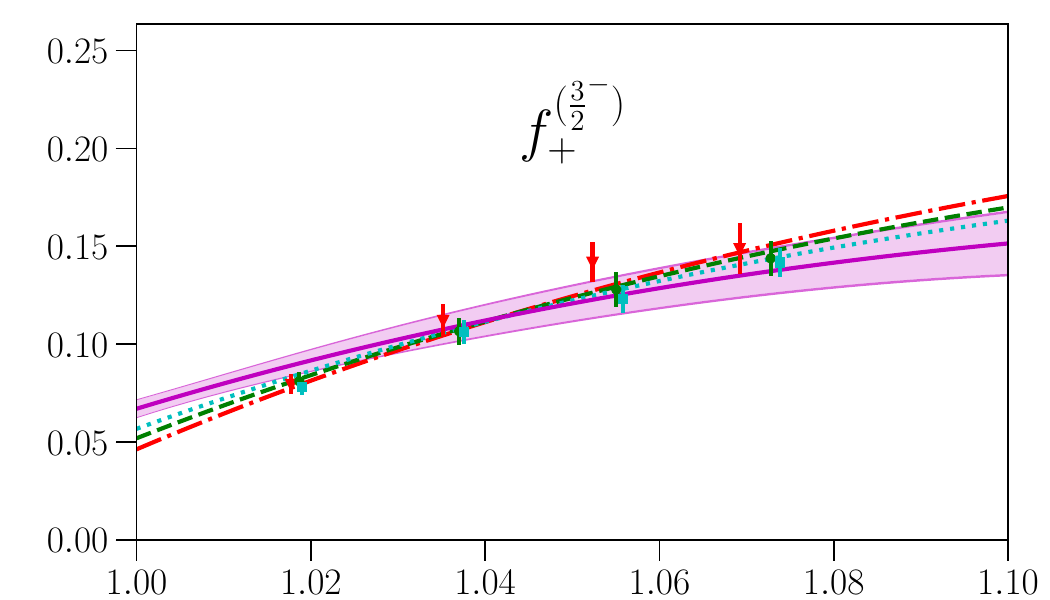}  \includegraphics[width=0.43\linewidth]{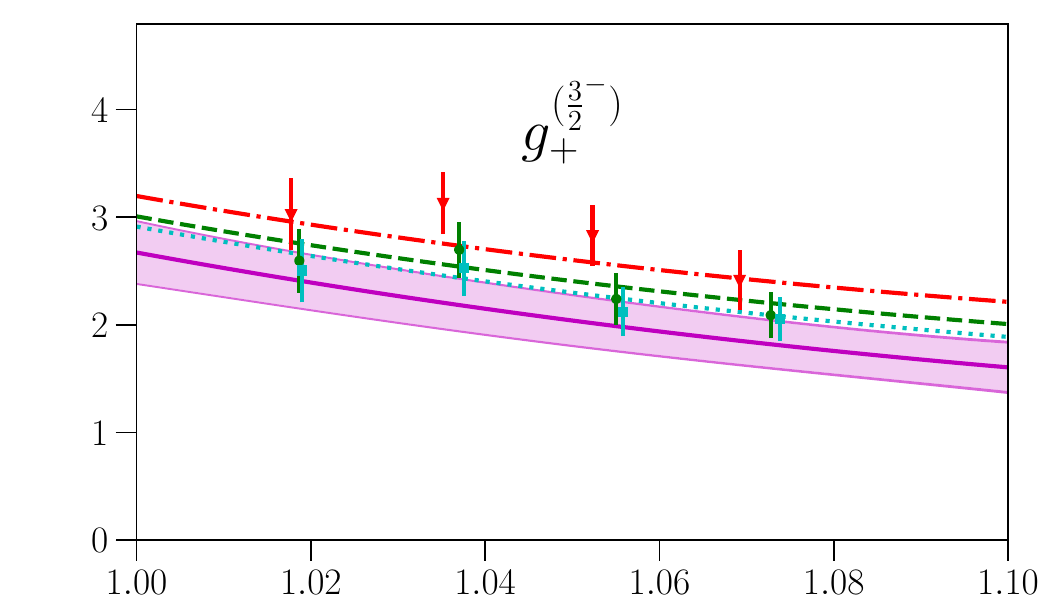} \\

 \includegraphics[width=0.43\linewidth]{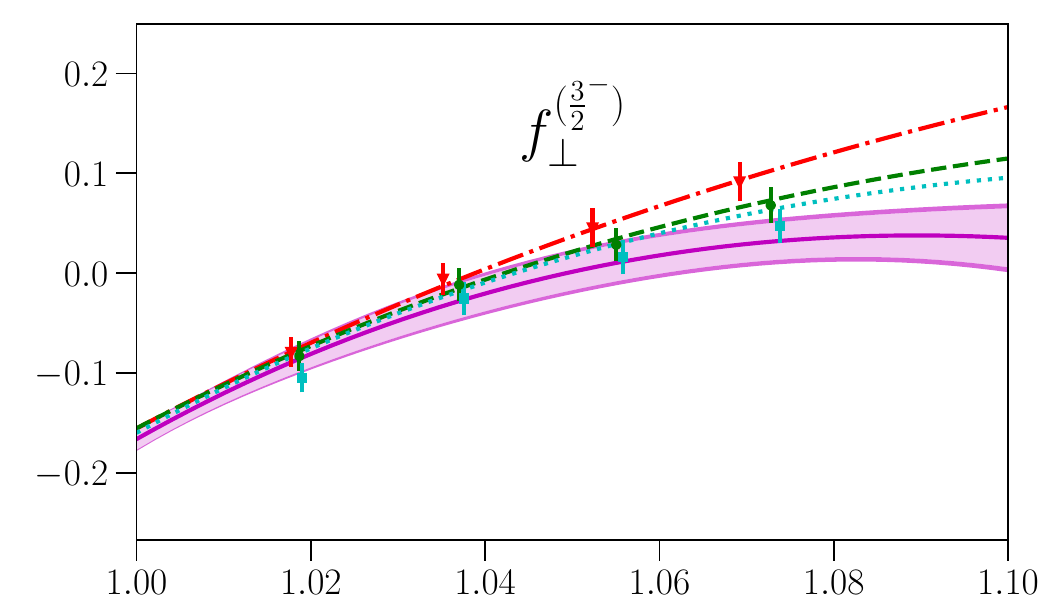}  \includegraphics[width=0.43\linewidth]{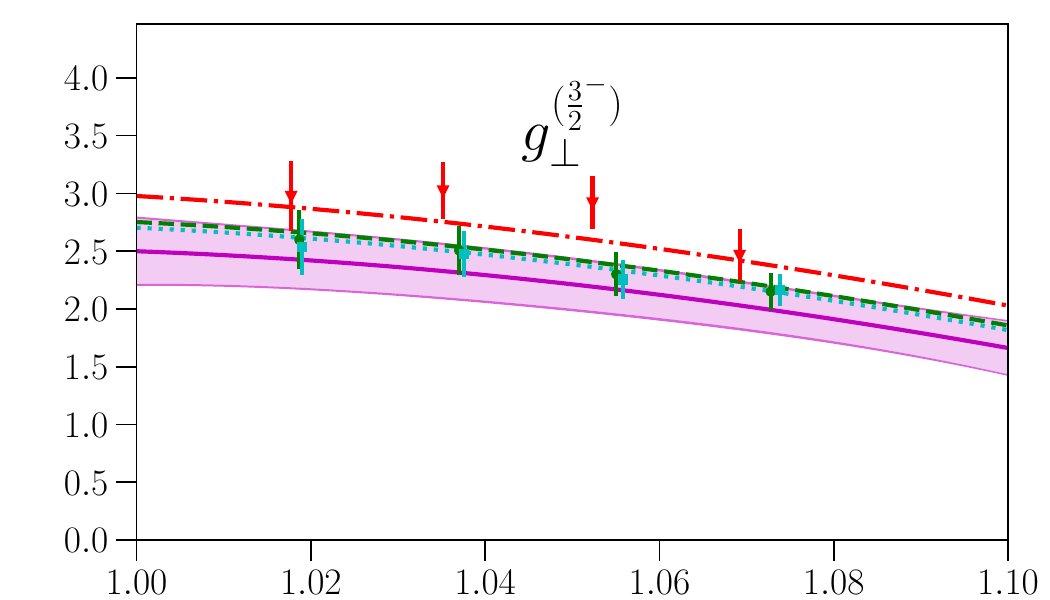} \\
 \includegraphics[width=0.43\linewidth]{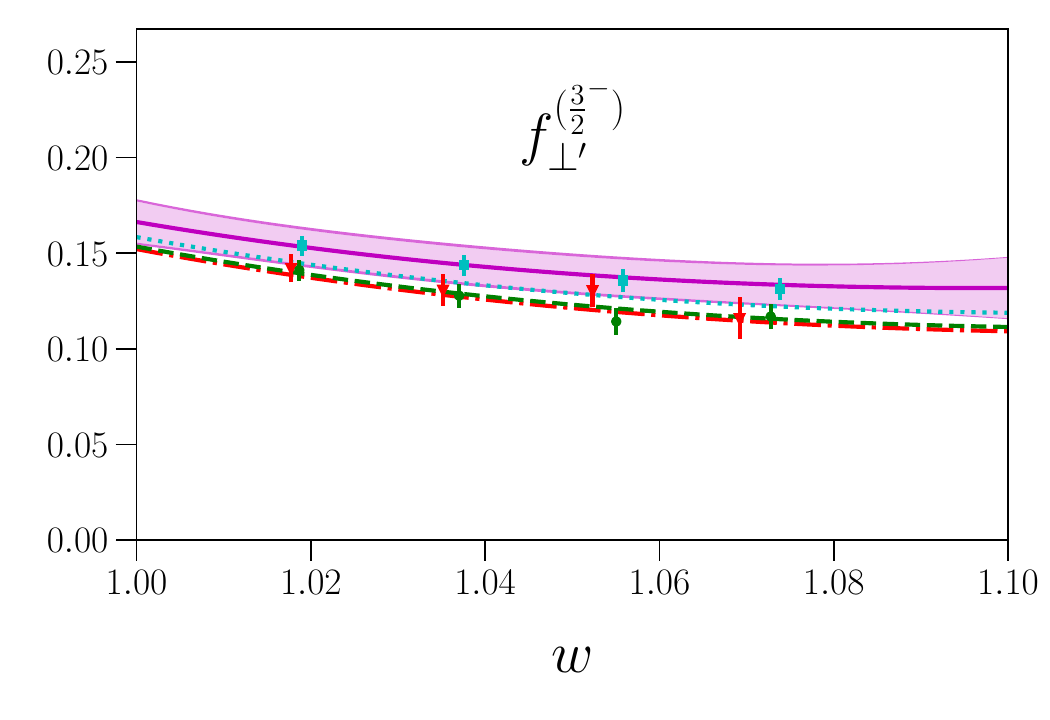}  \includegraphics[width=0.43\linewidth]{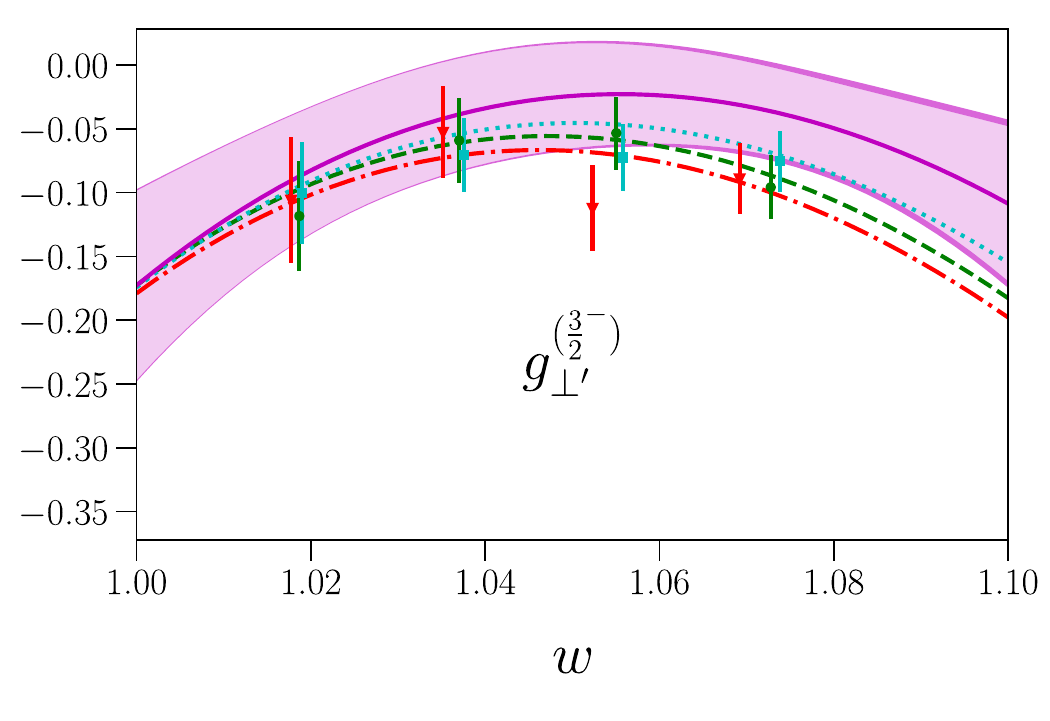} \\

 \caption{\label{fig:LcLstarVA}Chiral and continuum extrapolations of the $\Lambda_c \to \Lambda^*(1520)$ vector and axial vector form factors. The solid magenta curves show the form factors in the physical limit $a=0$, $m_\pi=135\:{\rm MeV}$, with inner light magenta bands indicating the statistical uncertainties and outer dark magenta bands indicating the total uncertainties. The dashed-dotted, dashed, and dotted curves show the fit functions evaluated at the lattice spacings and pion masses of the individual data sets C01, C005, and F004, respectively, with uncertainty bands omitted for clarity. Note that for physical baryon masses, $q^2=0$ corresponds to $w\approx 1.085$.}
 
 \end{figure*}

\begin{figure*}
 
\includegraphics[width=0.55\linewidth]{figures/legend_datasets.pdf}
 
 \vspace{2ex}
 
  \includegraphics[width=0.43\linewidth]{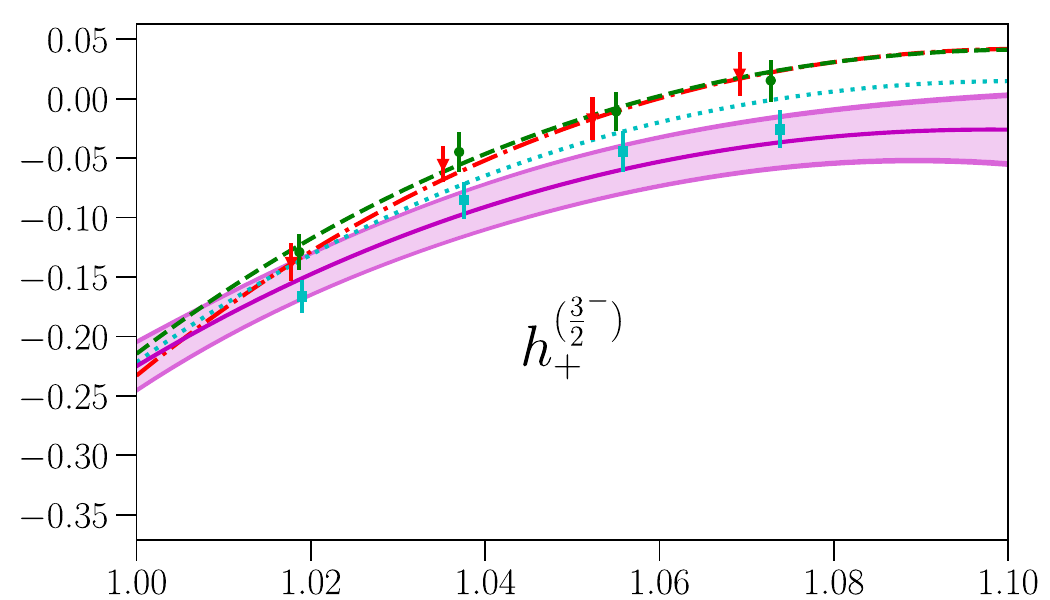}  \includegraphics[width=0.43\linewidth]{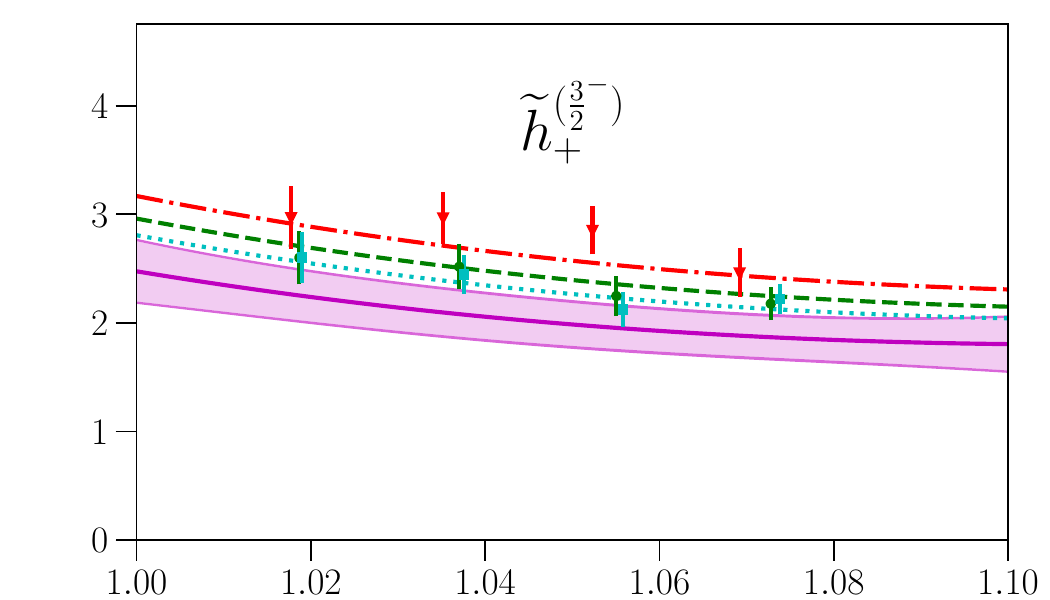} \\

 \includegraphics[width=0.43\linewidth]{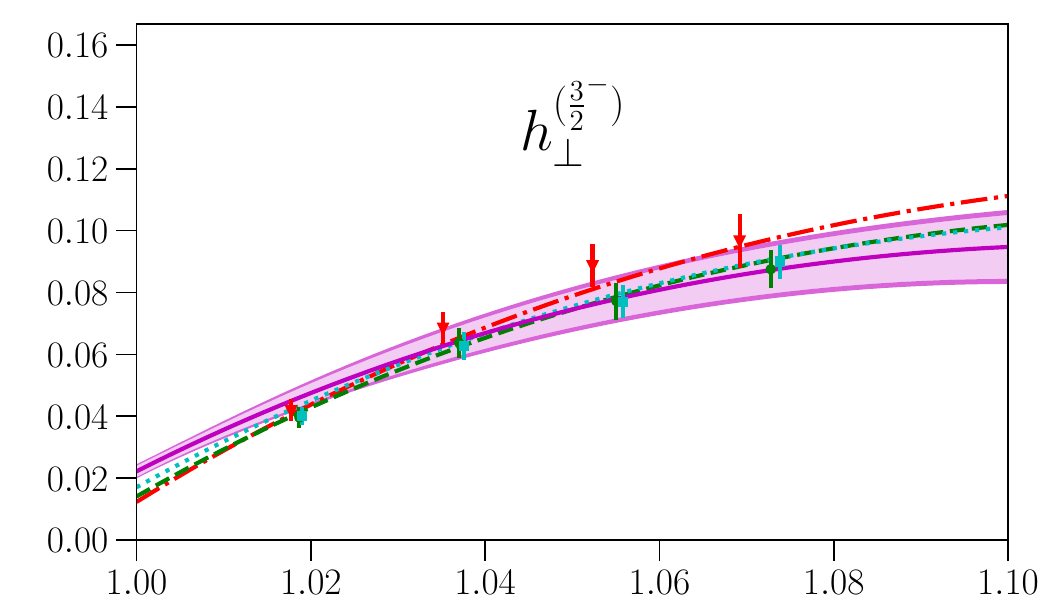}  \includegraphics[width=0.43\linewidth]{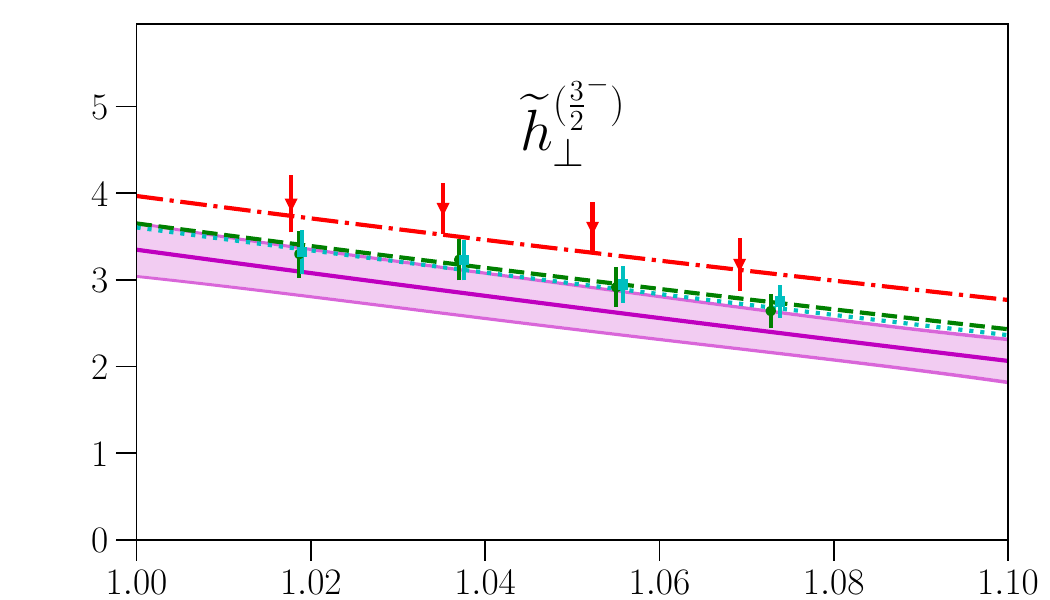} \\
 \includegraphics[width=0.43\linewidth]{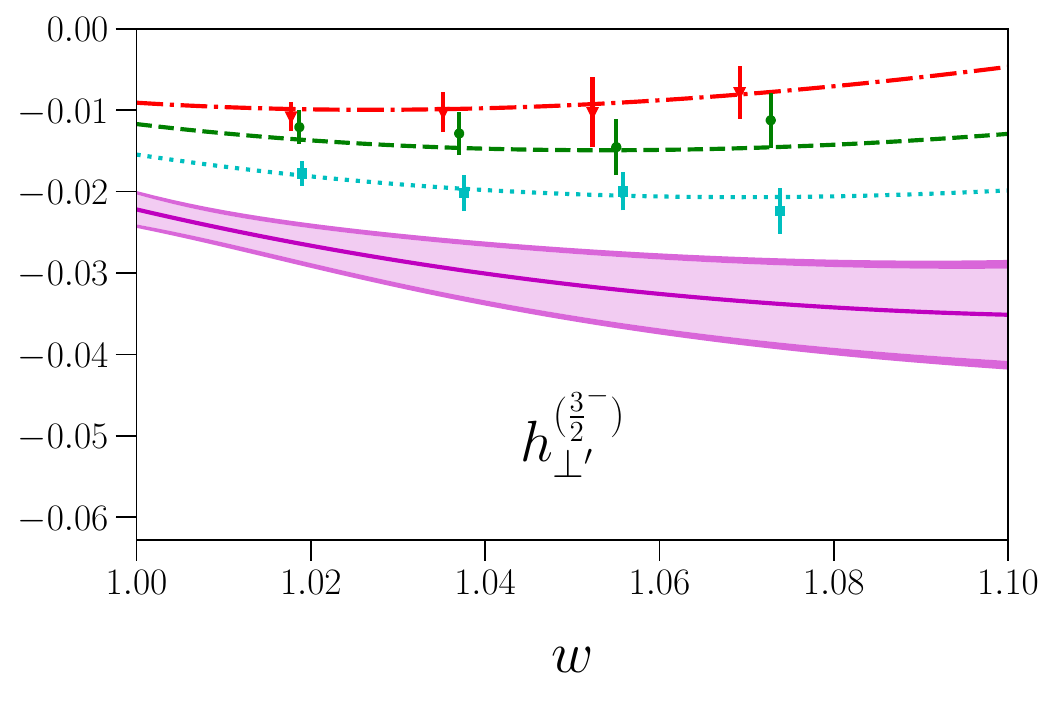}  \includegraphics[width=0.43\linewidth]{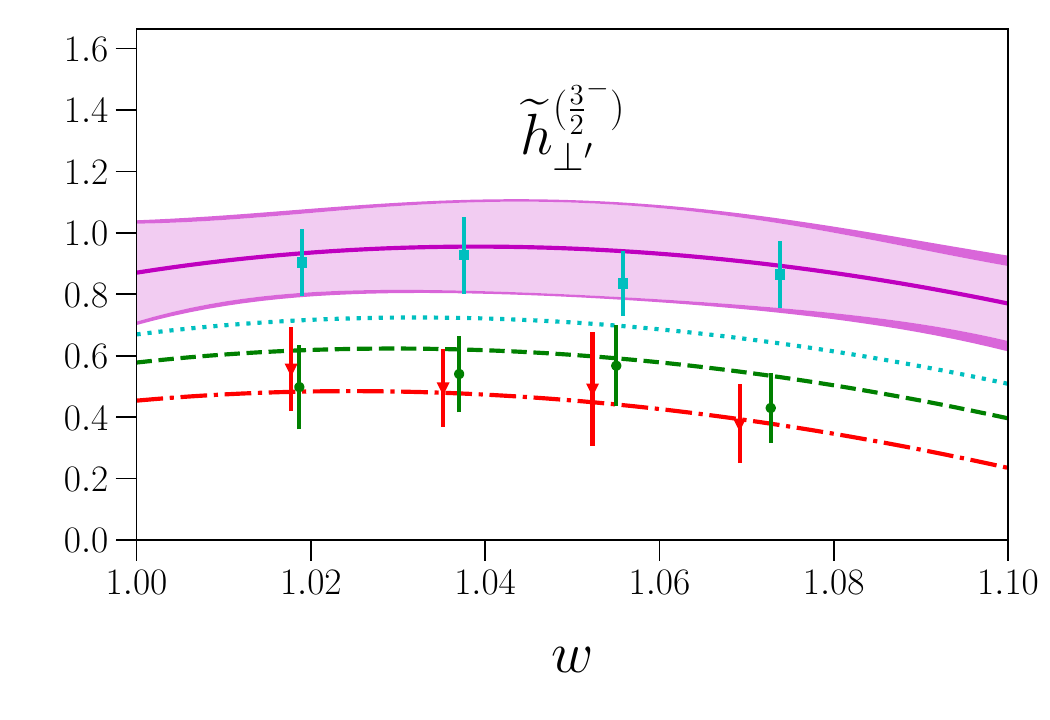} \\

  \caption{\label{fig:LcLstarT}Like Fig.~\protect\ref{fig:LcLstarVA}, but for the $\Lambda_c \to \Lambda^*(1520)$ tensor form factors. Note that the renormalization of the tensor form factors is incomplete at order $\alpha_s$, but our estimates of systematic uncertainties are appropriate for an assumed renormalization scale of $\mu=m_c$.}
 
 \end{figure*}

By construction, $L_n^f=1$ in the physical limit $a=0$, $m_{\pi}=m_{\pi,\rm phys}=135\:{\rm MeV}$, such that only the parameters $a_n^f$ (along with the constant pole massses) are needed to describe the form factors in that limit. No priors are used for the parameters $a_n^f$, while Gaussian priors with central values 0 and widths 10 are used for the coefficients $C_n^{f}$ and $D_n^{f}$, following Refs.~\cite{Meinel:2020owd,Meinel:2021rbm}. To ensure that the physical-limit form factors satisfy the endpoint relations of Sec.~\ref{sec:endpointrelationsJ32}, we eliminate the following $a_0^f$ parameters using Eqs.~(\ref{eq:J32qsqrmaxconstraintfirst})-(\ref{eq:J32qsqrmaxconstraintlast}),
\begin{eqnarray}
 a_0^{f_{\perp}^{(\frac{3}{2}^-)}}&=&-a_0^{f_{\perp^\prime}^{(\frac{3}{2}^-)}}, \label{eq:a0fstart} \\
 a_0^{f_{+}^{(\frac{3}{2}^-)}}&=& 2\frac{m_{\Lambda_Q}-m_{\Lambda_{q,3/2}^*}}{m_{\Lambda_Q}+m_{\Lambda_{q,3/2}^*}} a_0^{f_{\perp^\prime}^{(\frac{3}{2}^-)}}, \\
 a_0^{g_0^{(\frac{3}{2}^-)}}&=&0, \\
 a_0^{g_{+}^{(\frac{3}{2}^-)}}&=& a_0^{g_{\perp}^{(\frac{3}{2}^-)}}-a_0^{g_{\perp^\prime}^{(\frac{3}{2}^-)}}, \\
 a_0^{h_{\perp}^{(\frac{3}{2}^-)}}&=&-a_0^{h_{\perp^\prime}^{(\frac{3}{2}^-)}}, \\
 a_0^{h_{+}^{(\frac{3}{2}^-)}}&=& 2\frac{m_{\Lambda_Q}+m_{\Lambda_{q,3/2}^*}}{m_{\Lambda_Q}-m_{\Lambda_{q,3/2}^*}} a_0^{h_{\perp^\prime}^{(\frac{3}{2}^-)}}, \\
 a_0^{\widetilde{h}_{+}^{(\frac{3}{2}^-)}}&=& a_0^{\widetilde{h}_{\perp}^{(\frac{3}{2}^-)}}-a_0^{\widetilde{h}_{\perp^\prime}^{(\frac{3}{2}^-)}}, \label{eq:a0fend}
\end{eqnarray}
and the following $a_2^f$ parameters using using Eqs.~(\ref{eq:J32qsqr0constraintfirst})-(\ref{eq:J32qsqr0constraintlast}),

\begin{eqnarray}
\nonumber a_2^{f_0^{(\frac{3}{2}^-)}}&=& -\frac{1}{(w_0-1)^2}\Bigg[ a_0^{f_0^{(\frac{3}{2}^-)}}+ a_1^{f_0^{(\frac{3}{2}^-)}} (w_0-1) \\
 && \hspace{13ex} -\frac{(m_{\Lambda_Q}+m_{\Lambda_{q,3/2}^*})^2 }{(m_{\Lambda_Q}-m_{\Lambda_{q,3/2}^*})^2}\left(2 a_0^{f_{\perp^\prime}^{(\frac{3}{2}^-)}} \frac{m_{\Lambda_Q}-m_{\Lambda_{q,3/2}^*}}{m_{\Lambda_Q}+m_{\Lambda_{q,3/2}^*}}+(w_0-1) 
   \Big(a_1^{f_{+}^{(\frac{3}{2}^-)}}+a_2^{f_{+}^{(\frac{3}{2}^-)}} (w_0-1)\Big)\right)\Bigg], \hspace{2ex} \label{eq:a2fstart} \\
  \nonumber a_2^{g_0^{(\frac{3}{2}^-)}}&=&  -\frac{1}{(w_0-1)^2}\Bigg[a_1^{g_0^{(\frac{3}{2}^-)}} (w_0-1) \\
  && \hspace{13ex}-\frac{(m_{\Lambda_Q}-m_{\Lambda_{q,3/2}^*})^2
   }{(m_{\Lambda_Q}+m_{\Lambda_{q,3/2}^*})^2}\left(a_0^{g_{\perp}^{(\frac{3}{2}^-)}}-a_0^{g_{\perp^\prime}^{(\frac{3}{2}^-)}}+(w_0-1)\Big(
   a_1^{g_{+}^{(\frac{3}{2}^-)}}+a_2^{g_{+}^{(\frac{3}{2}^-)}} (w_0-1)\Big)\right)\Bigg], \\
  \nonumber a_2^{\widetilde{h}_{\perp}^{(\frac{3}{2}^-)}}&=& -\frac{1}{{(w_0-1)^2}}\Bigg[a_0^{\widetilde{h}_{\perp}^{(\frac{3}{2}^-)}}+a_1^{\widetilde{h}_{\perp}^{(\frac{3}{2}^-)}}
   (w_0-1)\\
   &&\hspace{13ex}+\frac{(m_{\Lambda_Q}+m_{\Lambda_{q,3/2}^*})^2 }{(m_{\Lambda_Q}-m_{\Lambda_{q,3/2}^*})^2}\left(a_0^{h_{\perp^\prime}^{(\frac{3}{2}^-)}}-(w_0-1)
   \Big(a_1^{h_{\perp}^{(\frac{3}{2}^-)}}+a_2^{h_{\perp}^{(\frac{3}{2}^-)}} (w_0-1)\Big)\right)\Bigg],  \\
  \nonumber a_2^{\widetilde{h}_{\perp^\prime}^{(\frac{3}{2}^-)}}&=&  -\frac{1}{(w_0-1)^2}\Bigg[a_0^{\widetilde{h}_{\perp^\prime}^{(\frac{3}{2}^-)}}+a_1^{\widetilde{h}_{\perp^\prime}^{(\frac{3}{2}^-)}}
   (w_0-1)  \\
   &&\hspace{13ex}+\frac{(m_{\Lambda_Q}+m_{\Lambda_{q,3/2}^*})^2 }{(m_{\Lambda_Q}-m_{\Lambda_{q,3/2}^*})^2}\left(a_0^{h_{\perp^\prime}^{(\frac{3}{2}^-)}}+(w_0-1)
   \Big(a_1^{h_{\perp^\prime}^{(\frac{3}{2}^-)}}+a_2^{h_{\perp^\prime}^{(\frac{3}{2}^-)}} (w_0-1)\Big)\right)\Bigg].  \label{eq:a2fend}
\end{eqnarray}
Here, $\Lambda_Q=\Lambda_c$ and $\Lambda_{q,3/2}^*=\Lambda^*(1520)$, and
\begin{equation}
 w_0\equiv w(q^2=0)=(m_{\Lambda_c}^2+m_{\Lambda^*}^2)/(2m_{\Lambda_c}m_{\Lambda^*}).
\end{equation}
To report the value of $\chi^2/{\rm dof}$ of the fit, we need to make a choice for the number of free parameters to be subtracted from the number of data points to obtain the number of degrees of freedom. If we count all parameters as free, the nominal fit has $\chi^2/{\rm dof}\approx 1.22$. However, the results for the coefficients $C_2^{f}$ and $D_2^{f}$ are all consistent with zero and their uncertainty is approximately equal to the prior width, suggesting that these parameters have little effect on the quality of the fit and should not be counted. With that choice, we find $\chi^2/{\rm dof}\approx 0.80$. The values of the physical-limit parameters are given in the first three columns of Table \ref{tab:LcLStarFFparams}, and the full covariance matrix is available as Supplemental Material \cite{Supplemental}. Plots of the fits are shown in Figs.~\ref{fig:LcLstarVA} and \ref{fig:LcLstarT}.

In the higher-order fit, the data for each form factor $f$ are fitted with
\begin{equation}
 f_{\rm HO}(q^2) = \frac{1}{1-q^2/(m_{\rm pole}^f)^2} \sum_{n=0}^{2} a_{n,{\rm HO}}^f L_{n,{\rm HO}}^f \:(w-1)^n, \label{eq:LcLparamHO}
\end{equation}
where
\begin{equation}
 L_{n,{\rm HO}}^f = \left[1+C_{n,{\rm HO}}^{f}\frac{m_{\pi}^2-m_{\pi,\rm phys}^2}{(4\pi f_{\pi})^2}+H_{n,{\rm HO}}^{f}\frac{m_{\pi}^3-m_{\pi,\rm phys}^3}{(4\pi f_{\pi})^3}+D_{n,{\rm HO}}^{f}a^2\Lambda^2+E_{n,{\rm HO}}^{f}a\Lambda +G_{n,{\rm HO}}^{f}a^3\Lambda^3 \right].
\end{equation}
We use Gaussian priors for the parameters $C_{n,{\rm HO}}^f$, $H_{n,{\rm HO}}^f$, $D_{n,{\rm HO}}^f$, $G_{n,{\rm HO}}^f$ with central values equal to 0 and widths equal to 10. The terms with coefficients $E_{n,{\rm HO}}^{f}$ allow for effects resulting from the incomplete $\mathcal{O}(a)$ improvement (done at tree level only) of the heavy-light currents \cite{Meinel:2020owd}. Because the largest momentum used here is only $2/3$ times the one in Ref.~\cite{Meinel:2020owd}, we reduce the prior widths of $E_{n,{\rm HO}}^{f}$ by the same factor, to $0.2$ (with central values $0$). This allows for the effect of the missing radiative corrections to the $\mathcal{O}(a)$ improvement to be as large as 3.3 percent at the coarse lattice spacing, which is substantially larger than observed in Ref.~\cite{Detmold:2015aaa}, where a numerical comparison between full one-loop and incomplete $\mathcal{O}(a)$ improvement was performed.  We also incorporate the systematic uncertainties associated with the matching of the heavy-light currents by multiplying each form factor with Gaussian random distributions of central value 1 and width corresponding to the estimated uncertainty. The residual matching factors computed at one loop for the $c\to s$ vector and axial-vector currents are very close to their tree-level values of 1 (see Table \ref{tab:matching}). We include a 1\% matching uncertainty for the  vector and axial-vector form factors, which would allow for two-loop corrections with coefficients  of $\alpha_s^2$ that are about six times the size of the one-loop coefficients of $\alpha_s$. This estimate also allows for some small changes in the matching coefficients due to the slightly different tuning of the charm-quark-action parameters. For the tensor currents, the residual matching factors were set equal to their tree-level values because a one-loop calculation was not available. The procedure used in Ref.~\cite{Meinel:2020owd} to estimate the resulting systematic uncertainty would yield an unrealistically small value in the $c\to s$ case, and we instead include a 5\% uncertainty (this estimate should be viewed as corresponding to a renormalization scale $\mu=m_c$). Also recall that the tensor form factors are not needed to describe  $\Lambda_c \to \Lambda^*(1520)\ell^+\nu_\ell$ in the Standard-Model. In the higher-order fit, we furthermore include the estimated uncertainty from the missing isospin-breaking/QED corrections, also by multiplying with further Gaussian random distributions of central value 1 and width corresponding to the estimated uncertainty (0.9\%),  and we include the scale-setting uncertainty by promoting the lattice spacings to fit parameters, constrained to have the known values and uncertainties.

The parameters $a_{n,{\rm HO}}^f$ obtained from the higher-order fit are listed in the last three columns of Table \ref{tab:LcLStarFFparams}, and again their full covariance matrix is available as a supplemental file. As in Refs.~\cite{Detmold:2015aaa,Detmold:2016pkz,Meinel:2020owd,Meinel:2021rbm}, we evaluate the systematic form-factor uncertainty of any observable $O$ through
\begin{equation}
 \sigma_{O,{\rm syst}} = {\rm max}\left( |O_{\rm HO}-O|,\: \sqrt{|\sigma_{O,{\rm HO}}^2-\sigma_O^2|}  \right), \label{eq:sigmasyst}
\end{equation}
where $O$, $\sigma_O$ denote the central value and uncertainty obtained using the parameter values and covariance matrix of the nominal fit and $O_{\rm HO}$, $\sigma_{O,{\rm HO}}^2$ denote the central value and uncertainty obtained using the parameter values and covariance matrix of the higher-order fit. The systematic and statistical uncertainties are then added in quadrature to obtain the total uncertainties, which are shown as the darker bands in Figs.~\ref{fig:LcLstarVA} and \ref{fig:LcLstarT}.

To discuss the uncertainties in a representative observable, we consider the $\Lambda_c \to \Lambda^*(1520)e^+\nu_e$ total ({\it i.e.}, integrated over the full $q^2$ range) decay rate. Using the nominal fit to compute the central value and statistical uncertainty and the higher-order fit to compute the total systematic uncertainty, we find
\begin{equation}
 \frac{\Gamma(\Lambda_c\to\Lambda^*(1520) e^+\nu_e)}{|V_{cs}|^2}= (0.00267 \pm 0.00039_{\rm stat.}\pm0.00018_{\rm syst.})\:{\rm ps}^{-1}. \label{eq:GammaLcL}
\end{equation}
The relative uncertainties are 14.6\% statistical and 6.7\% systematic.\footnote{We also performed additional form-factor fits (both nominal and higher-order) in which we either doubled or halved all prior widths. This has the following effect on the results for the integrated decay rate: half prior widths: $\Gamma(\Lambda_c\to\Lambda^*(1520) e^+\nu_e)/|V_{cs}|^2=0.00296 \pm 0.00032 \pm 0.00011$; double prior widths: $\Gamma(\Lambda_c\to\Lambda^*(1520) e^+\nu_e)/|V_{cs}|^2=0.00241 \pm 0.00044 \pm 0.00033$. The changes [relative to Eq.~(\protect\ref{eq:GammaLcL})] in the central value are below 1$\sigma$. The changes in the estimated systematic uncertainty are entirely expected since many of our prior widths correspond to our estimates of the maximal sizes of higher-order terms that are neglected in the nominal fit. We consider our original prior widths to be the most appropriate.} To assess the breakdown of systematic uncertainties into individual sources, we also performed five additional fits that differ from the nominal fit by including only subsets of the higher-order terms or only selected additional uncertainties:
\begin{enumerate}
 \item To estimate the uncertainty associated with the continuum extrapolation, we add the terms with the coefficients $E_{n,{\rm HO}}^{f}$ and $G_{n,{\rm HO}}^{f}$.
 \item To estimate the uncertainty associated with the chiral extrapolation, we add the terms with the coefficients $H_{n,{\rm HO}}^{f}$.
 \item To estimate the uncertainty associated with the matching of the heavy-light currents, we multiply each form factor with Gaussian random distributions of central value 1 and width corresponding to the estimated uncertainty (1\% for the vector and axial-vector form factors, 5\% for the tensor form factors).
 \item To estimate the uncertainty associated with the missing isospin-breaking/QED corrections, we multiply each form factor with Gaussian random distributions of central value 1 and width corresponding to the estimated uncertainty (0.9\%).
 \item To estimate the uncertainty associated with the scale setting, we promote the lattice spacings to fit parameters, constrained to have the known values and uncertainties.
\end{enumerate}
By comparing the decay rates calculated using each of these fits with that calculated using the nominal fit, we obtain the error budget shown in Table \ref{tab:errorbudget}. While not quantified explicitly, we expect finite-volume errors to be much smaller than the statistical uncertainties, give that (i) all of our data sets have $m_\pi L>4$ and (ii) for a narrow non-$S$-wave resonance like the $\Lambda^*(1520)$, at zero total momentum the energy level in the finite volume caused by the resonance will be far below all multi-hadron scattering states with the same quantum numbers, and the narrow-width approximation is expected to be very accurate.

Our full Standard-Model predictions for the $\Lambda_c \to \Lambda^*(1520)\ell^+\nu_\ell$ differential and integrated decay rates and angular observables, along with a comparison to the quark-model calculation of Ref.~\cite{Hussain:2017lir}, are presented in an accompanying Letter \cite{Meinel:2021grq}.

\begin{table}
 \begin{tabular}{lllllllllllll}
  \hline\hline
  $f$                & \hspace{2ex} & \hspace{6ex}$a_0^f$ & & \hspace{4ex}$a_1^f$ & & \hspace{4ex}$a_2^f$  & \hspace{2ex} & \hspace{4ex}$a_{0,{\rm HO}}^f$ & & \hspace{4ex}$a_{1,{\rm HO}}^f$  & & \hspace{4ex}$a_{2,{\rm HO}}^f$  \\
  \hline
  $f_0^{(\frac{3}{2}^-)}$                                 &&   $\wm    4.71(40)$   &&   $   -26.2(9.5)$   &&                  &&   $\wm    4.78(44)$   &&   $   -27(10)$   &&   \\
  $f_+^{(\frac{3}{2}^-)}$                                 &&       &&   $\wm    1.28(24)$   &&   $   -    3.1(2.4)$                      &&      &&   $\wm    1.27(25)$   &&   $   -    2.7(2.5)$\\
  $f_{\perp}^{(\frac{3}{2}^-)}$                           &&      &&   $\wm    4.02(76)$   &&   $   -   22.1(8.7)$                       &&      &&   $\wm    3.92(86)$   &&   $   -20.4(9.6)$\\
  $f_{\perp^{\prime}}^{(\frac{3}{2}^-)}$                  &&   $\wm  0.1444(94)$   &&   $   -   0.42(32)$   &&   $\wm     3.2(3.8)$      &&   $\wm   0.145(10)$   &&   $   -   0.41(33)$   &&   $\wm     3.1(3.9)$\\
  $g_0^{(\frac{3}{2}^-)}$                                 &&     &&   $\wm    1.16(14)$   &&                        &&    &&   $\wm    1.19(17)$   &&  \\
  $g_+^{(\frac{3}{2}^-)}$                                 &&   &&   $   -   10.4(4.9)$   &&   $\wm 26(44)$                               &&    &&   $   -    9.9(5.1)$   &&   $\wm 25(45)$\\
  $g_{\perp}^{(\frac{3}{2}^-)}$                           &&   $\wm    2.26(25)$   &&   $\wm     0.4(5.6)$   &&   $   -60(58)$           &&   $\wm    2.27(27)$   &&   $   -    0.2(6.0)$   &&   $   -53(61)$\\
  $g_{\perp^{\prime}}^{(\frac{3}{2}^-)}$                  &&   $   -  0.156(67)$   &&   $\wm     4.9(2.3)$   &&   $   -44(24)$           &&   $   -  0.155(68)$   &&   $\wm     4.9(2.4)$   &&   $   -44(24)$\\
  $h_+^{(\frac{3}{2}^-)}$                                 &&     &&   $\wm    3.49(66)$   &&   $   -   18.0(6.8)$                        &&     &&   $\wm    3.58(72)$   &&   $   -   18.2(7.2)$\\
  $h_{\perp}^{(\frac{3}{2}^-)}$                           &&    &&   $\wm    1.27(15)$   &&   $   -    4.9(1.5)$                         &&    &&   $\wm    1.30(18)$   &&   $   -    4.9(1.6)$\\
  $h_{\perp^{\prime}}^{(\frac{3}{2}^-)}$                  &&   $   - 0.0193(16)$   &&   $   -   0.25(10)$   &&   $\wm    0.84(74)$       &&   $   - 0.0197(20)$   &&   $   -   0.27(11)$   &&   $\wm    0.85(75)$\\
  $\widetilde{h}_+^{(\frac{3}{2}^-)}$                     &&  &&   $   -    9.1(6.0)$   &&   $\wm 52(61)$                                  &&    &&   $   -   10.1(6.3)$   &&   $\wm 62(64)$\\
  $\widetilde{h}_{\perp}^{(\frac{3}{2}^-)}$               &&   $\wm    3.02(26)$   &&   $   -    8.5(6.0)$   &&          &&   $\wm    3.05(29)$   &&   $   -    8.1(6.2)$   && \\
  $\widetilde{h}_{\perp^{\prime}}^{(\frac{3}{2}^-)}$      &&   $\wm    0.79(14)$   &&   $\wm     5.0(6.1)$   &&           &&   $\wm    0.82(15)$   &&   $\wm     4.9(5.6)$   &&  \\
  \hline\hline
 \end{tabular}
 \caption{\label{tab:LcLStarFFparams} $\Lambda_c \to \Lambda^*(1520)$ form-factor parameters. The unlisted parameters should be determined using Eqs.~(\ref{eq:a0fstart})-(\ref{eq:a2fend}) with $\Lambda_Q=\Lambda_c$ and $\Lambda_{q,3/2}^*=\Lambda^*(1520)$.  Machine-readable files containing the parameter values and the covariance matrices are provided as supplemental material. The nominal and higher-order form factors are given by Eqs.~(\ref{eq:LcLparam}) and (\ref{eq:LcLparamHO}), respectively, with $L_{n}^f=L_{n,{\rm HO}}^f=1$ in the physical limit. Systematic uncertainties are evaluated using Eq.~(\ref{eq:sigmasyst}).}
\end{table}

\begin{table}
\begin{tabular}{llc}
\hline\hline
Source   & & Relative uncertainty [\%]  \\
\hline
Statistics                         && 14.6  \\
Chiral extrapolation && 5.0 \\
Continuum extrapolation && 3.9 \\
Matching of the $c\to s$ currents && 1.8 \\
Isospin breaking/QED && 1.5 \\
Scale setting && 0.4 \\
\hline\hline
\end{tabular}
\caption{\label{tab:errorbudget}Approximate breakdown of relative uncertainties (in \%) in the integrated decay rate $\Gamma(\Lambda_c\to\Lambda^*(1520) e^+\nu_e)/|V_{cs}|^2$.}
\end{table}

\FloatBarrier
\section{\label{sec:LbLstar}Improved determination of the \texorpdfstring{$\bm{\Lambda_b \to \Lambda^*(1520)}$}{Lambdab to Lambda*(1520)} form factors}
\FloatBarrier

Our new fits to the $\Lambda_b \to \Lambda^*(1520)$ lattice-QCD results differ from those in Ref.~\cite{Meinel:2020owd} in the following ways: (i), we perform simultaneous, fully correlated fits with a single $\chi^2$ function to all form factors; (ii), we enforce the seven constraints at $q^2_{\rm max}$ [Eqs.(\ref{eq:J32qsqrmaxconstraintfirst}-\ref{eq:J32qsqrmaxconstraintlast})] by eliminating redundant parameters before the fit using Eqs.~(\ref{eq:a0fstart})-(\ref{eq:a0fend}); and (iii), we include pole factors in the parametrizations, which now read
\begin{equation}
 f(q^2) = \frac{1}{1-q^2/(m_{\rm pole}^f)^2} \sum_{n=0}^{1} a_n^f L_n^f \:(w-1)^n, \label{eq:LbL1520nominal}
\end{equation}
\begin{equation}
 f_{\rm HO}(q^2) = \frac{1}{1-q^2/(m_{\rm pole}^f)^2} \sum_{n=0}^{1} a_{n,{\rm HO}}^f L_{n,{\rm HO}}^f \:(w-1)^n. \label{eq:LbL1520HO}
\end{equation}
The factors $L_n^f$ and $L_{n,{\rm HO}}^f$ describe the lattice-spacing and pion-mass dependence and are identical to Ref.~\cite{Meinel:2020owd}; in the physical limit, $L_n^f=L_{n,{\rm HO}}^f=1$. The pole masses used are given in Table \ref{tab:polemassesLbLstar}. We find that including the pole factors in the fits has negligible impact on the values of the form factors in the kinematic region $1\leq w\leq 1.05$, but there is no harm in doing so and it could potentially slightly improve the description farther away from this region. Counting $a_0^f$, $a_1^f$, and the two parameters in $L_0^f$ as free parameters \cite{Meinel:2020owd}, the nominal fit has $\chi^2/{\rm dof}\approx 0.75$.

\begin{figure}

\includegraphics[width=0.5\linewidth]{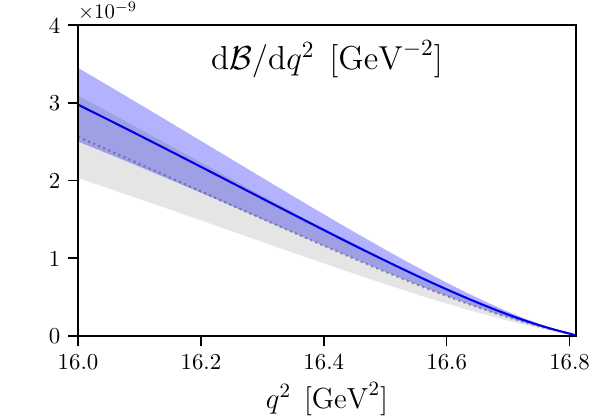} 
 
 \caption{\label{fig:LbL1520dBcomparison}The $\Lambda_b \to \Lambda^*(1520)\ell^+\ell^-$ differential branching fraction in the high-$q^2$ region calculated in the Standard Model. The blue solid curve is obtained using the improved form factor results with the exact endpoint constraints, while the gray dashed curve shows the previous results without these constraints from Ref.~\cite{Meinel:2020owd}.}
\end{figure}

Tables and plots of the fit results are shown in Appendix \ref{sec:LbLstarApp}. The uncertainties of some of the form factors are reduced noticeably compared to Ref.~\cite{Meinel:2020owd} as a result of the additional constraints at $q^2_{\rm max}$. The impact of these improvements on the Standard-Model predictions for $\Lambda_b \to \Lambda^*(1520)\ell^+\ell^-$ is illustrated for the differential branching fraction in Fig.~\ref{fig:LbL1520dBcomparison} and for two of the angular observables (defined in Refs.~\cite{Meinel:2020owd,Descotes-Genon:2019dbw}) in Fig.~\ref{fig:LbL1520angularcomparison}. Updated plots of additional observables are given in Appendix \ref{sec:LbLstarApp}. The uncertainties of all the angular observables considered here now vanish at the endpoint $q^2=q^2_{\rm max}$, where these observables take on the exact values
\begin{eqnarray}
 S_{1c} &\to & 0, \\
 S_{1cc} &\to & \frac16, \\
 S_{1ss} &\to & \frac{5}{12}, \\
 S_{2c} &\to & 0, \\
 S_{2cc} &\to & \frac{5}{12}, \\
 S_{2ss} &\to & \frac{5}{12}, \\
 S_{3ss} &\to & -\frac{1}{4}, \\
 S_{5s} &\to & 0, \\
 S_{5sc} &\to & -\frac{1}{2}, \\
 F_L &\to & \frac13, \\
 A_{FB} &\to & 0.
\end{eqnarray}
The uncertainties near the endpoint are also reduced substantially, as expected. Our previous predictions in Ref.~\cite{Meinel:2020owd} are mostly consistent with the new results within the (old) uncertainties, with deviations at the $2\sigma$ level seen in some angular observables at the endpoint, such as $F_L$ and $A_{FB}^\ell$ as shown in Fig.~\ref{fig:LbL1520angularcomparison}.

\begin{table}
\begin{tabular}{ccccc}
\hline\hline
 $f$     & \hspace{1ex} & $J^P$ & \hspace{1ex}  & $m_{\rm pole}^f$ [GeV]  \\
\hline
$f_+^{(\frac32^-)}$, $f_\perp^{(\frac32^-)}$, $f_{\perp^\prime}^{(\frac32^-)}$, $h_+$, $h_\perp^{(\frac32^-)}$, $h_{\perp^\prime}^{(\frac32^-)}$                         && $1^-$   && $5.416$  \\
$f_0^{(\frac32^-)}$                                                      && $0^+$   && $5.711$  \\
$g_+^{(\frac32^-)}$, $g_\perp^{(\frac32^-)}$, $g_{\perp^\prime}^{(\frac32^-)}$, $\widetilde{h}_+^{(\frac32^-)}$, $\widetilde{h}_\perp^{(\frac32^-)}$, $\widetilde{h}_{\perp^\prime}^{(\frac32^-)}$ && $1^+$   && $5.750$  \\
$g_0^{(\frac32^-)}$                                                      && $0^-$   && $5.367$  \\
\hline\hline
\end{tabular}
\caption{\label{tab:polemassesLbLstar}Pole masses used in the parametrizations of the $\Lambda_b\to\Lambda^*(1520)$ form factors. The $0^-$ and $1^-$ masses are from the Particle Data Group \cite{Zyla:2020zbs}, while the $0^+$ and $1^+$ masses are taken from the lattice QCD calculation of Ref.~\cite{Lang:2015hza}.}
\end{table}

\begin{figure}

\includegraphics[width=0.47\linewidth]{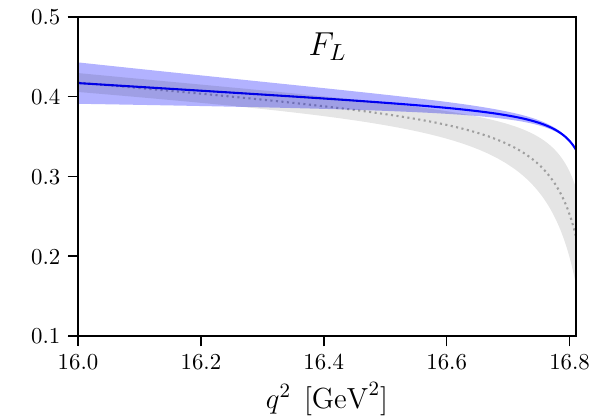} \hfill  \includegraphics[width=0.47\linewidth]{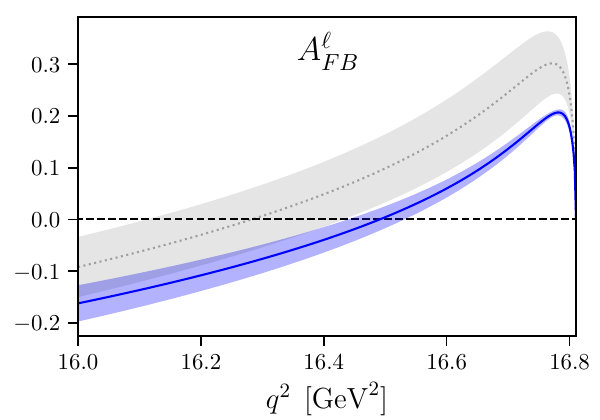}
 
 \caption{\label{fig:LbL1520angularcomparison}The $\Lambda_b \to \Lambda^*(1520)\ell^+\ell^-$ fraction of longitudinally polarized dileptons (left) and lepton-side forward-backward asymmetry (right) in the high-$q^2$ region calculated in the Standard Model. The blue solid curves are obtained using the improved form factor results with the exact endpoint constraints, while the gray dashed curves show the previous results without these constraints from Ref.~\cite{Meinel:2020owd}.}
\end{figure}

\FloatBarrier
\section{\label{sec:LbLcstar}Improved determination of the \texorpdfstring{$\bm{\Lambda_b \to \Lambda_c^*(2595)}$}{Lambdab to Lambdac*(2595)} and \texorpdfstring{$\bm{\Lambda_b \to \Lambda_c^*(2625)}$}{Lambdab to Lambdac*(2625)} form factors}
\FloatBarrier

\begin{figure}

\includegraphics[width=0.47\linewidth]{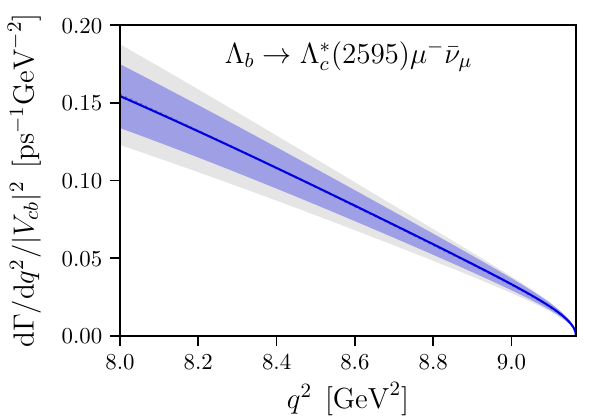} \hfill  \includegraphics[width=0.47\linewidth]{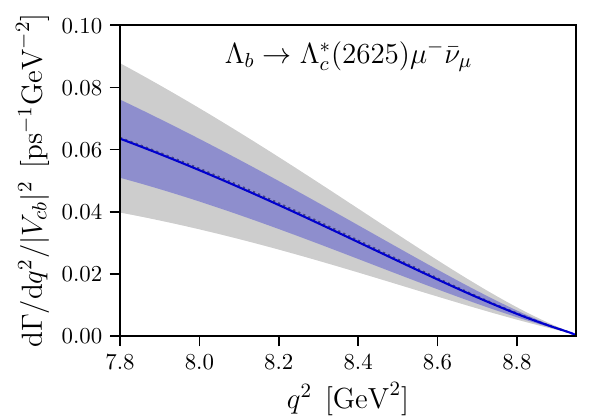}

\includegraphics[width=0.47\linewidth]{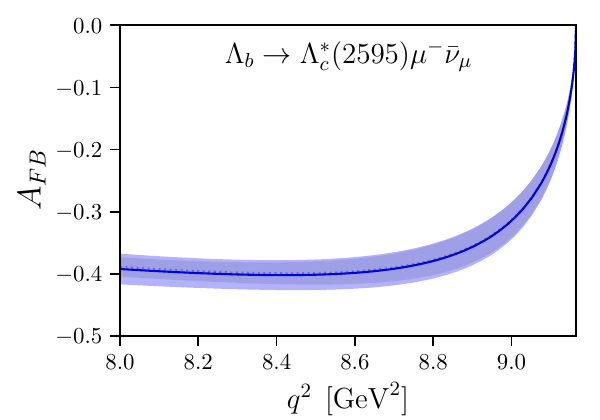} \hfill  \includegraphics[width=0.47\linewidth]{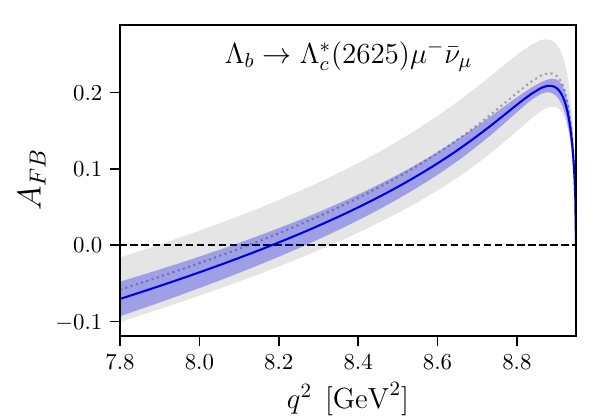}

\includegraphics[width=0.47\linewidth]{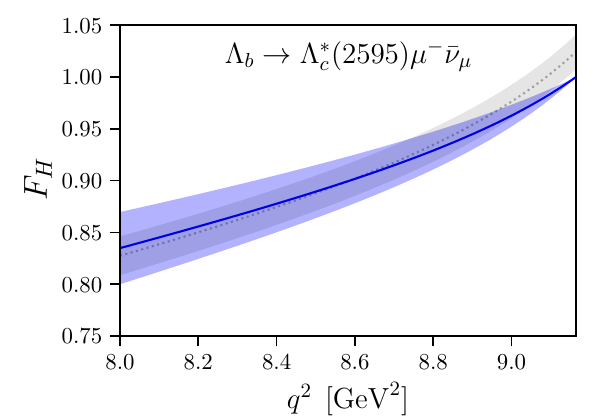} \hfill  \includegraphics[width=0.47\linewidth]{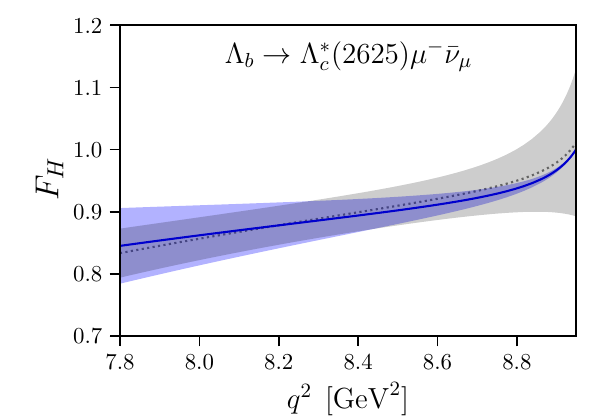}
 
 \caption{\label{fig:LbLcstarcomparison}The $\Lambda_b \to \Lambda_c^*(2595)\mu^-\bar{\nu}$ (left) and $\Lambda_b \to \Lambda_c^*(2625)\mu^-\bar{\nu}$ (right) observables, defined in Ref.~\cite{Boer:2018vpx}, in the high-$q^2$ region calculated in the Standard Model. From top to bottom: the differential decay rate divided by $|V_{cb}|^2$, the forward-backward asymmetry, and the flat term. The blue solid curves are obtained using the improved form factor results with the exact endpoint constraints, while the gray dashed curves show the previous results without these constraints from Ref.~\cite{Meinel:2021rbm}. The new results for the decays to tau leptons are given in Appendix \ref{sec:LbLcstarApp}. }
\end{figure}

In the case of $\Lambda_b \to \Lambda_c^*$, the updates to the fits are the following: (i), we perform simultaneous, fully correlated fits with a single $\chi^2$ function to all form factors for both final states; (ii), we enforce the constraints at $q^2_{\rm max}$ [Eqs.~(\ref{eq:J12qsqrmaxconstraintfirst}) and (\ref{eq:J12qsqrmaxconstraintlast}) for the $ \Lambda_c^*(2595)$ final state and Eqs.~(\ref{eq:J32qsqrmaxconstraintfirst}-\ref{eq:J32qsqrmaxconstraintlast}) for the $\Lambda_c^*(2625)$ final state] by eliminating redundant parameters before the fit. The fit functions thus still have the same form as in Eqs.~(67) and (70) of Ref.~\cite{Meinel:2021rbm} (the poles in $q^2$ caused by $B_c$ bound states are very far away from the physical region, and we do not include them in our form-factor parametrizations.) In the limit of zero lattice spacing and physical pion mass, these functions reduce to
\begin{eqnarray}\label{eq:LbLcstarphysicalFF}
f(q^2)&=&F^{f}+A^{f}(w-1), \\
f_{\rm HO}(q^2)&=&F_{\rm HO}^{f}+A_{\rm HO}^{f}(w-1). \label{eq:LbLcstarphysicalFFHO}
\end{eqnarray}
For the $J^P=\frac32^-$ final-state form factors, the redundant parameters $F^f$ are eliminated using Eqs.~(\ref{eq:a0fstart})-(\ref{eq:a0fend}), where the parameters $a_0^f$ are now renamed to $F^f$. For the $J^P=\frac12^-$ final-state form factors, we use the endpoint relations (\ref{eq:J12qsqrmaxconstraintfirst}) and (\ref{eq:J12qsqrmaxconstraintlast}) to eliminate the parameters $F^{f_\perp^{(\frac12^-)}}$ and $F^{h_\perp^{(\frac12^-)}}$,
\begin{eqnarray}
 F^{f_\perp^{(\frac12^-)}}&=&F^{f_+^{(\frac12^-)}}, \label{eq:J12Ffirst} \\
 F^{h_\perp^{(\frac12^-)}}&=&F^{h_+^{(\frac12^-)}}. \label{eq:J12Flast}
\end{eqnarray}
Tables and plots of the fit results are shown in Appendix \ref{sec:LbLcstarApp}. A comparison of the $\Lambda_b \to \Lambda_c^*(2595)\mu^-\bar{\nu}_\ell$ and $\Lambda_b \to \Lambda_c^*(2625)\mu^-\bar{\nu}_\ell$ observables computed with and without the endpoint constraints is shown in Fig.~\ref{fig:LbLcstarcomparison} (see Ref.~\cite{Boer:2018vpx} for the definitions). We see that imposing the endpoint constraints in the form-factor fits has substantially increased the precision near $q^2_{\rm max}$, compared to Ref.~\cite{Meinel:2021rbm}. The angular observable $F_H$ now become exactly equal to 1 at $q^2=q^2_{\rm max}$, with uncertainty vanishing toward that point. All of our previous predictions are consistent with the new, more precise results. The updated results for the tau-lepton final states are shown in Appendix \ref{sec:LbLcstarApp}. Finally, note that in Ref.~\cite{Meinel:2021rbm}, we had evaluated the combinations of $\Lambda_b \to \Lambda_c^*(2595)$ and $\Lambda_b \to \Lambda_c^*(2625)$ form factors that appear in zero-recoil sum rules \cite{Boer:2018vpx}. The updated results for these combinations are
\begin{eqnarray}
 F_{\rm inel,1/2}+F_{\rm inel,3/2}&=& 0.0942\pm0.0075_{\rm stat}\pm0.0081_{\rm syst} ,\\
 G_{\rm inel,1/2}+G_{\rm inel,3/2}&=& 0.0162\pm0.0015_{\rm stat}\pm0.0019_{\rm syst},
\end{eqnarray}
which are consistent with the previous results and slightly more precise. As before, our result for the axial current falls within the range given in Ref.~\cite{Boer:2018vpx}, while our result for the vector current is slightly above the upper limit.

\FloatBarrier
\section{Conclusions}
\FloatBarrier

In summary, here we have extended our lattice studies of heavy-baryon semileptonic decays to negative-parity baryons in two ways: (i), we performed the first calculation of the $\Lambda_c \to \Lambda^*(1520)$ form factors describing the charm-to-strange decays $\Lambda_c \to \Lambda^*(1520)\ell^+\nu_\ell$, and (ii), we improved our determinations of the $\Lambda_b \to \Lambda^*(1520)$ and $\Lambda_b \to \Lambda_c^*(2595,2625)$ form factors such that the required relations between different helicity form factors at the kinematic endpoint $q^2=q^2_{\rm max}$ are satisfied exactly.

In contrast to the $\Lambda_b$ decays, for $\Lambda_c \to \Lambda^*(1520)$ it is possible to determine the form factors in the full kinematic range occuring in the semileptonic decays using just moderately-sized initial-baryon momenta---this is a consequence of the much lower mass of the $\Lambda_c$. This allows us to predict the total $\Lambda_c \to \Lambda^*(1520)\ell^+\nu_\ell$ decay rates in the Standard Model with 15.9\% uncertainty, of which 14.6\% are statistical and 6.7\% are systematic. As in our previous study of $\Lambda_b \to \Lambda^*(1520)\mu^+\mu^-$, the estimate of systematic uncertainties does not include finite-volume effects / effects associated with the unstable nature of the $\Lambda^*(1520)$. While we believe these effects to be small in our case (for a narrow, non-$S$-wave resonance and at zero spatial momentum, the energy level caused by the resonance will be far below all scattering states for typical lattice sizes), only a new, more complicated and more expensive calculation using the proper multi-hadron formalism \cite{Briceno:2014uqa,Briceno:2015csa,Hansen:2021ofl} would be able to fully control this issue. An experimental measurement of the $\Lambda_c \to \Lambda^*(1520)\ell^+\nu_\ell$ branching fraction would of course provide a valuable check of our methodology, which is largely shared also with the $\Lambda_b$-decay calculations.

When performing the combined chiral/continuum/kinematic extrapolations of the $\Lambda_c \to \Lambda^*(1520)$ form factors, we have enforced exact relations among the different helicity form factors in the physical limit at the kinematic endpoints $q^2=0$ and $q^2_{\rm max}$. These relations ensure that angular observables approach exactly the values predicted by rotational symmetry at the endpoints, and also ensure that transforming the helicity form factors to a non-helicity basis (if desired) does not introduce singularities at the endpoints. In our previous analyses of the $\Lambda_b \to \Lambda^*(1520)$ and $\Lambda_b \to \Lambda_c^*(2595,2625)$ helicity form factors \cite{Meinel:2020owd,Meinel:2021rbm}, we did not explicitly impose the endpoint relations when fitting the lattice results, resulting in them being satisfied only approximately (this statement refers only to the relations at $q^2=q^2_{\rm max}$, since the results for the $\Lambda_b$ decays are limited to the vicinity of that endpoint). In the present work, we have updated the fits to the $\Lambda_b \to \Lambda^*(1520)$ and $\Lambda_b \to \Lambda_c^*(2595,2625)$ form factors by imposing the endpoint relations at $q^2=q^2_{\rm max}$, and we have re-calculated the differential decay rates and angular observables of $\Lambda_b \to \Lambda^*(1520)(\to p K^-)\mu^+\mu^-$ and $\Lambda_b \to \Lambda_c^*(2595,2625)\ell^-\bar{\nu}_\ell$ in the Standard Model. The predictions are now more precise, and the angular observables exactly approach the values predicted by rotational symmetry at $q^2=q^2_{\rm max}$.

As already mentioned in Ref.~\cite{Meinel:2021rbm} and further analyzed in Ref.~\cite{Papucci:2021pmj}, our lattice results for the $\Lambda_b \to \Lambda_c^*(2595,2625)$ form factors imply large higher-order corrections in heavy-quark effective theory near $q^2=q^2_{\rm max}$, in particular for the $\Lambda_c^*(2595)$ final state with $J^P=\frac12^-$. Our improved form-factor results are more precise but are consistent with the previous results and therefore do not alter this conclusion. As before, near $q^2_{\rm max}$ we find the $\Lambda_b \to \Lambda_c^*(2595)\mu^-\bar{\nu}_\mu$ differential decay rate to be significantly larger than the $\Lambda_b \to \Lambda_c^*(2625)\mu^-\bar{\nu}_\mu$ differential decay rate, whereas the \emph{total} decay rates measured in experiment \cite{Aaltonen:2008eu} have the opposite order. We therefore expect the differential decay rates to cross at some value of $q^2$ lower than covered by our lattice results. Such a crossing is in fact seen in the quark-model predictions of Ref.~\cite{Pervin:2005ve}. Also note that the authors of Ref.~\cite{Nieves:2019kdh,Nieves:2019nol} suggested an exotic structure of the $\Lambda_c^*(2595)$, possibly with two resonance poles of which only one is a heavy-quark symmetry partner of the $\Lambda_c^*(2625)$. This warrants further investigation.

\begin{acknowledgments}
We thank M.~Bordone, S.~Descotes-Genon, G.~Hiller, C.~Marin-Benito, J.~Toelstede, D.~van Dyk, and R.~Zwicky for discussions. We are grateful to the RBC and UKQCD Collaborations for making their gauge field ensembles available. SM is supported by the U.S. Department of Energy, Office of Science, Office of High Energy Physics under Award Number D{E-S}{C0}009913. GR is supported by the U.S. Department of Energy, Office of Science, Office of Nuclear Physics, under Contract No.~D{E-S}C0012704 (BNL). The computations for this work were carried out on facilities at the National Energy Research Scientific Computing Center, a DOE Office of Science User Facility supported by the Office of Science of the U.S. Department of Energy under Contract No. DE-AC02-05CH1123, and on facilities of the Extreme Science and Engineering Discovery Environment (XSEDE) \cite{XSEDE}, which is supported by National Science Foundation grant number ACI-1548562. We acknowledge the use of Chroma \cite{Edwards:2004sx,Chroma}, QPhiX \cite{JOO2015139,QPhiX}, QLUA \cite{QLUA}, MDWF \cite{MDWF}, and related USQCD software \cite{USQCD}.
\end{acknowledgments}

\FloatBarrier

\appendix

\FloatBarrier
\section{\label{sec:LbLstarApp}Plots and tables of the improved results for \texorpdfstring{$\bm{\Lambda_b \to \Lambda^*(1520)\ell^+\ell^-}$}{Lambdab to Lambda*(1520) l+l-}}
\FloatBarrier

\begin{figure}
 \centerline{\includegraphics[width=0.6\linewidth]{figures/legend_datasets.pdf}}
 
 \vspace{1ex}

 \includegraphics[width=0.47\linewidth]{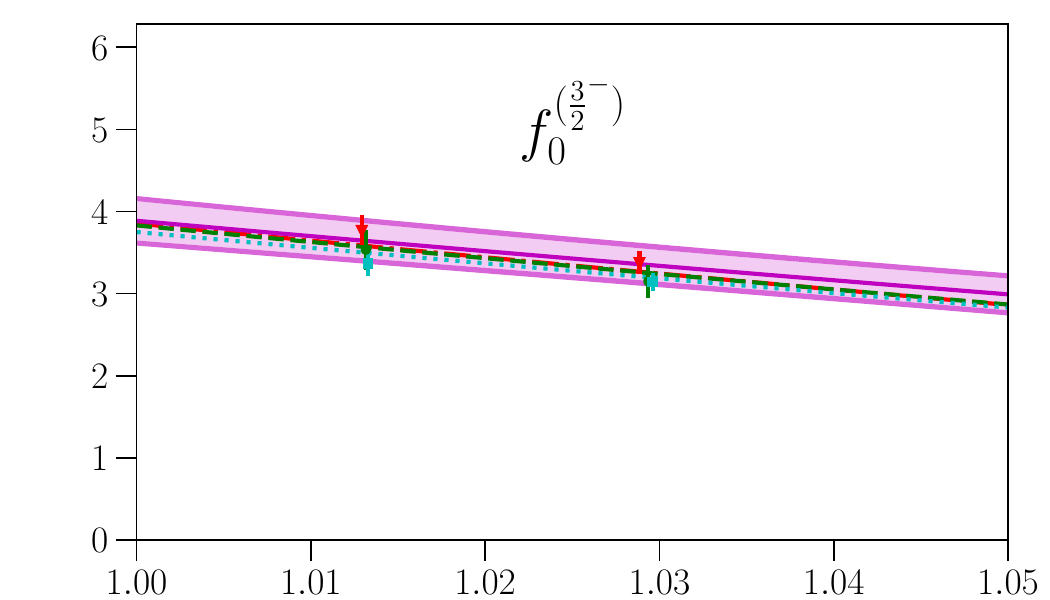} \hfill \includegraphics[width=0.47\linewidth]{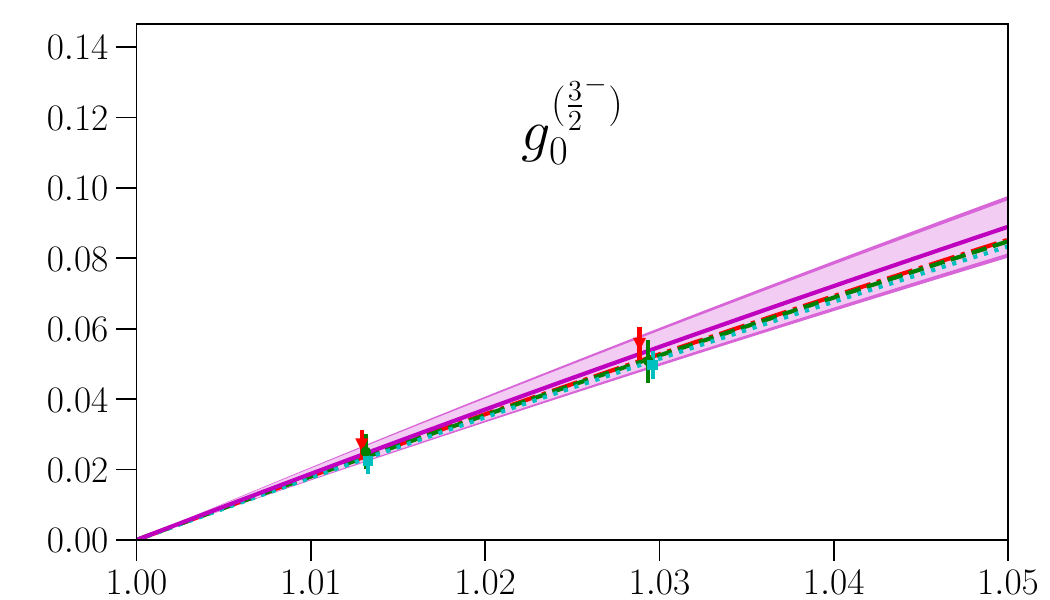} \\
 \includegraphics[width=0.47\linewidth]{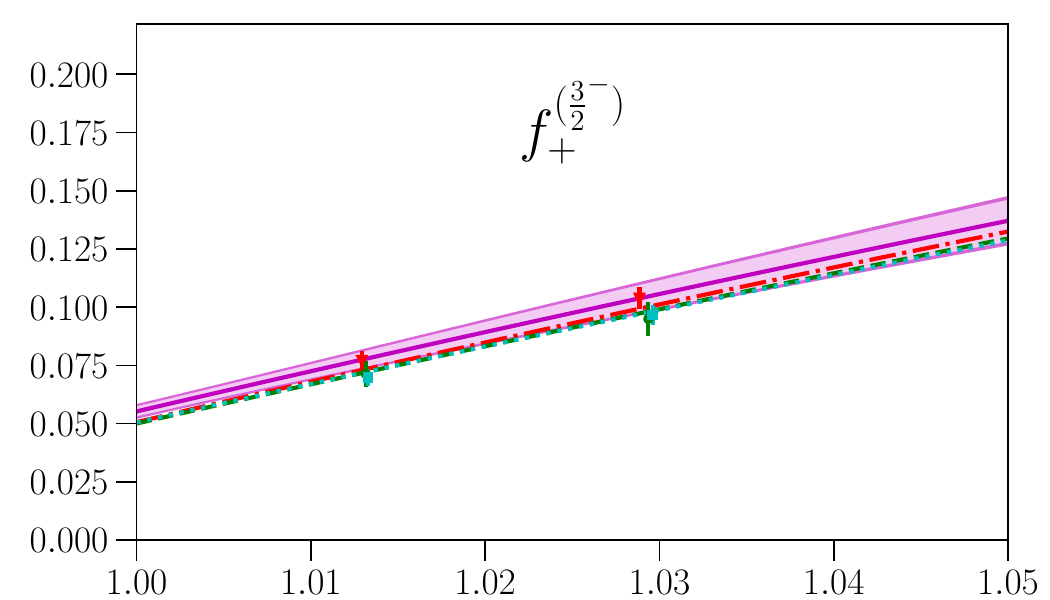} \hfill \includegraphics[width=0.47\linewidth]{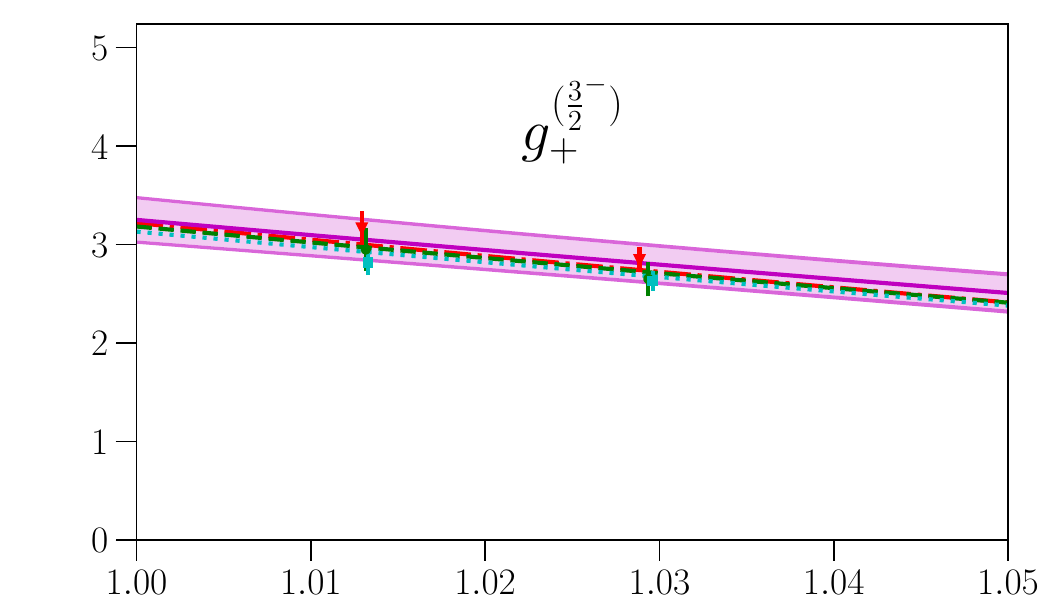} \\
 \includegraphics[width=0.47\linewidth]{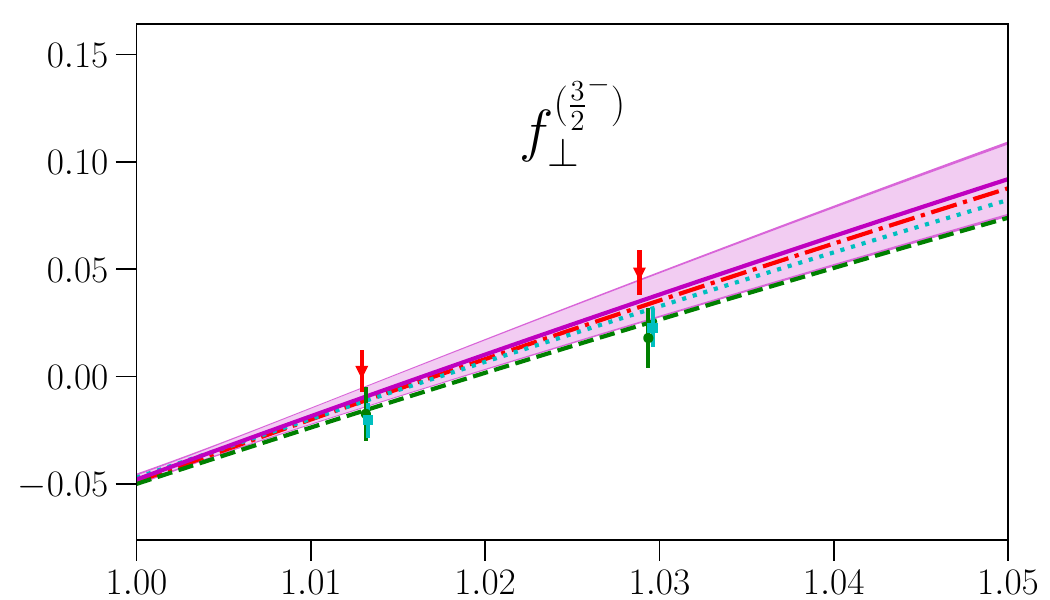} \hfill \includegraphics[width=0.47\linewidth]{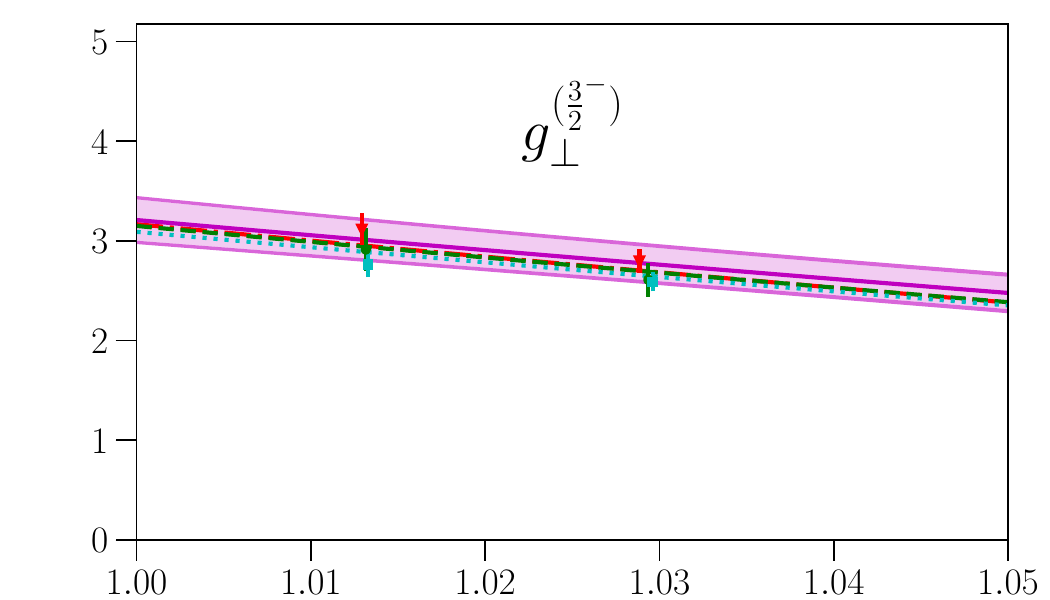} \\
 \includegraphics[width=0.47\linewidth]{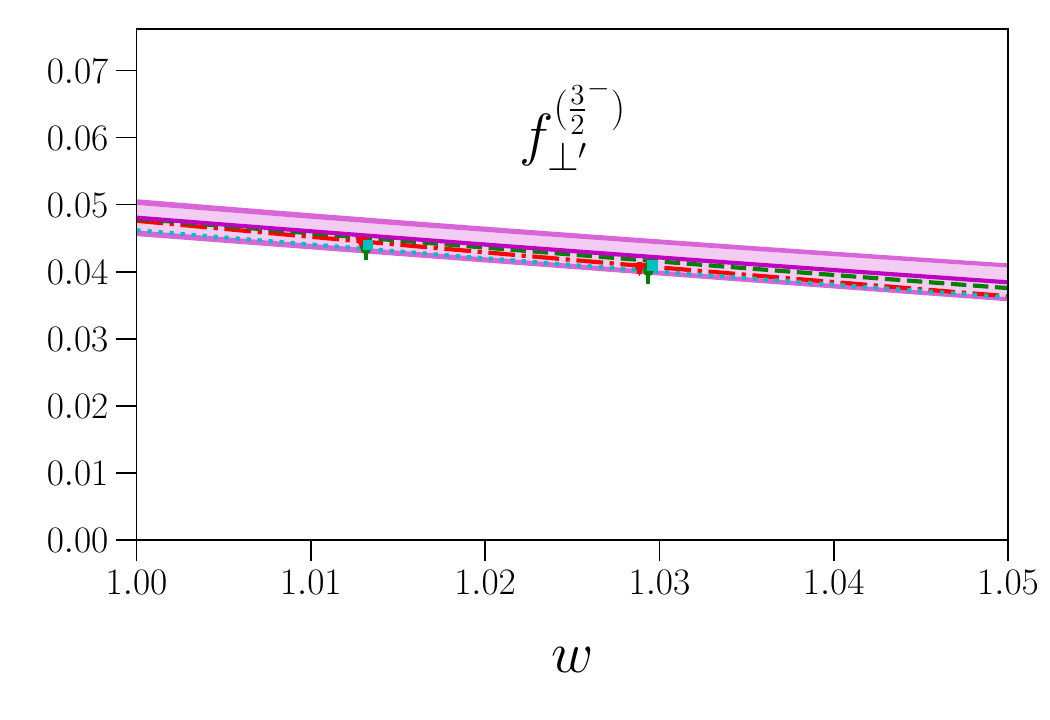} \hfill \includegraphics[width=0.47\linewidth]{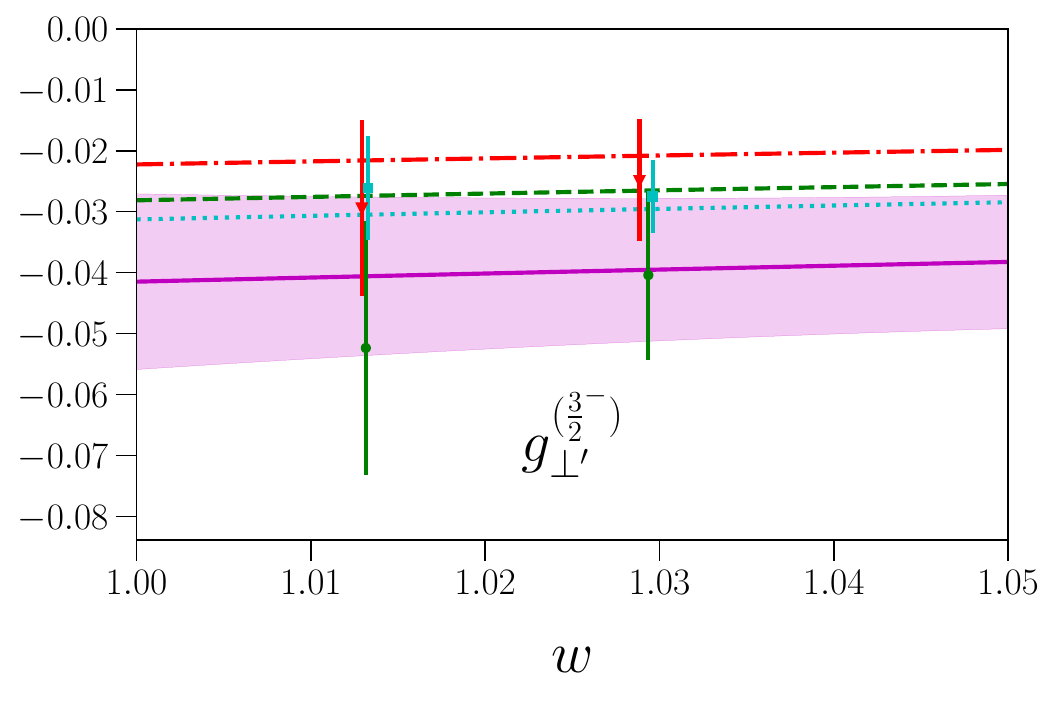} 

 \caption{\label{fig:LbLstarVA}Updated chiral-continuum-kinematic extrapolation fits for the $\Lambda_b\to\Lambda^*(1520)$ vector and axial-vector form factors.}
\end{figure}

\begin{figure}
 \centerline{\includegraphics[width=0.6\linewidth]{figures/legend_datasets.pdf}}
 
 \vspace{1ex}
 
 \includegraphics[width=0.47\linewidth]{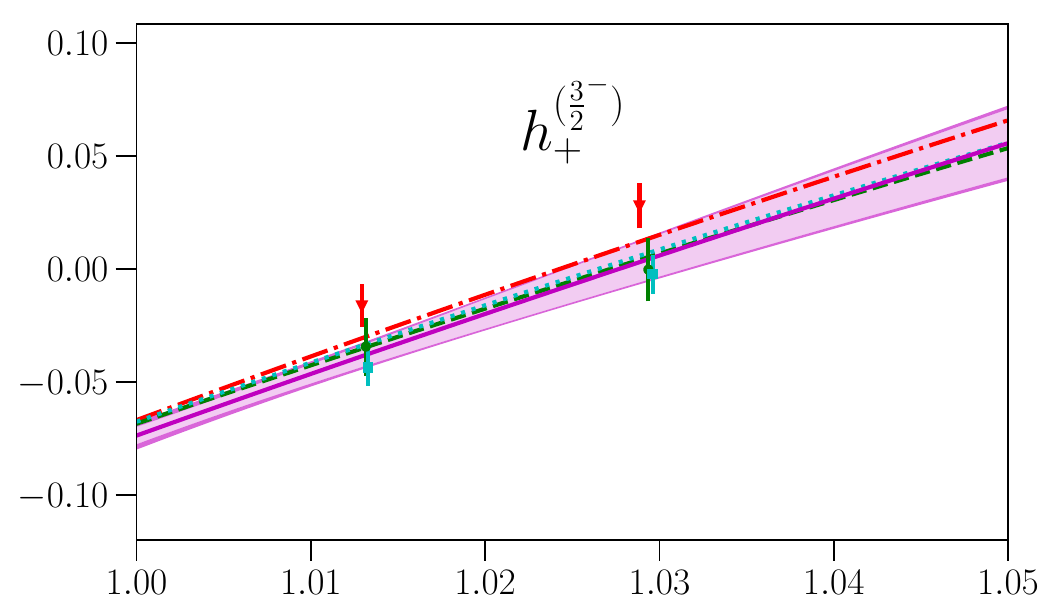} \hfill \includegraphics[width=0.47\linewidth]{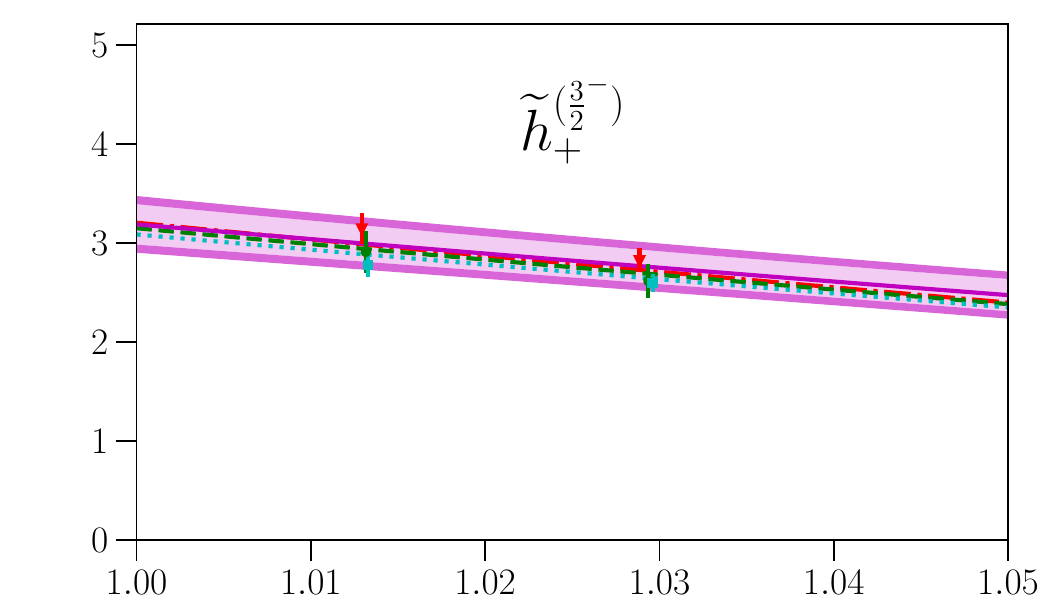} \\
 \includegraphics[width=0.47\linewidth]{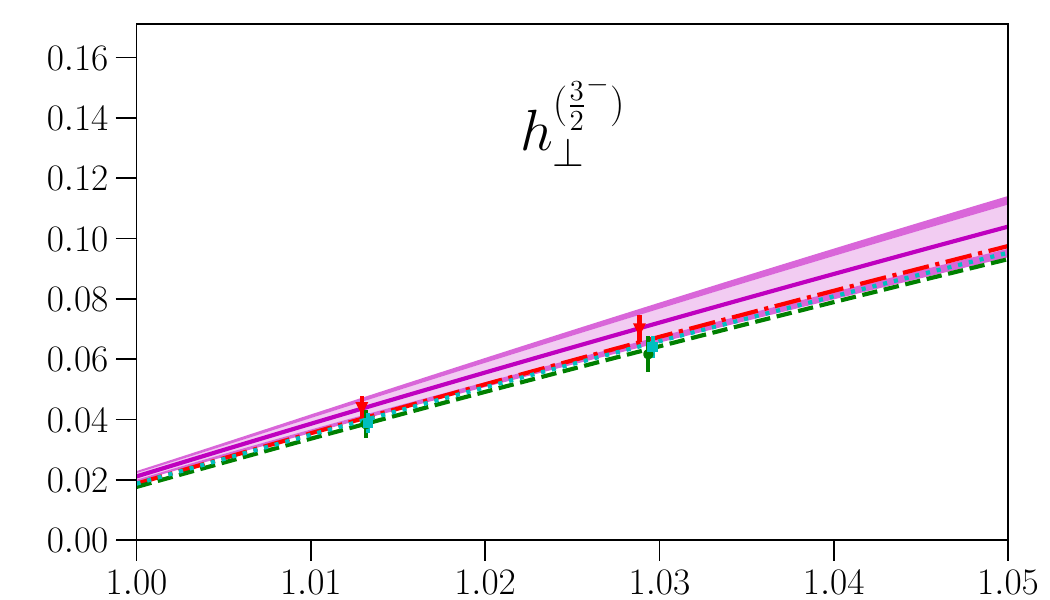} \hfill \includegraphics[width=0.47\linewidth]{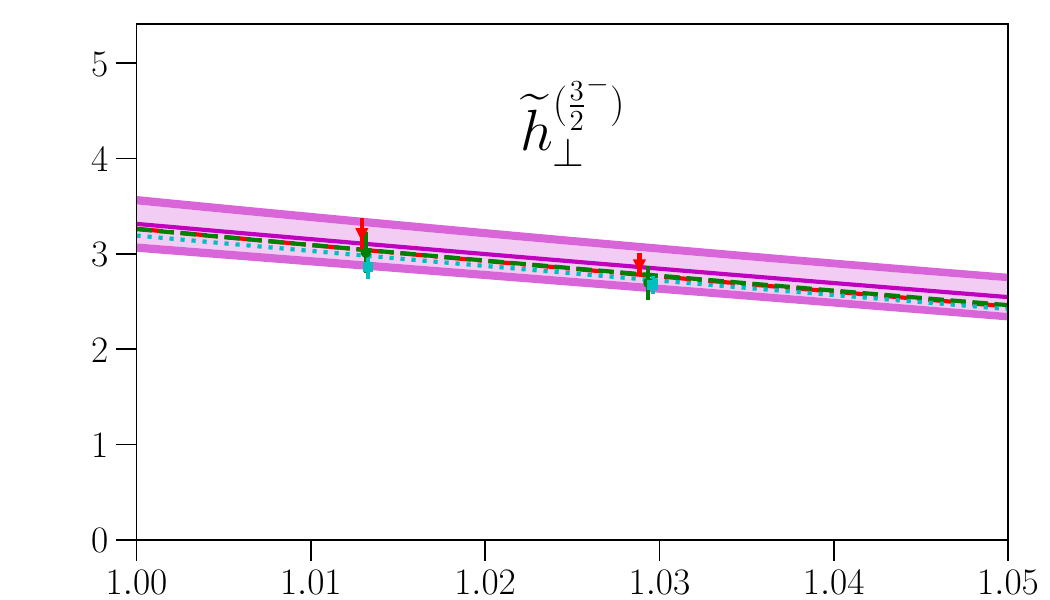} \\
 \includegraphics[width=0.47\linewidth]{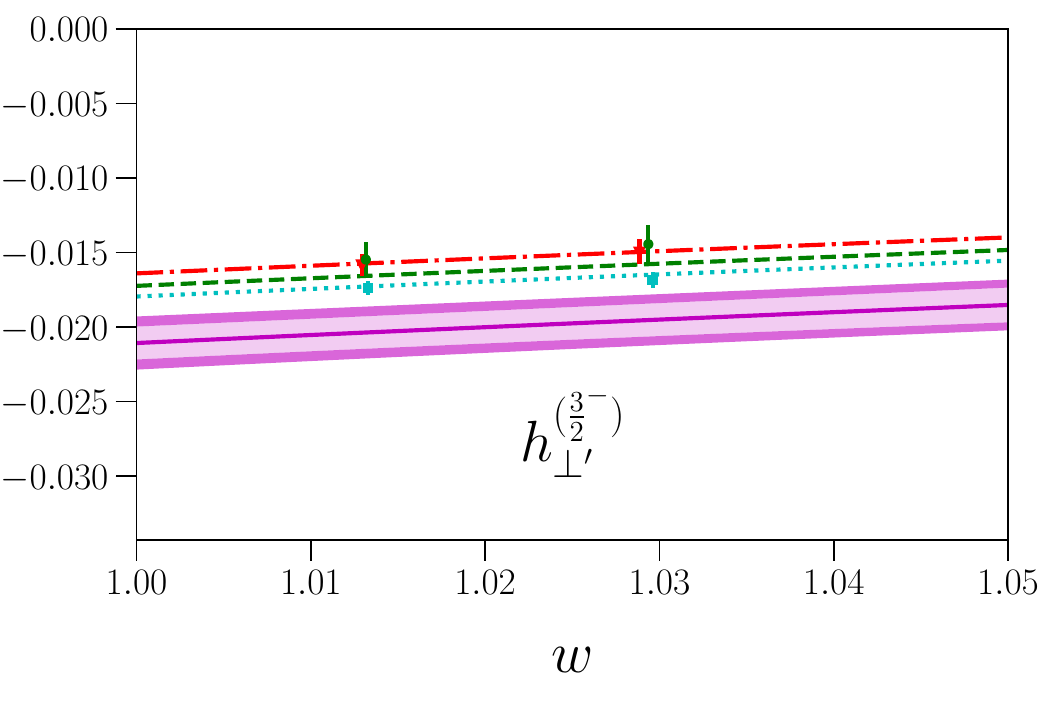} \hfill \includegraphics[width=0.47\linewidth]{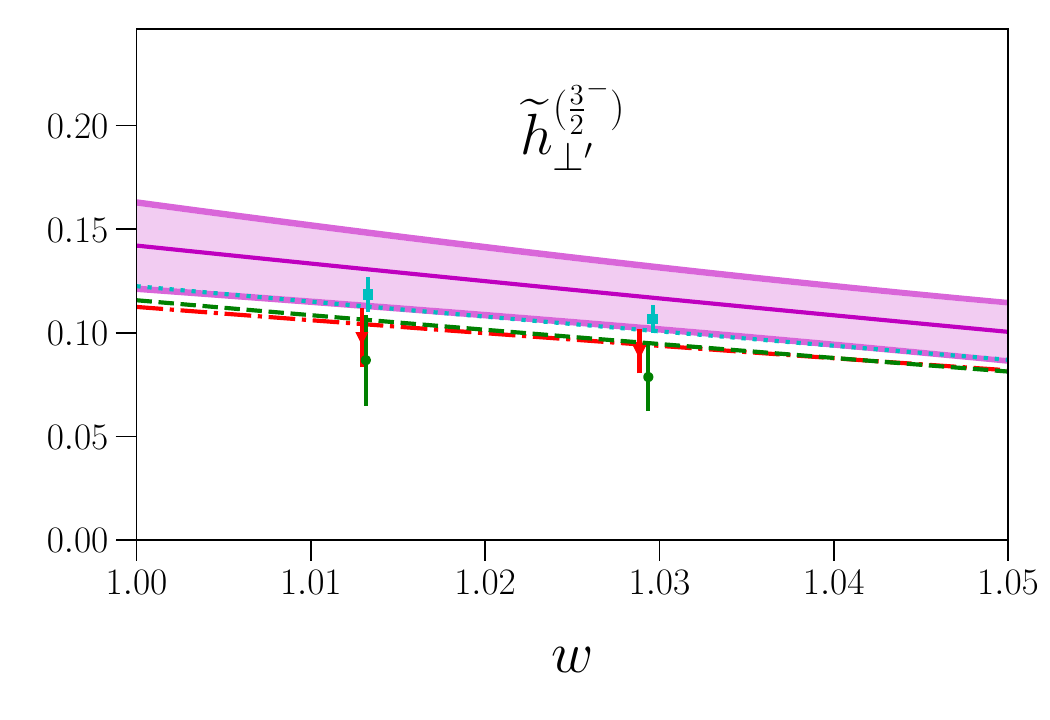} 

 \caption{\label{fig:LbLstarT}Updated chiral-continuum-kinematic extrapolation fits for the $\Lambda_b\to\Lambda^*(1520)$ tensor form factors.}
\end{figure}

\begin{figure}

\includegraphics[width=0.45\linewidth]{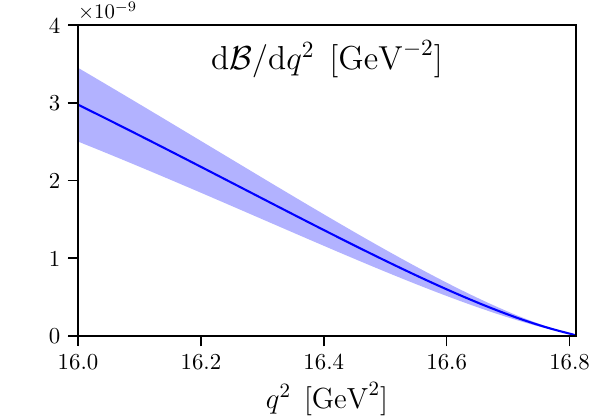} \hfill  \includegraphics[width=0.45\linewidth]{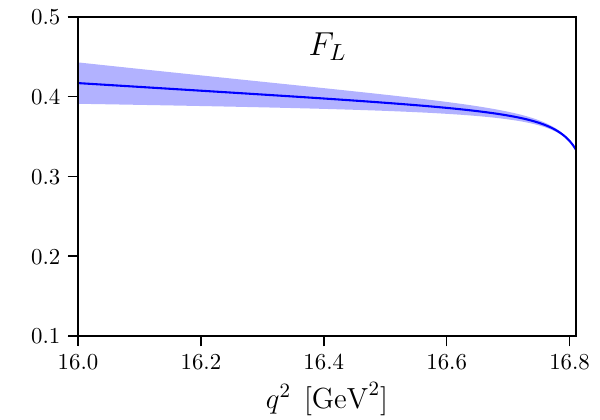}

\includegraphics[width=0.45\linewidth]{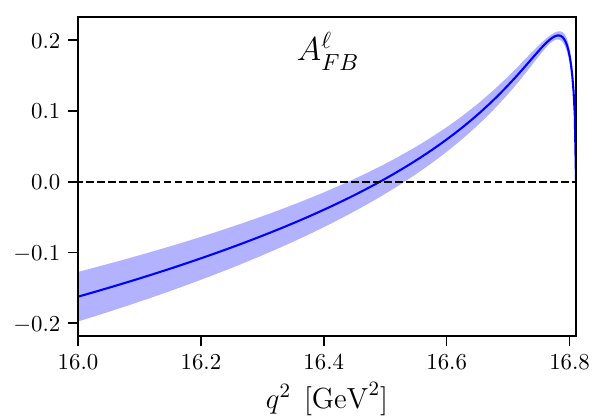} \hfill \includegraphics[width=0.45\linewidth]{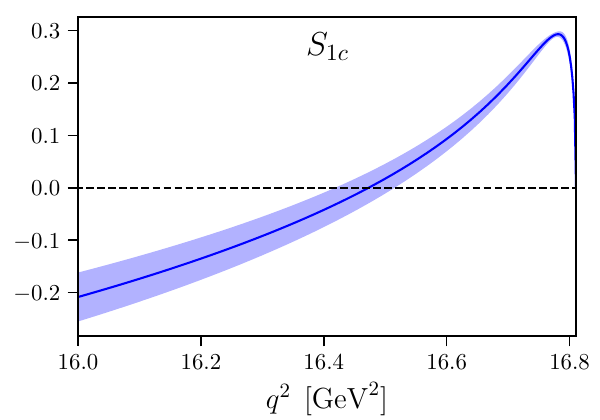}

\includegraphics[width=0.45\linewidth]{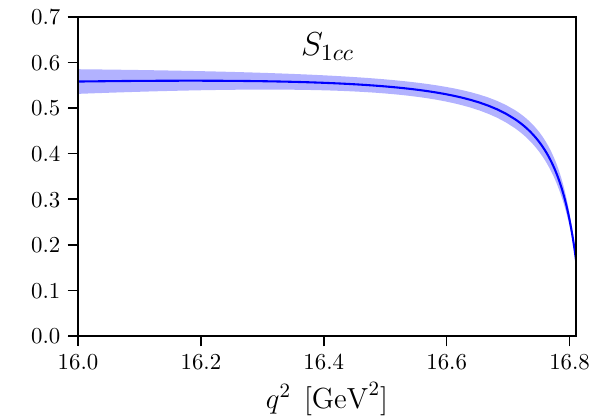}
 \hfill \includegraphics[width=0.45\linewidth]{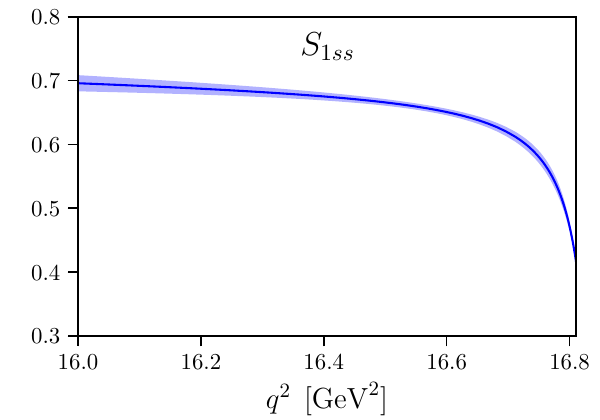}

\includegraphics[width=0.45\linewidth]{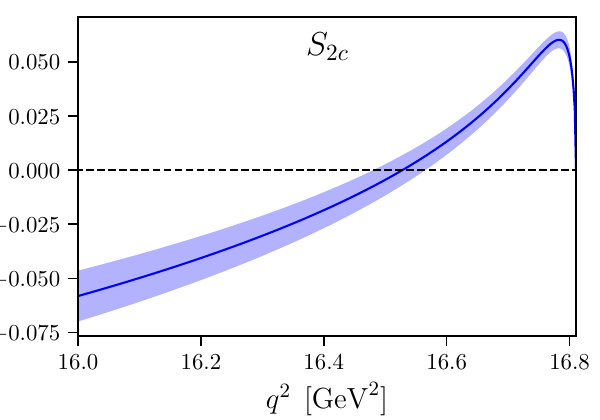}  \hfill \includegraphics[width=0.45\linewidth]{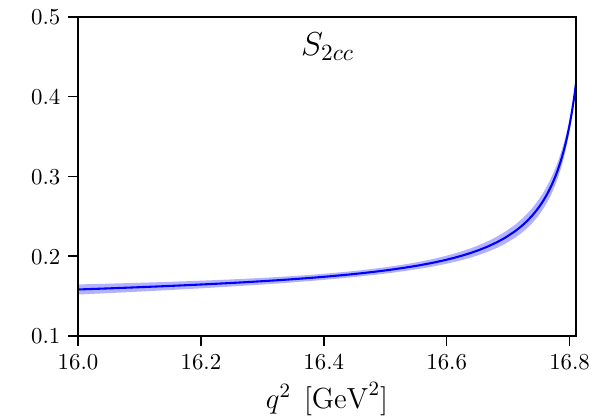}

 \caption{\label{fig:LbLobs1}Updated Standard-Model predictions of the $\Lambda_b \to \Lambda^*(1520)(\to p K^-)\ell^+\ell^-$ observables in the high-$q^2$ region (continued in Fig.~\protect\ref{fig:LbLobs2}). See Refs.~\cite{Meinel:2020owd,Descotes-Genon:2019dbw} for the definitions.}
\end{figure}

\begin{figure}

\includegraphics[width=0.45\linewidth]{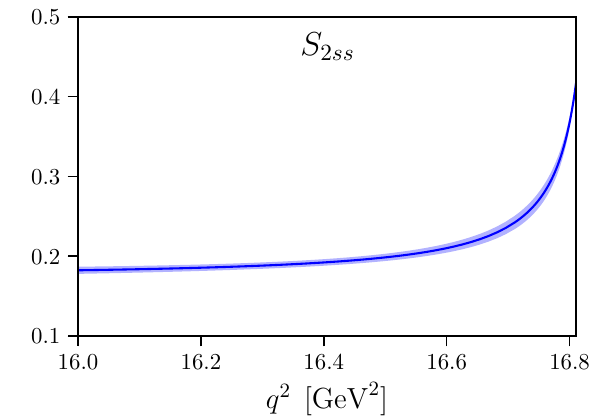} \hfill \includegraphics[width=0.45\linewidth]{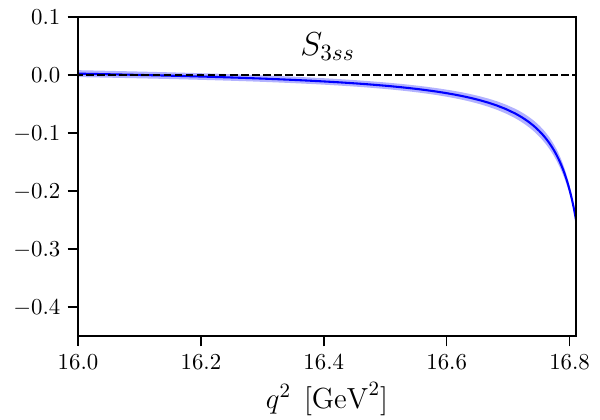}

\includegraphics[width=0.45\linewidth]{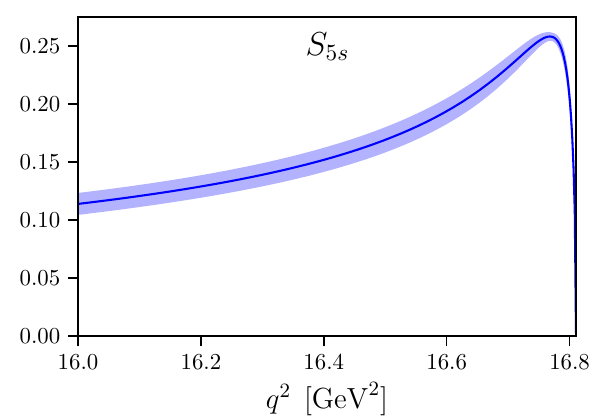} \hfill \includegraphics[width=0.45\linewidth]{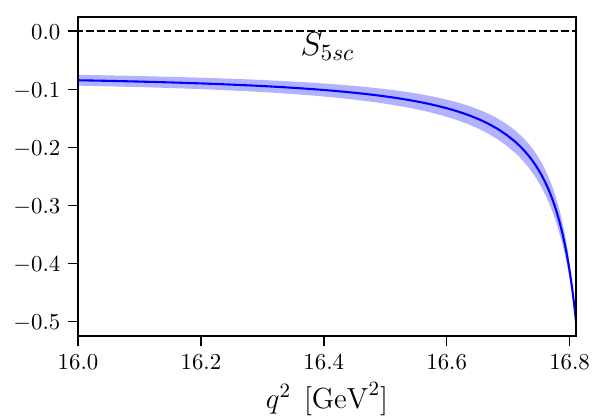}
 \caption{\label{fig:LbLobs2}Updated Standard-Model predictions of the $\Lambda_b \to \Lambda^*(1520)(\to p K^-)\ell^+\ell^-$ observables in the high-$q^2$ region (continuation of Fig.~\protect\ref{fig:LbLobs1}). See Refs.~\cite{Meinel:2020owd,Descotes-Genon:2019dbw} for the definitions.}
\end{figure}

The parameters obtained from the new nominal and higher-order fits are listed in Table \ref{tab:LbLStarFFparams}, and the covariance matrices are provided as supplemental files \cite{Supplemental}. Plots of the fits are shown in Figs.~\ref{fig:LbLstarVA} and \ref{fig:LbLstarT}. The updated Standard-Model predictions of the $\Lambda_b \to \Lambda^*(1520)\ell^+\ell^-$ differential branching fraction and angular observables are shown in Figs.~\ref{fig:LbLobs1} and \ref{fig:LbLobs2} (see Ref.~\cite{Meinel:2020owd} and references therein for the definitions).

\begin{table}
 \begin{tabular}{lllllllll}
  \hline\hline
  $f$                & \hspace{2ex} & \hspace{6ex}$a_0^f$ & & \hspace{4ex}$a_1^f$  & \hspace{2ex} & \hspace{4ex}$a_{0,{\rm HO}}^f$ & & \hspace{4ex}$a_{1,{\rm HO}}^f$  \\
  \hline
  $f_0^{(\frac{3}{2}^-)}$                                 &&   $\wm    1.88(12)$   &&   $   -    7.1(1.8)$    &&   $\wm    1.87(15)$   &&   $   -    7.0(1.8)$     \\
  $f_+^{(\frac{3}{2}^-)}$                                 &&    &&   $\wm   0.779(77)$    &&     &&   $\wm   0.791(88)$                        \\
  $f_{\perp}^{(\frac{3}{2}^-)}$                           &&   &&   $\wm    1.25(15)$    &&      &&   $\wm    1.23(17)$                       \\
  $f_{\perp^{\prime}}^{(\frac{3}{2}^-)}$                  &&   $\wm 0.02052(81)$   &&   $   -  0.060(14)$    &&   $\wm  0.0202(12)$   &&   $   -  0.060(14)$      \\
  $g_0^{(\frac{3}{2}^-)}$                                 &&    &&   $\wm   0.793(67)$    &&    &&   $\wm   0.790(78)$                         \\
  $g_+^{(\frac{3}{2}^-)}$                                 &&    &&   $   -    6.0(1.5)$    &&    &&   $   -    6.0(1.6)$                        \\
  $g_{\perp}^{(\frac{3}{2}^-)}$                           &&   $\wm    1.58(10)$   &&   $   -    5.9(1.4)$    &&   $\wm    1.57(12)$   &&   $   -    5.8(1.6)$     \\
  $g_{\perp^{\prime}}^{(\frac{3}{2}^-)}$                  &&   $   - 0.0204(71)$   &&   $\wm   0.012(84)$    &&   $   - 0.0199(71)$   &&   $\wm   0.012(84)$      \\
  $h_+^{(\frac{3}{2}^-)}$                                 &&    &&   $\wm    1.14(14)$    &&     &&   $\wm    1.13(17)$                        \\
  $h_{\perp}^{(\frac{3}{2}^-)}$                           &&   &&   $\wm   0.767(66)$    &&     &&   $\wm   0.757(86)$                        \\
  $h_{\perp^{\prime}}^{(\frac{3}{2}^-)}$                  &&   $   -0.00900(47)$   &&   $\wm  0.0110(53)$    &&   $   -0.00904(76)$   &&   $\wm  0.0107(53)$      \\
  $\widetilde{h}_+^{(\frac{3}{2}^-)}$                     &&    &&   $   -    5.7(1.4)$    &&    &&   $   -    6.0(1.7)$                        \\
  $\widetilde{h}_{\perp}^{(\frac{3}{2}^-)}$               &&   $\wm    1.63(10)$   &&   $   -    6.2(1.4)$    &&   $\wm    1.64(14)$   &&   $   -    6.0(1.6)$     \\
  $\widetilde{h}_{\perp^{\prime}}^{(\frac{3}{2}^-)}$      &&   $\wm  0.0698(95)$   &&   $   -   0.36(16)$    &&   $\wm   0.071(11)$   &&   $   -   0.36(17)$      \\
  \hline\hline
 \end{tabular}
 \caption{\label{tab:LbLStarFFparams}Updated  $\Lambda_b \to \Lambda^*(1520)$ form-factor parameters. The unlisted parameters should be determined using Eqs.~(\ref{eq:a0fstart})-(\ref{eq:a0fend}) with with $\Lambda_Q=\Lambda_b$ and $\Lambda_{q,3/2}^*=\Lambda^*(1520)$. Also note that the fit functions now include pole factors as shown in Eqs.~(\ref{eq:LbL1520nominal}) and (\ref{eq:LbL1520HO}). Systematic uncertainties in the form factors and derived quantities are evaluated using Eq.~(\ref{eq:sigmasyst}). Machine-readable files containing the parameter values and the covariance matrices are provided as supplemental material \cite{Supplemental}.}
\end{table}

\FloatBarrier
\section{\label{sec:LbLcstarApp}Plots and tables of the improved results for \texorpdfstring{$\bm{\Lambda_b \to \Lambda_c^*(2595,2625)\ell^-\bar{\nu}_\ell}$}{Lambdab to Lambdac* l nu}}
\FloatBarrier

\begin{figure}
 \centerline{\includegraphics[width=0.6\linewidth]{figures/legend_datasets.pdf}}
 
 \vspace{1ex}

 \includegraphics[width=0.47\linewidth]{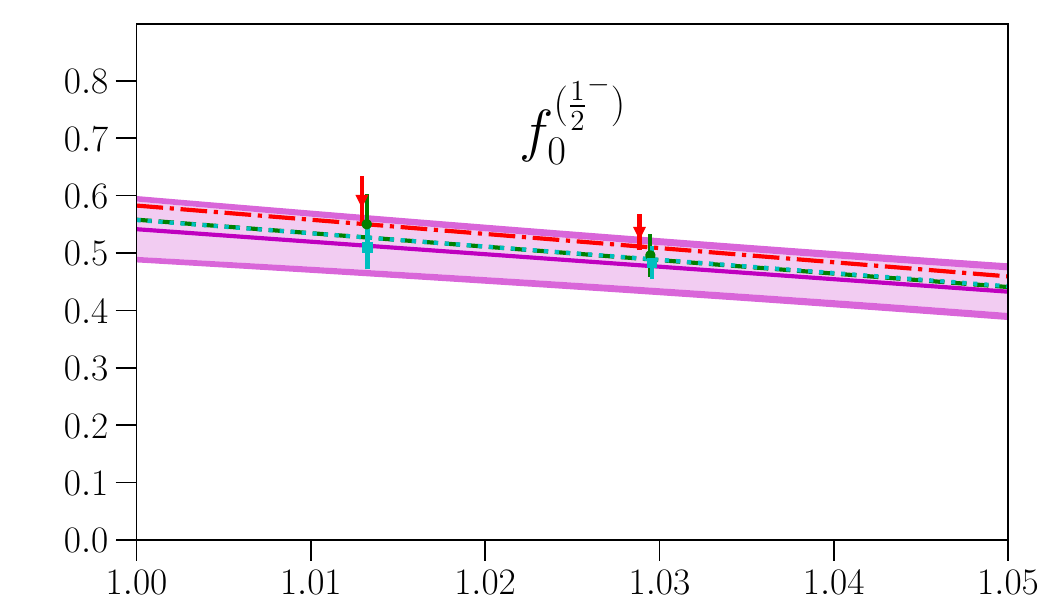} \hfill \includegraphics[width=0.47\linewidth]{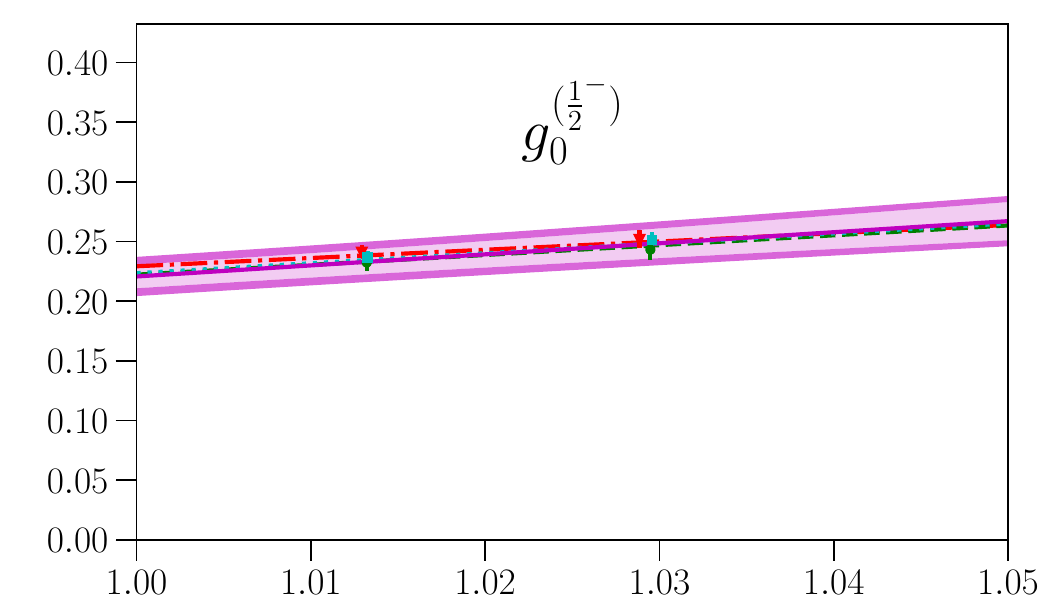} \\
 \includegraphics[width=0.47\linewidth]{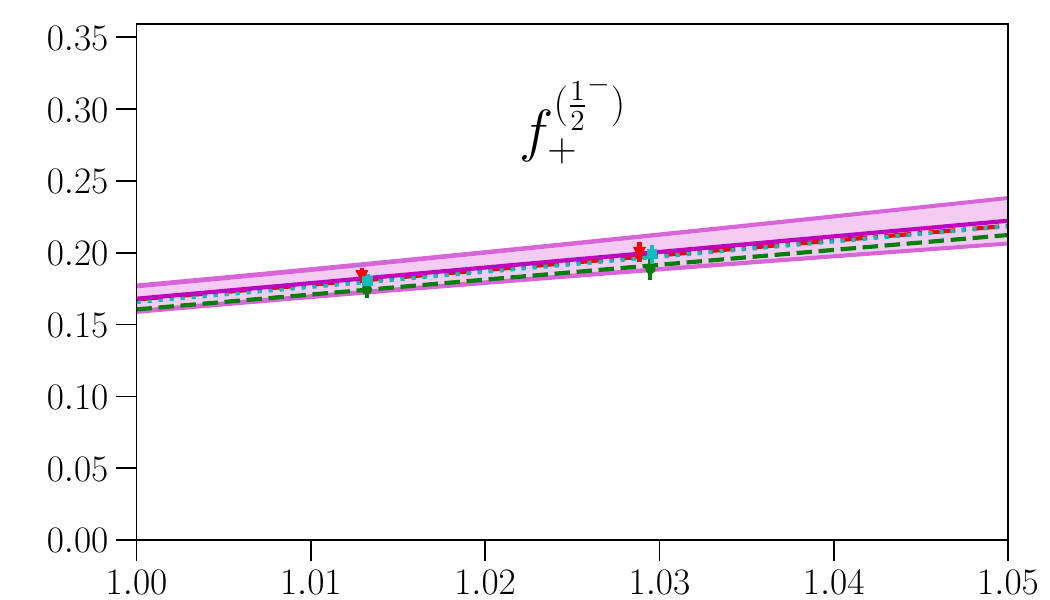} \hfill \includegraphics[width=0.47\linewidth]{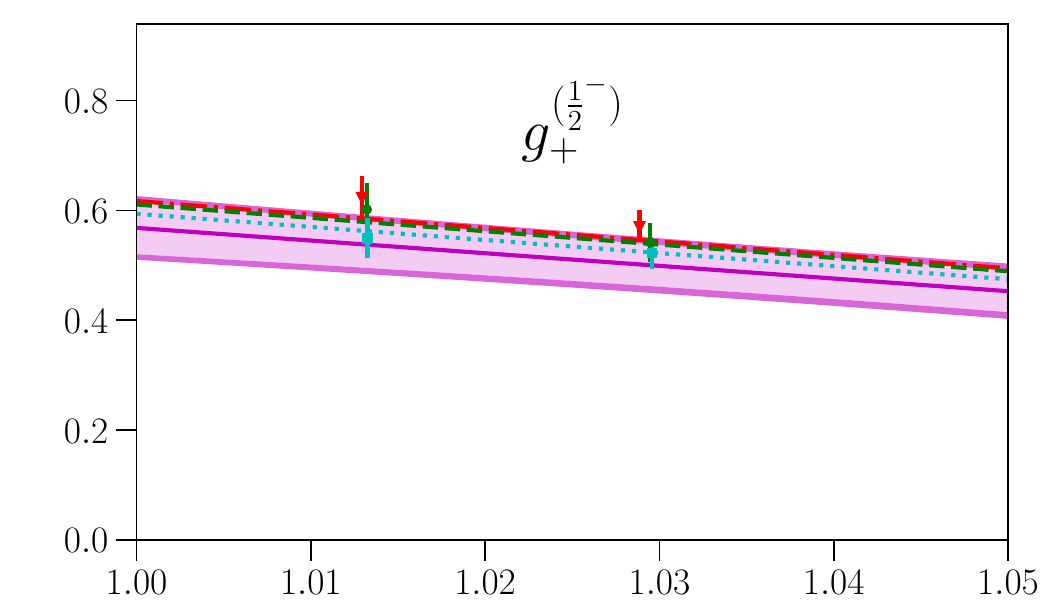} \\
 \includegraphics[width=0.47\linewidth]{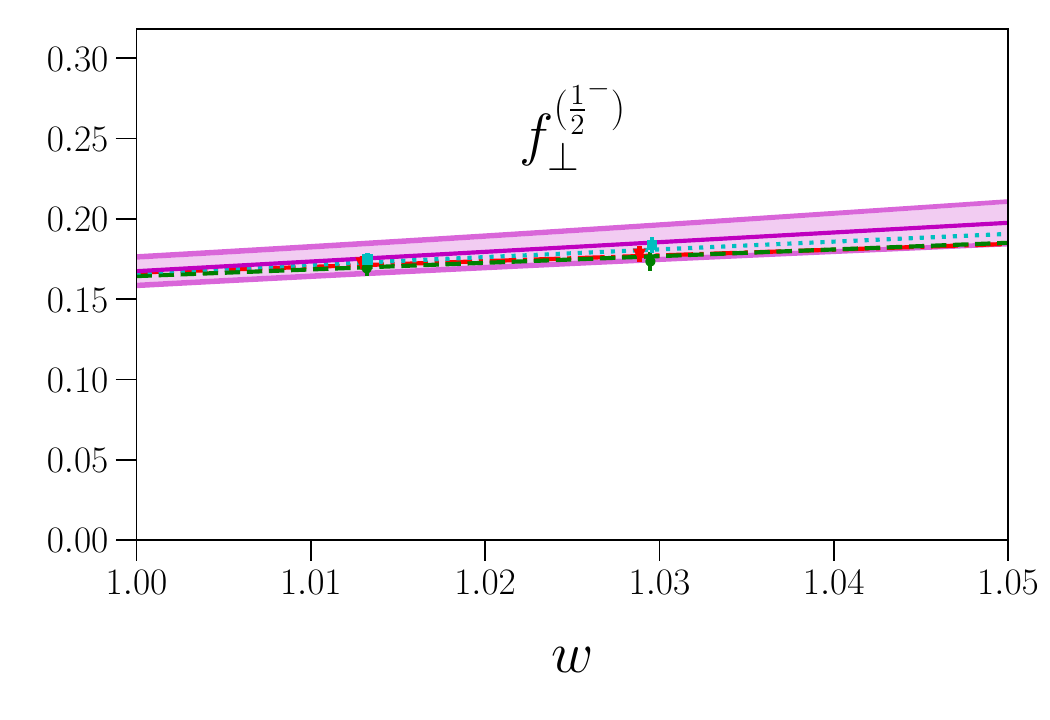} \hfill \includegraphics[width=0.47\linewidth]{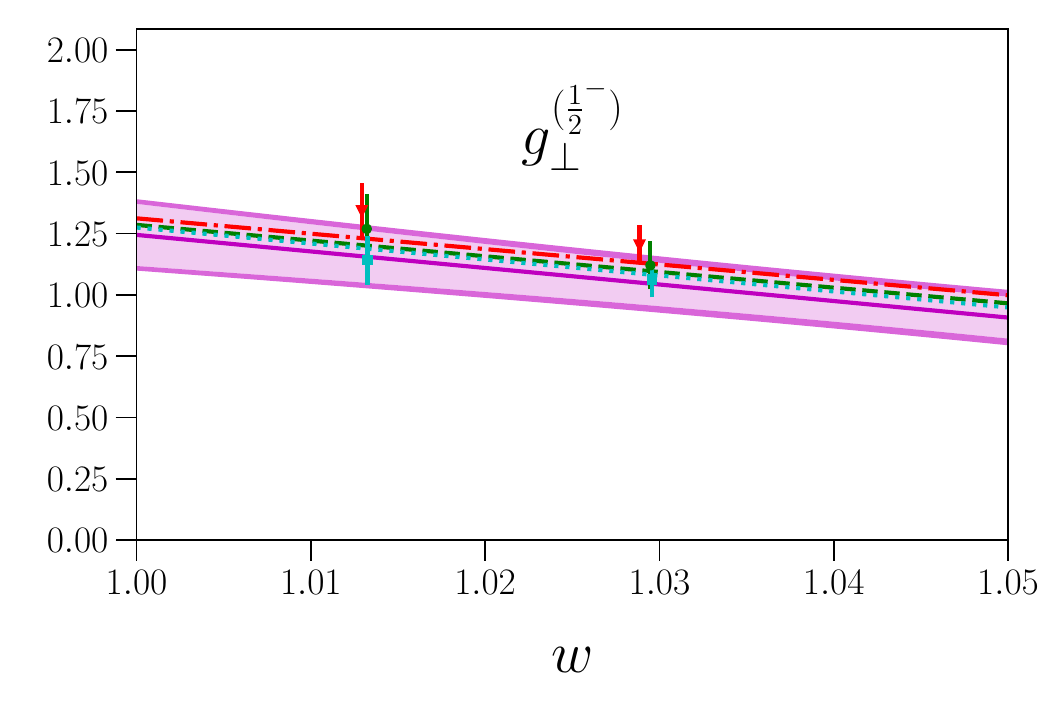} \\

 \caption{\label{fig:FFextrapJ12VA}Updated chiral and continuum extrapolations of the $\Lambda_b \to \Lambda_c^*(2595)$ vector and axial vector form factors.}
\end{figure}

\begin{figure}
 \centerline{\includegraphics[width=0.6\linewidth]{figures/legend_datasets.pdf}}
 
 \vspace{1ex}
 
 \includegraphics[width=0.47\linewidth]{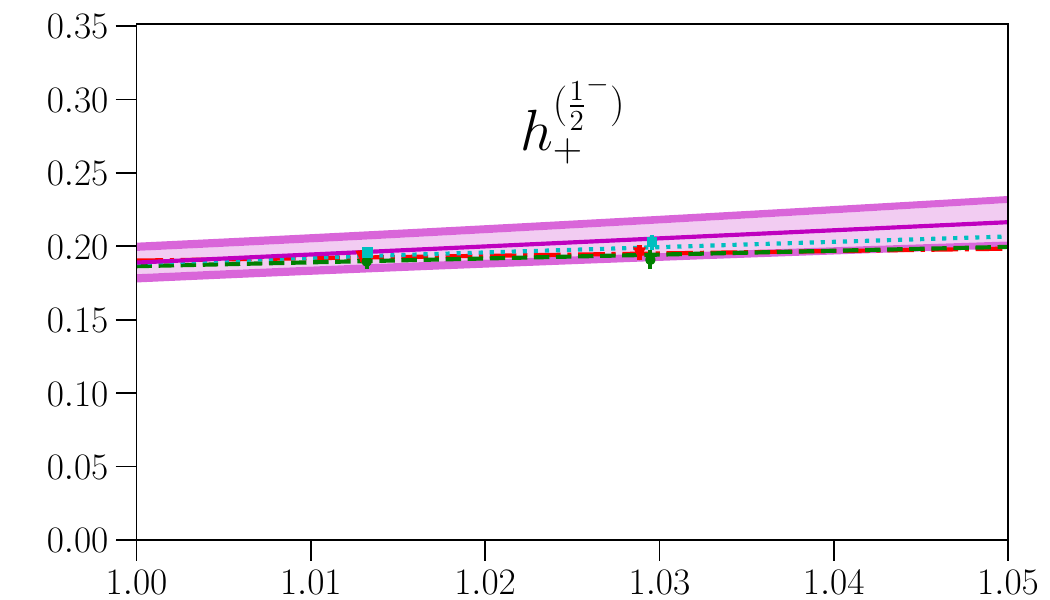} \hfill \includegraphics[width=0.47\linewidth]{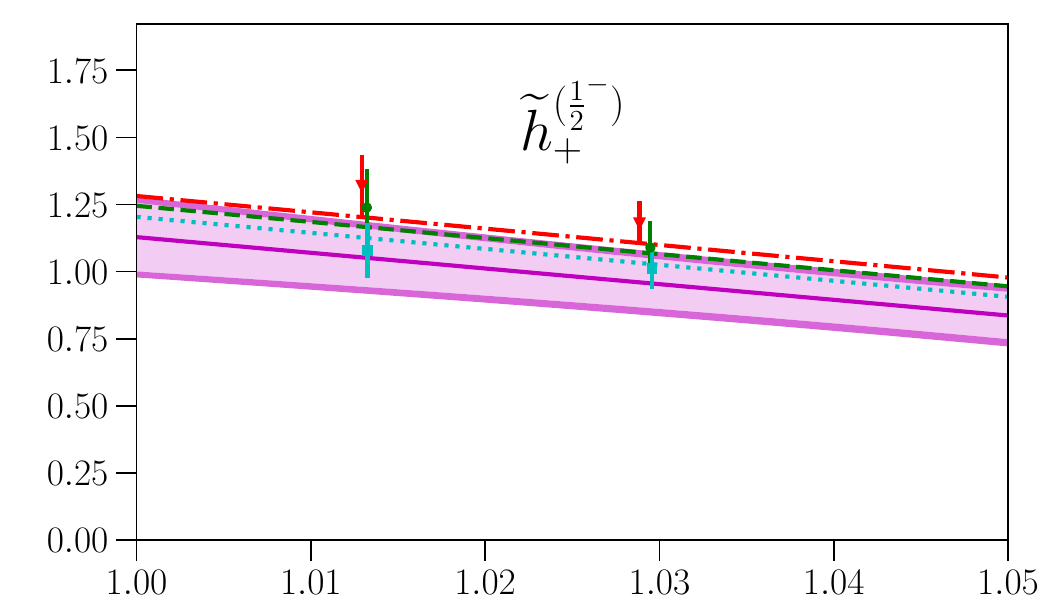} \\
 \includegraphics[width=0.47\linewidth]{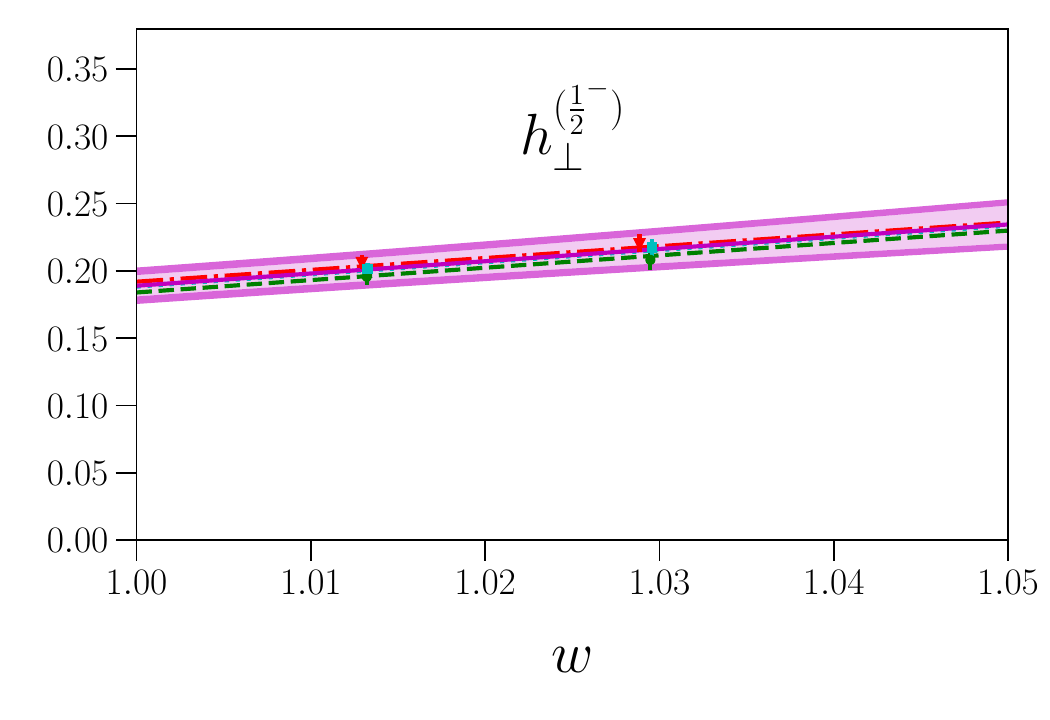} \hfill \includegraphics[width=0.47\linewidth]{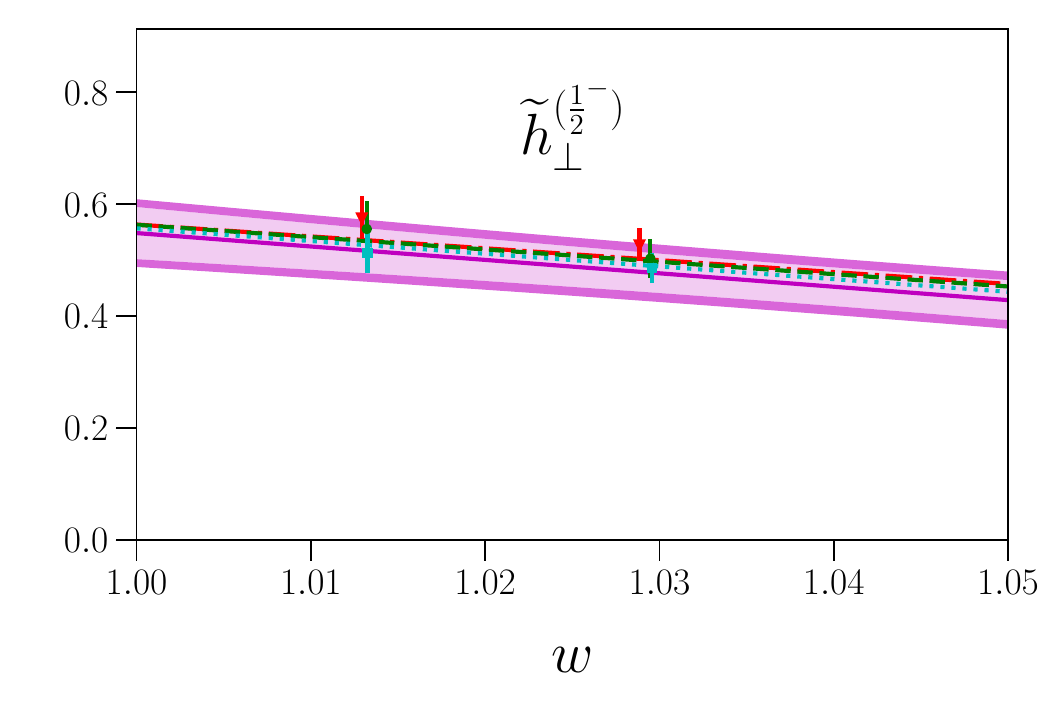} \\

 \caption{\label{fig:FFextrapJ12T}Updated chiral and continuum extrapolations of the $\Lambda_b \to \Lambda_c^*(2595)$ tensor form factors. }
\end{figure}

\begin{figure}
 \centerline{\includegraphics[width=0.6\linewidth]{figures/legend_datasets.pdf}}
 
 \vspace{1ex}
 
 \includegraphics[width=0.47\linewidth]{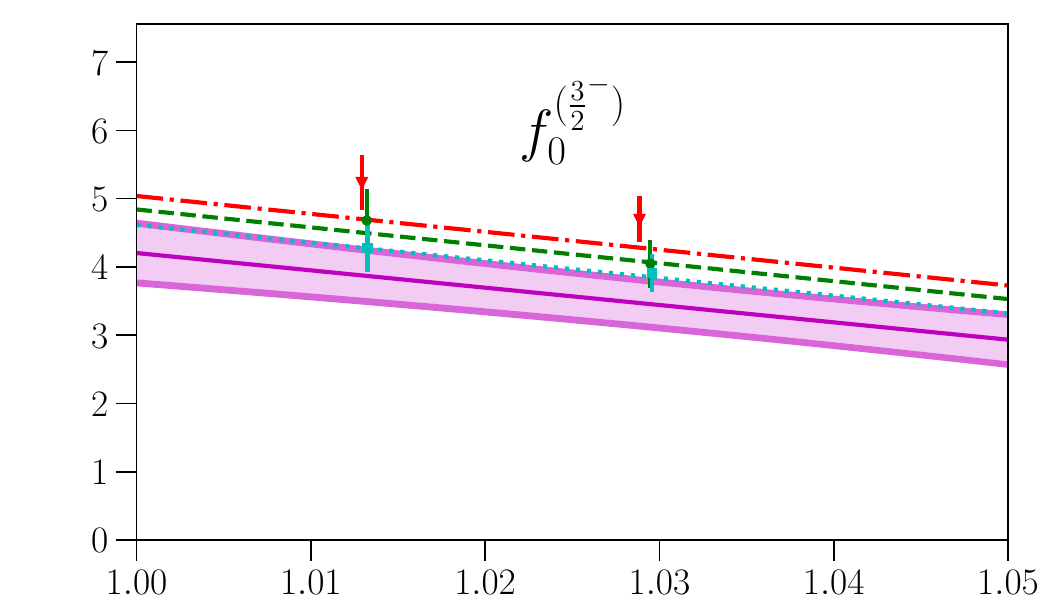} \hfill \includegraphics[width=0.47\linewidth]{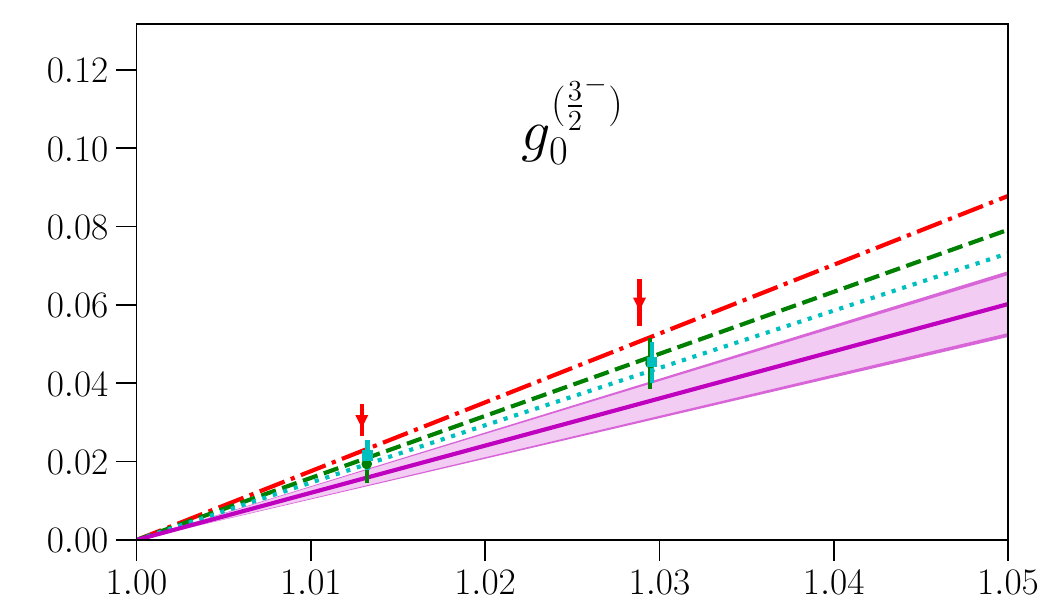} \\
 \includegraphics[width=0.47\linewidth]{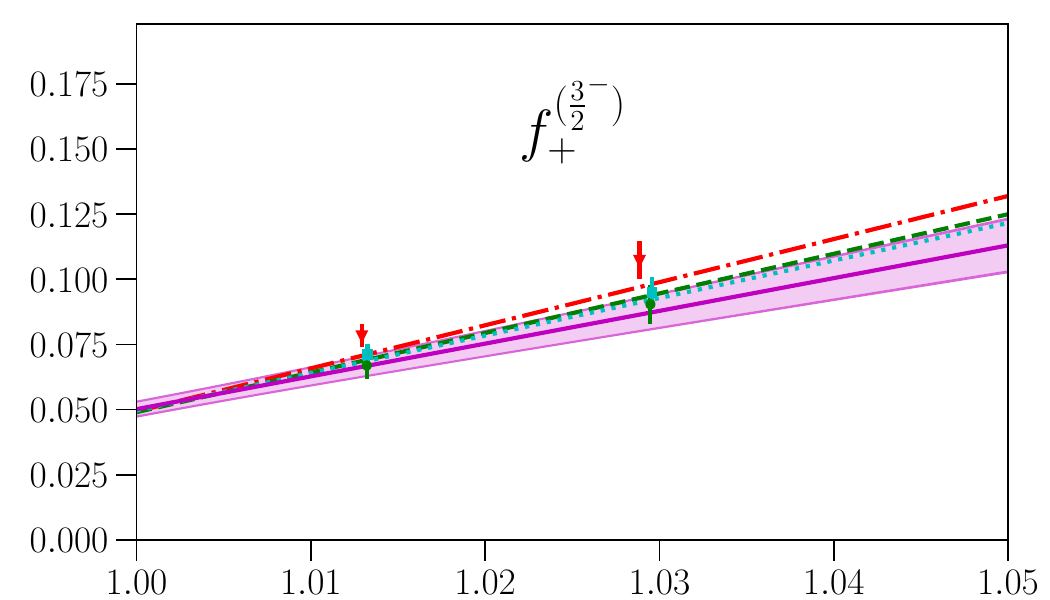} \hfill \includegraphics[width=0.47\linewidth]{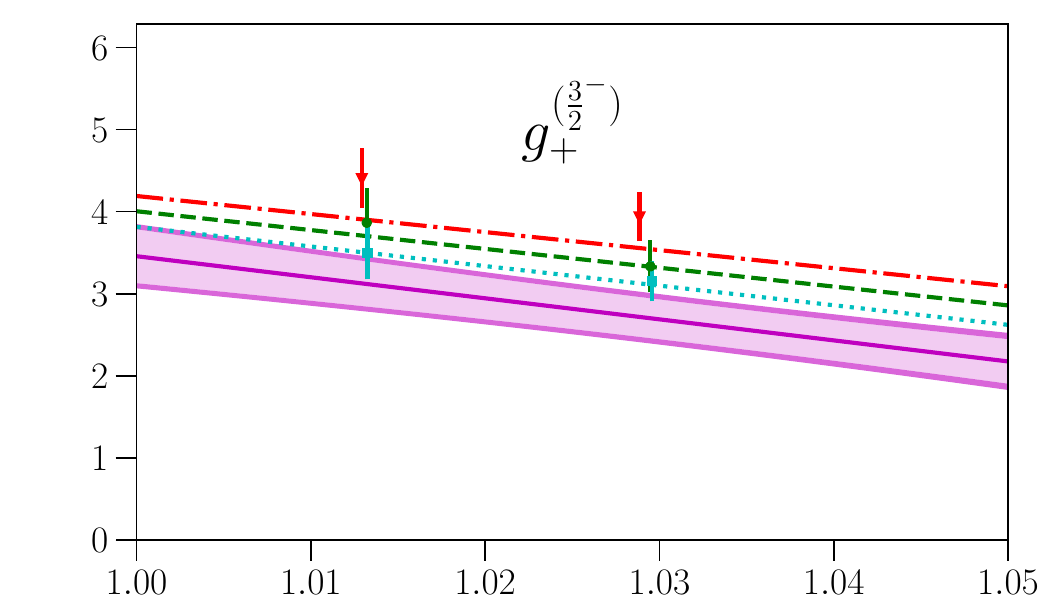} \\
 \includegraphics[width=0.47\linewidth]{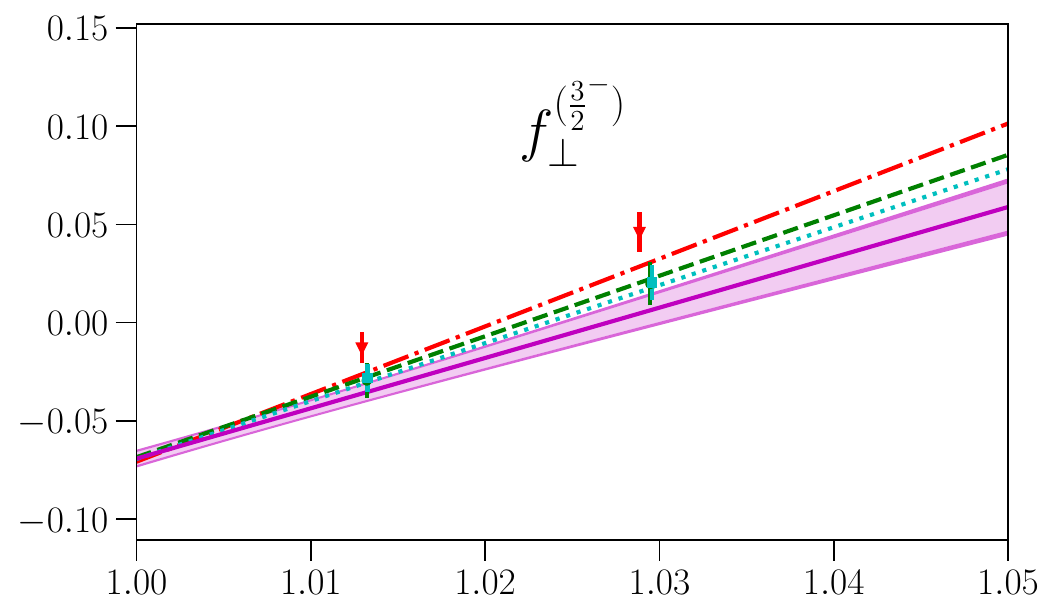} \hfill \includegraphics[width=0.47\linewidth]{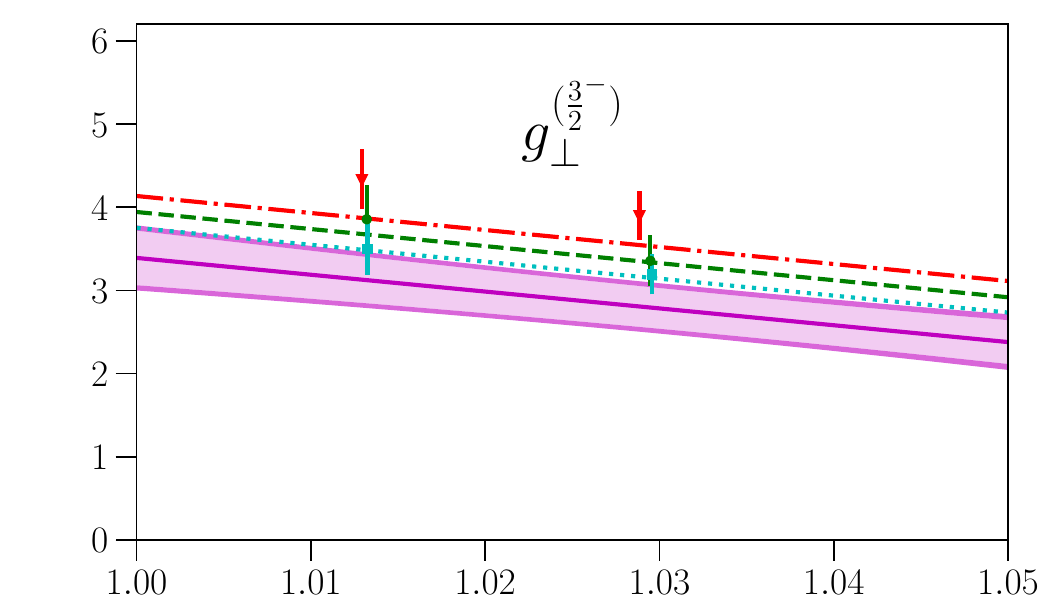} \\
 \includegraphics[width=0.47\linewidth]{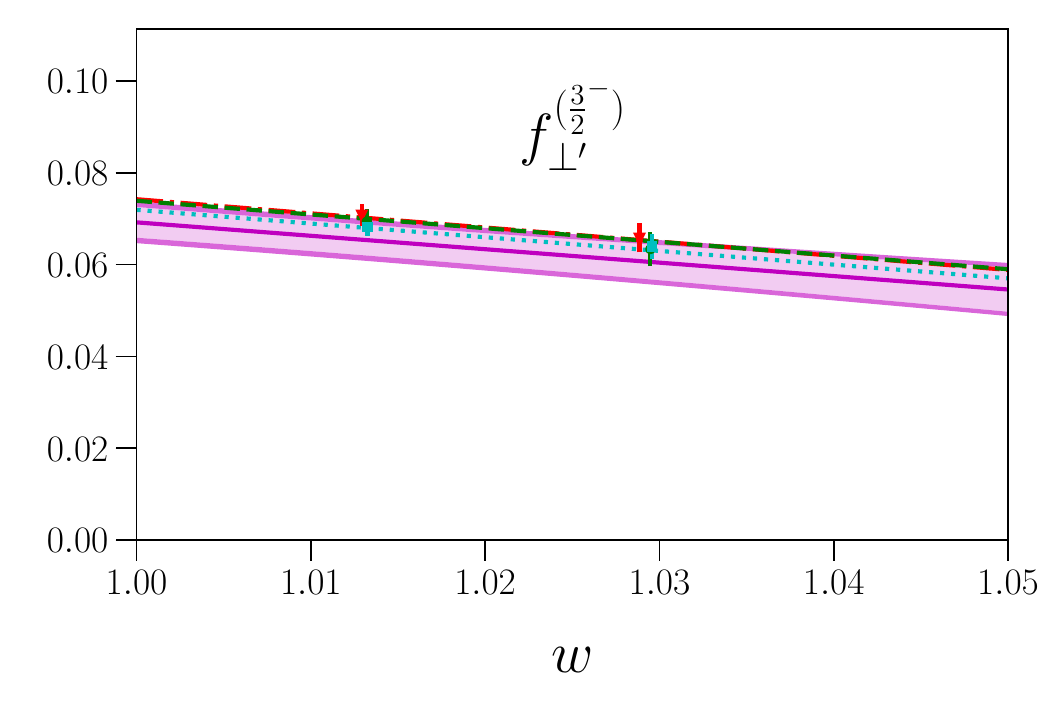} \hfill \includegraphics[width=0.47\linewidth]{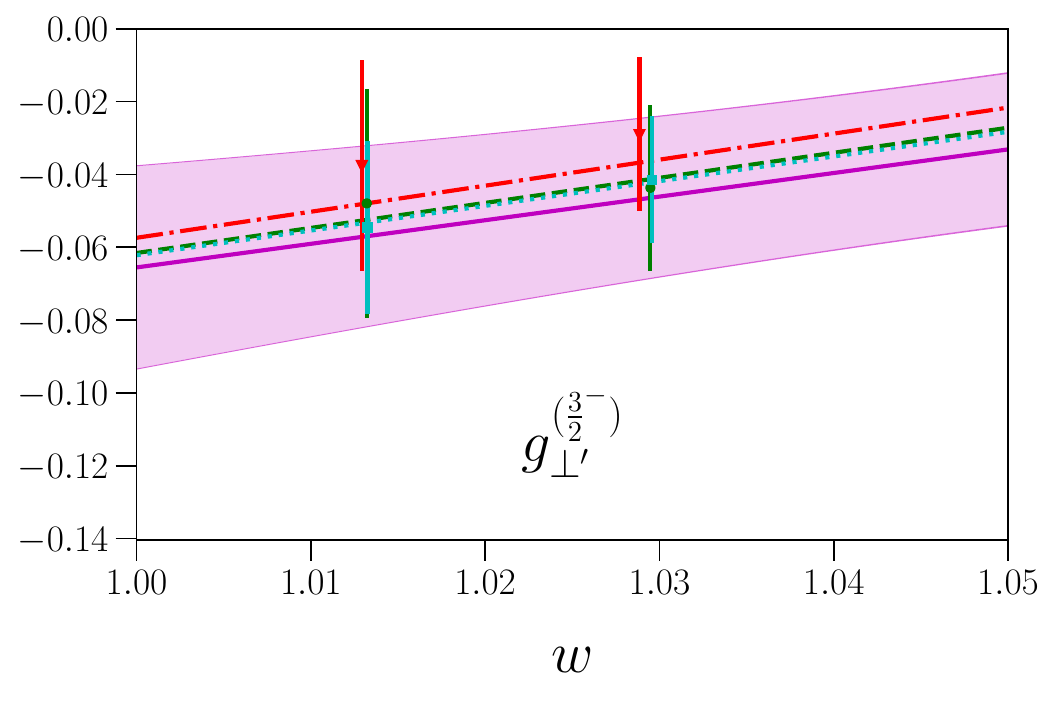} \\

 \caption{\label{fig:FFextrapJ32VA}Updated chiral and continuum extrapolations of the $\Lambda_b \to \Lambda_c^*(2625)$ vector and axial vector form factors. }
\end{figure}

\begin{figure}
 \centerline{\includegraphics[width=0.6\linewidth]{figures/legend_datasets.pdf}}
 
 \vspace{1ex}

 \includegraphics[width=0.47\linewidth]{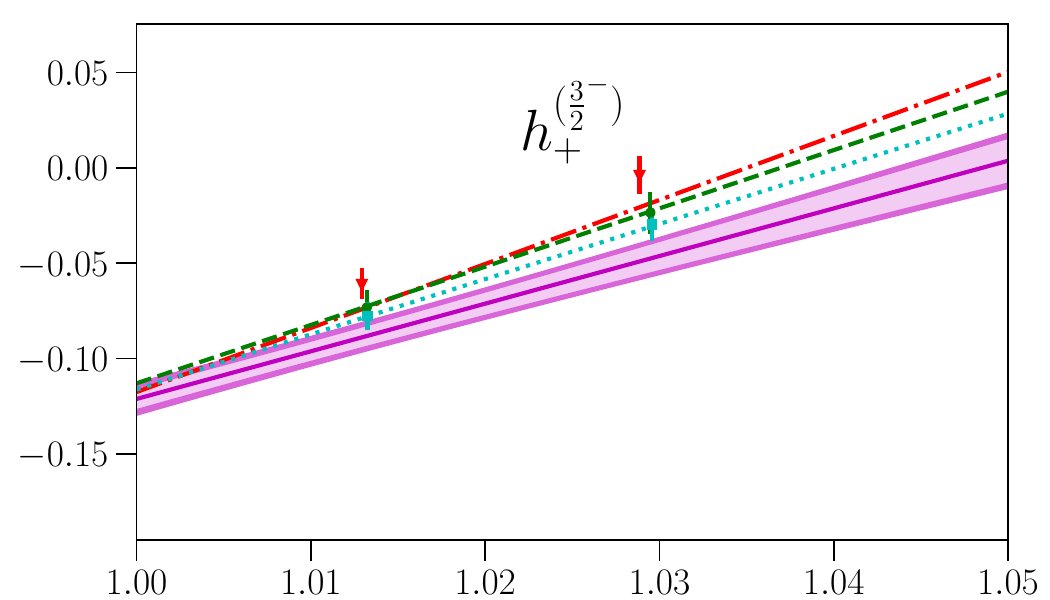} \hfill \includegraphics[width=0.47\linewidth]{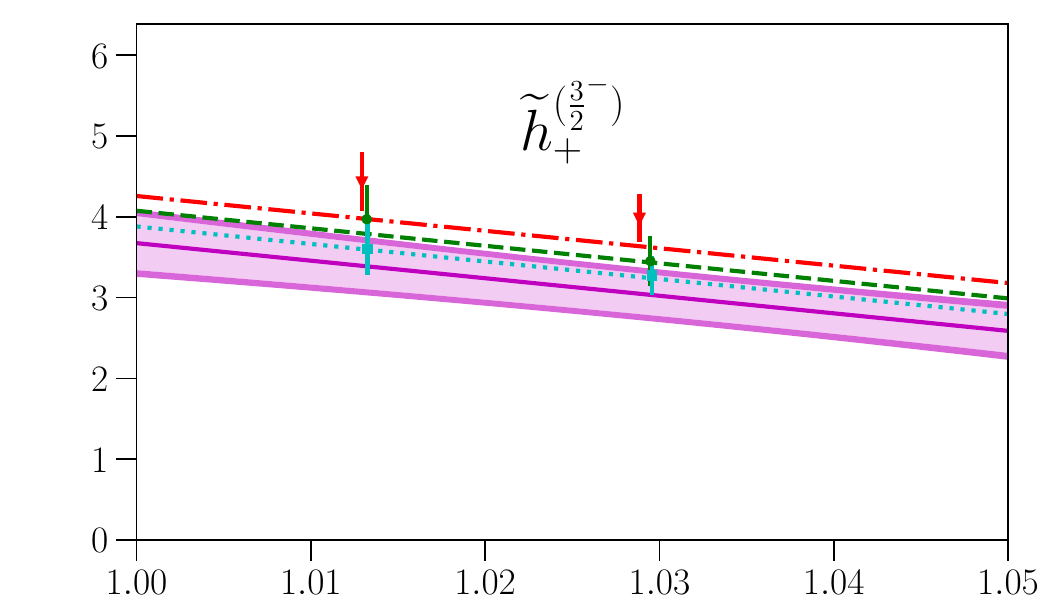} \\
 \includegraphics[width=0.47\linewidth]{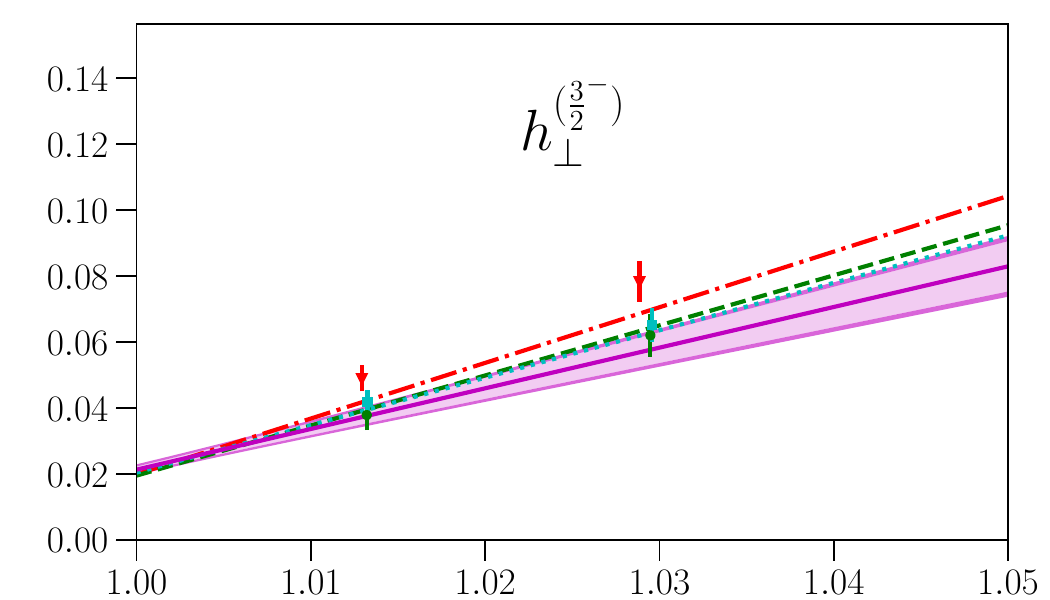} \hfill \includegraphics[width=0.47\linewidth]{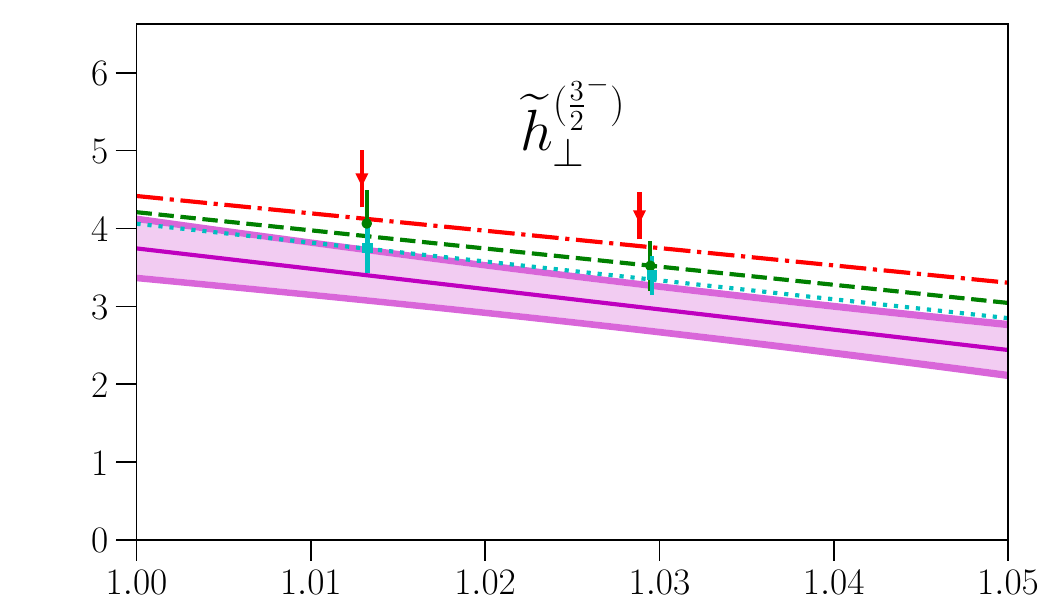} \\
 \includegraphics[width=0.47\linewidth]{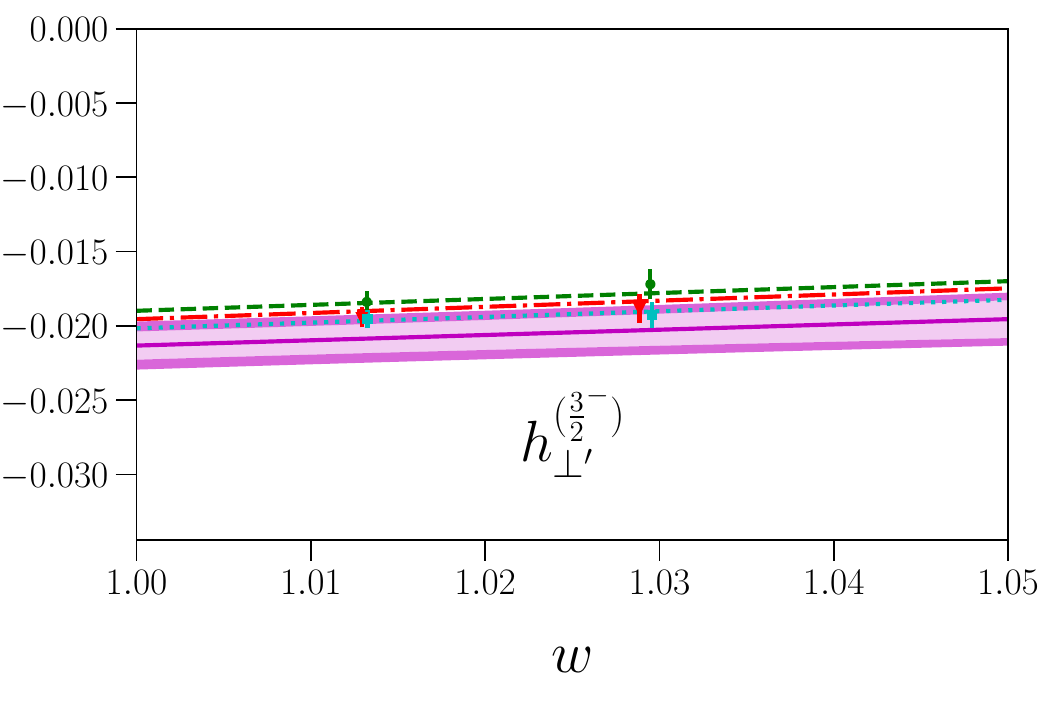} \hfill \includegraphics[width=0.47\linewidth]{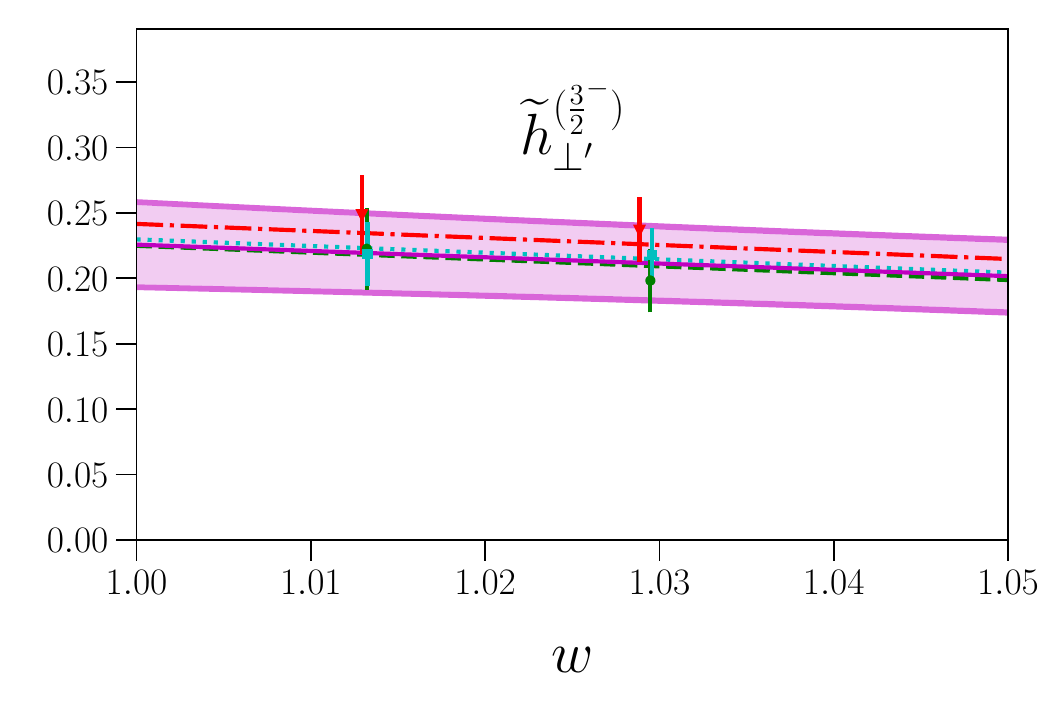} \\
 
 \caption{\label{fig:FFextrapJ32T}Updated chiral and continuum extrapolations of the $\Lambda_b \to \Lambda_c^*(2625)$ tensor form factors. }
\end{figure}

\begin{figure}
 \includegraphics[width=0.6\linewidth]{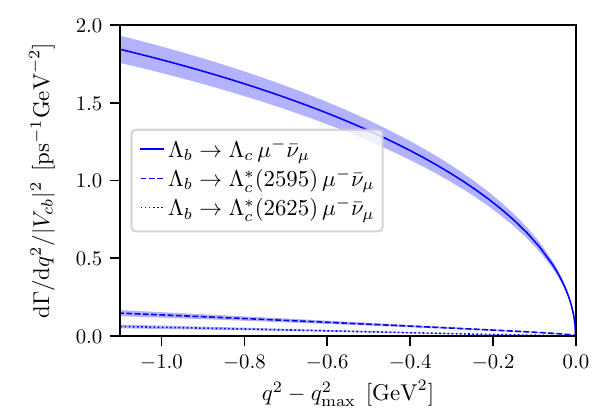}
 
 \caption{\label{fig:LbLcstarobservables2}Updated comparison of the $\Lambda_b \to \Lambda_c\,\mu^-\bar{\nu}_\mu$, $\Lambda_b \to \Lambda_c^*(2595)\mu^-\bar{\nu}_\mu$, and $\Lambda_b \to \Lambda_c^*(2625)\mu^-\bar{\nu}_\mu$ differential decay rates near $q^2_{\rm max}$.}
\end{figure}

\begin{figure}
 \includegraphics[width=0.47\linewidth]{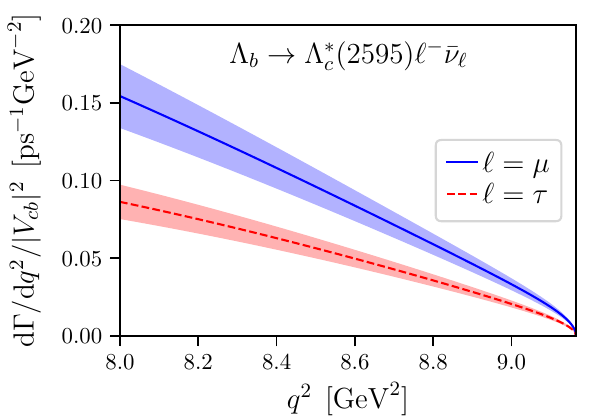} \hfill \includegraphics[width=0.47\linewidth]{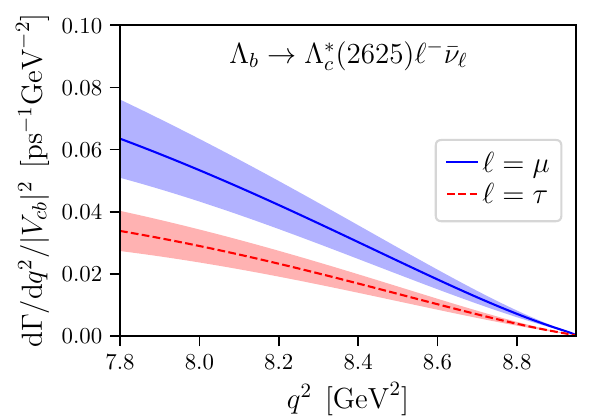}
 
 \includegraphics[width=0.47\linewidth]{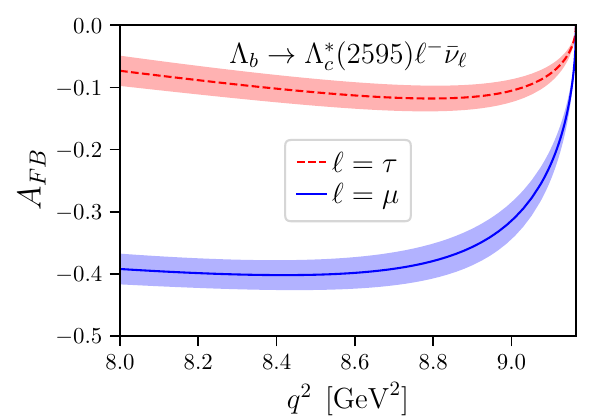} \hfill \includegraphics[width=0.47\linewidth]{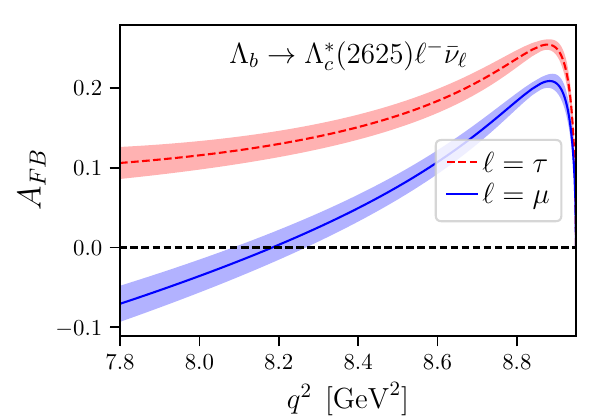}
 
 \includegraphics[width=0.47\linewidth]{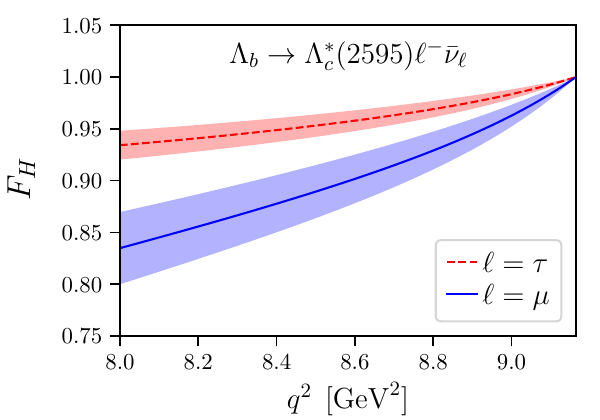} \hfill \includegraphics[width=0.47\linewidth]{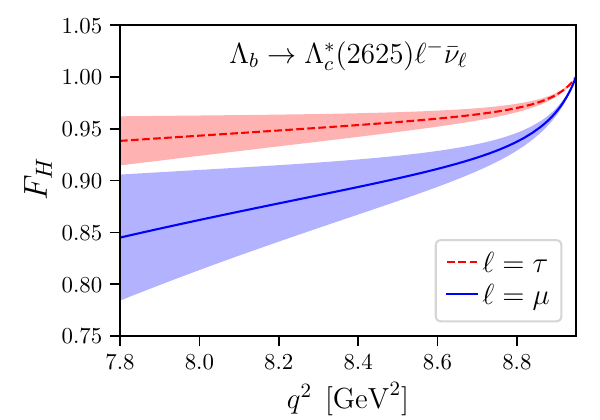}
 
 \caption{\label{fig:LbLcstarobservables}Updated Standard-Model predictions of the $\Lambda_b \to \Lambda_c^*(2595)\ell^-\bar{\nu}_\ell$ (left) and $\Lambda_b \to \Lambda_c^*(2625)\ell^-\bar{\nu}_\ell$ (right) observables in the high-$q^2$ region.}
\end{figure}

The new results for the parameters $F^f$ and $A^f$ (nominal fit) and $F^f_{\rm HO}$, $A^f_{\rm HO}$ (higher-order fit) are listed in Table \ref{tab:LbLcStarFFparams}, and plots of the fits are shown in Figs.~\ref{fig:FFextrapJ12VA}-\ref{fig:FFextrapJ32T}. The covariance matrices of the fit parameters are provided as supplemental material. Updated plots of the $\Lambda_b \to \Lambda_c^*(2595)\ell^-\bar{\nu}_\ell$ and $\Lambda_b \to \Lambda_c^*(2625)\ell^-\bar{\nu}_\ell$ differential decay rates and angular observables near $q^2_{\rm max}$ are shown in Figs.~\ref{fig:LbLcstarobservables2} and \ref{fig:LbLcstarobservables}.

\begin{table}
 \begin{tabular}{lllllllll}
  \hline\hline
  $f$                & \hspace{2ex} & \hspace{6ex}$F^f$ & & \hspace{4ex}$A^f$  & \hspace{2ex} & \hspace{4ex}$F^f_{\rm HO}$ & & \hspace{4ex}$A^f_{\rm HO}$  \\
  \hline
  $f_0^{(\frac{1}{2}^-)}$                                 &&   $\wm   0.541(48)$   &&   $   -   2.18(76)$      &&   $\wm   0.528(58)$   &&   $   -   2.10(74)$\\
  $f_+^{(\frac{1}{2}^-)}$                                 &&   $\wm  0.1680(72)$   &&   $\wm    1.09(24)$      &&   $\wm   0.167(11)$   &&   $\wm    1.10(25)$\\
  $f_{\perp}^{(\frac{1}{2}^-)}$                           &&      &&   $\wm    0.60(18)$                       &&     &&   $\wm    0.53(19)$\\
  $g_0^{(\frac{1}{2}^-)}$                                 &&   $\wm   0.2207(99)$   &&   $\wm    0.93(25)$     &&   $\wm   0.221(16)$   &&   $\wm    0.86(25)$\\
  $g_+^{(\frac{1}{2}^-)}$                                 &&   $\wm   0.568(48)$   &&   $   -   2.31(76)$      &&   $\wm   0.561(58)$   &&   $   -   2.27(76)$\\
  $g_{\perp}^{(\frac{1}{2}^-)}$                           &&   $\wm    1.24(13)$   &&   $   -    6.7(2.2)$     &&   $\wm    1.22(15)$   &&   $   -    6.5(2.2)$\\
  $h_+^{(\frac{1}{2}^-)}$                                 &&   $\wm  0.1889(80)$   &&   $\wm    0.55(21)$      &&   $\wm   0.190(14)$   &&   $\wm    0.49(22)$\\
  $h_{\perp}^{(\frac{1}{2}^-)}$                           &&    &&   $\wm    0.91(23)$                         &&     &&   $\wm    0.89(23)$\\
  $\widetilde{h}_+^{(\frac{1}{2}^-)}$                     &&   $\wm    1.13(13)$   &&   $   -    5.8(2.1)$     &&   $\wm    1.12(15)$   &&   $   -    5.8(2.1)$\\
  $\widetilde{h}_{\perp}^{(\frac{1}{2}^-)}$               &&   $\wm   0.548(47)$   &&   $   -   2.40(81)$      &&   $\wm   0.543(60)$   &&   $   -   2.36(82)$\\
  $f_0^{(\frac{3}{2}^-)}$                                 &&   $\wm    4.20(39)$   &&   $   -   25.4(8.0)$       &&   $\wm    4.32(49)$   &&   $   -   27.4(8.9)$\\
  $f_+^{(\frac{3}{2}^-)}$                                 &&     &&   $\wm    1.26(18)$                        &&     &&   $\wm    1.23(20)$\\
  $f_{\perp}^{(\frac{3}{2}^-)}$                           &&   &&   $\wm    2.56(25)$                          &&    &&   $\wm    2.58(30)$\\
  $f_{\perp^{\prime}}^{(\frac{3}{2}^-)}$                  &&   $\wm  0.0692(34)$   &&   $   -  0.292(80)$      &&   $\wm  0.0701(45)$   &&   $   -  0.295(83)$\\
  $g_0^{(\frac{3}{2}^-)}$                                 &&   &&   $\wm    1.20(15)$                          &&   &&   $\wm    1.21(17)$\\
  $g_+^{(\frac{3}{2}^-)}$                                 &&     &&   $   -   25.6(6.9)$                       &&   &&   $   -   26.3(8.1)$\\
  $g_{\perp}^{(\frac{3}{2}^-)}$                           &&   $\wm    3.39(33)$   &&   $   -   20.2(6.4)$     &&   $\wm    3.50(39)$   &&   $   -   23.1(7.8)$\\
  $g_{\perp^{\prime}}^{(\frac{3}{2}^-)}$                  &&   $   -  0.066(28)$   &&   $\wm    0.65(38)$      &&   $   -  0.065(28)$   &&   $\wm    0.63(39)$\\
  $h_+^{(\frac{3}{2}^-)}$                                 &&    &&   $\wm    2.50(24)$                         &&    &&   $\wm    2.63(32)$\\
  $h_{\perp}^{(\frac{3}{2}^-)}$                           &&    &&   $\wm    1.23(15)$                         &&   &&   $\wm    1.22(18)$\\
  $h_{\perp^{\prime}}^{(\frac{3}{2}^-)}$                  &&   $   - 0.02133(95)$   &&   $\wm   0.036(17)$     &&   $   - 0.0220(16)$   &&   $\wm   0.035(17)$\\
  $\widetilde{h}_+^{(\frac{3}{2}^-)}$                     &&   &&   $   -   21.7(6.5)$                         &&    &&   $   -   25.0(8.2)$\\
  $\widetilde{h}_{\perp}^{(\frac{3}{2}^-)}$               &&   $\wm    3.74(34)$   &&   $   -   26.1(7.1)$     &&   $\wm    3.90(42)$   &&   $   -   27.2(8.4)$\\
  $\widetilde{h}_{\perp^{\prime}}^{(\frac{3}{2}^-)}$      &&   $\wm   0.226(30)$   &&   $   -   0.48(36)$      &&   $\wm   0.224(35)$   &&   $   -   0.50(37)$\\
  \hline\hline
 \end{tabular}
 \caption{\label{tab:LbLcStarFFparams}Updated  $\Lambda_b \to \Lambda_c^*(2595)$ and $\Lambda_b \to \Lambda_c^*(2625)$ form-factor parameters. The nominal and higher-order form factors are given by Eqs.~(\ref{eq:LbLcstarphysicalFF}) and (\ref{eq:LbLcstarphysicalFFHO}), respectively. Systematic uncertainties are evaluated using Eq.~(\ref{eq:sigmasyst}). The unlisted parameters for the $J^P=\frac12^-$ final state are given by Eqs.~(\ref{eq:J12Ffirst}) and (\ref{eq:J12Flast}). The unlisted parameters for the $J^P=\frac32^-$ final state are given by Eqs.~(\ref{eq:a0fstart})-(\ref{eq:a0fend}) with $\Lambda_Q=\Lambda_b$, $\Lambda_{q,3/2}^*=\Lambda_c^*(2625)$ and with renamed parameters $a_0^f\mapsto F^f$.  Machine-readable files containing the parameter values and the covariance matrices are provided as supplemental material \cite{Supplemental}.}
\end{table}

\FloatBarrier

\providecommand{\href}[2]{#2}\begingroup\raggedright\endgroup


\begin{thebibliography}{10}

\bibitem{Aaij:2015bfa}
{\bfseries LHCb} Collaboration, R.~Aaij {\em et~al.}, ``{Determination of the
  quark coupling strength $|V_{ub}|$ using baryonic decays},''
  \href{https://dx.doi.org/10.1038/nphys3415}{Nature Phys. {\bfseries 11} (2015)
  743--747}, \href{https://arxiv.org/abs/1504.01568}{{\ttfamily arXiv:1504.01568
  [hep-ex]}}.

\bibitem{Blake:2019guk}
T.~Blake, S.~Meinel, and D.~van Dyk, ``{Bayesian Analysis of $b\to s\mu^+\mu^-$
  Wilson Coefficients using the Full Angular Distribution of $\Lambda_b\to
  \Lambda(\to p\, \pi^-)\mu^+\mu^-$ Decays},''
  \href{https://dx.doi.org/10.1103/PhysRevD.101.035023}{Phys. Rev. D {\bfseries
  101} no.~3, (2020) 035023}, \href{https://arxiv.org/abs/1912.05811}{{\ttfamily
  arXiv:1912.05811 [hep-ph]}}.

\bibitem{Bowler:1997ej}
{\bfseries UKQCD} Collaboration, K.~C. Bowler, R.~D. Kenway, L.~Lellouch,
  J.~Nieves, O.~Oliveira, D.~G. Richards, C.~T. Sachrajda, N.~Stella, and
  P.~Ueberholz, ``{First lattice study of semileptonic decays of $\Lambda_b$
  and $\Xi_b$ baryons},''
  \href{https://dx.doi.org/10.1103/PhysRevD.57.6948}{Phys. Rev. D {\bfseries 57}
  (1998) 6948--6974}, \href{https://arxiv.org/abs/hep-lat/9709028}{{\ttfamily
  arXiv:hep-lat/9709028}}.

\bibitem{Gottlieb:2003yb}
S.~A. Gottlieb and S.~Tamhankar, ``{A Lattice study of $\Lambda_b$ semileptonic
  decay},'' \href{https://dx.doi.org/10.1016/S0920-5632(03)01612-8}{Nucl. Phys.
  B Proc. Suppl. {\bfseries 119} (2003) 644--646},
  \href{https://arxiv.org/abs/hep-lat/0301022}{{\ttfamily
  arXiv:hep-lat/0301022}}.

\bibitem{Detmold:2015aaa}
W.~Detmold, C.~Lehner, and S.~Meinel, ``{$\Lambda_b \to p \ell^-
  \bar{\nu}_\ell$ and $\Lambda_b \to \Lambda_c \ell^- \bar{\nu}_\ell$ form
  factors from lattice QCD with relativistic heavy quarks},''
  \href{https://dx.doi.org/10.1103/PhysRevD.92.034503}{Phys. Rev. {\bfseries
  D92} no.~3, (2015) 034503},
\href{https://arxiv.org/abs/1503.01421}{{\ttfamily arXiv:1503.01421 [hep-lat]}}.

\bibitem{Datta:2017aue}
A.~Datta, S.~Kamali, S.~Meinel, and A.~Rashed, ``{Phenomenology of $
  {\Lambda}_b\to {\Lambda}_c\tau {\overline{\nu}}_{\tau } $ using lattice QCD
  calculations},'' \href{https://dx.doi.org/10.1007/JHEP08(2017)131}{JHEP
  {\bfseries 08} (2017) 131}, \href{https://arxiv.org/abs/1702.02243}{{\ttfamily
  arXiv:1702.02243 [hep-ph]}}.

\bibitem{Detmold:2013nia}
W.~Detmold, C.~J.~D. Lin, S.~Meinel, and M.~Wingate, ``{$\Lambda_b \to p l^-
  \bar{\nu}_\ell$ form factors from lattice QCD with static $b$ quarks},''
  \href{https://dx.doi.org/10.1103/PhysRevD.88.014512}{Phys. Rev. D {\bfseries
  88} no.~1, (2013) 014512}, \href{https://arxiv.org/abs/1306.0446}{{\ttfamily
  arXiv:1306.0446 [hep-lat]}}.

\bibitem{Detmold:2012vy}
W.~Detmold, C.~J.~D. Lin, S.~Meinel, and M.~Wingate, ``{$\Lambda_b\to \Lambda
  \ell^+\ell^-$ form factors and differential branching fraction from lattice
  QCD},'' \href{https://dx.doi.org/10.1103/PhysRevD.87.074502}{Phys. Rev.
  {\bfseries D87} no.~7, (2013) 074502},
\href{https://arxiv.org/abs/1212.4827}{{\ttfamily arXiv:1212.4827 [hep-lat]}}.

\bibitem{Detmold:2016pkz}
W.~Detmold and S.~Meinel, ``{$\Lambda_b \to \Lambda \ell^+ \ell^-$ form
  factors, differential branching fraction, and angular observables from
  lattice QCD with relativistic $b$ quarks},''
  \href{https://dx.doi.org/10.1103/PhysRevD.93.074501}{Phys. Rev. {\bfseries
  D93} no.~7, (2016) 074501},
\href{https://arxiv.org/abs/1602.01399}{{\ttfamily arXiv:1602.01399 [hep-lat]}}.

\bibitem{Meinel:2016dqj}
S.~Meinel, ``{$\Lambda_c \to \Lambda l^+ \nu_l$ form factors and decay rates
  from lattice QCD with physical quark masses},''
  \href{https://dx.doi.org/10.1103/PhysRevLett.118.082001}{Phys. Rev. Lett.
  {\bfseries 118} no.~8, (2017) 082001},
  \href{https://arxiv.org/abs/1611.09696}{{\ttfamily arXiv:1611.09696
  [hep-lat]}}.

\bibitem{Meinel:2017ggx}
S.~Meinel, ``{$\Lambda_c \to N$ form factors from lattice QCD and phenomenology
  of $\Lambda_c \to n \ell^+ \nu_\ell$ and $\Lambda_c \to p \mu^+ \mu^-$
  decays},'' \href{https://dx.doi.org/10.1103/PhysRevD.97.034511}{Phys. Rev. D
  {\bfseries 97} no.~3, (2018) 034511},
  \href{https://arxiv.org/abs/1712.05783}{{\ttfamily arXiv:1712.05783
  [hep-lat]}}.

\bibitem{Zhang:2021oja}
Q.-A. Zhang, J.~Hua, F.~Huang, R.~Li, Y.~Li, C.-D. Lu, P.~Sun, W.~Sun, W.~Wang,
  and Y.-B. Yang, ``{First lattice QCD calculation of semileptonic decays of
  charmed-strange baryons \ensuremath{\Xi}$_{c}$},''
  \href{https://dx.doi.org/10.1088/1674-1137/ac2b12}{Chin. Phys. C {\bfseries
  46} no.~7, (2022) 011002}, \href{https://arxiv.org/abs/2103.07064}{{\ttfamily
  arXiv:2103.07064 [hep-lat]}}.

\bibitem{Meinel:2020owd}
S.~Meinel and G.~Rendon, ``{$\Lambda_b \to \Lambda^*(1520)\ell^+\ell^-$ form
  factors from lattice QCD},''
  \href{https://dx.doi.org/10.1103/PhysRevD.103.074505}{Phys. Rev. D {\bfseries
  103} no.~7, (2021) 074505}, \href{https://arxiv.org/abs/2009.09313}{{\ttfamily
  arXiv:2009.09313 [hep-lat]}}.

\bibitem{Meinel:2021rbm}
S.~Meinel and G.~Rendon, ``{$\Lambda_b \to
  \Lambda_c^*(2595,2625)\ell^-\bar{\nu}$ form factors from lattice QCD},''
  \href{https://dx.doi.org/10.1103/PhysRevD.103.094516}{Phys. Rev. D {\bfseries
  103} no.~9, (2021) 094516}, \href{https://arxiv.org/abs/2103.08775}{{\ttfamily
  arXiv:2103.08775 [hep-lat]}}.

\bibitem{Legger:2006cq}
F.~Legger and T.~Schietinger, ``{Photon helicity in $\Lambda_b \to p K \gamma$
  decays},'' \href{https://dx.doi.org/10.1016/j.physletb.2006.12.011,
  10.1016/j.physletb.2007.02.044}{Phys. Lett. {\bfseries B645} (2007)
  204--212}, \href{https://arxiv.org/abs/hep-ph/0605245}{{\ttfamily
  arXiv:hep-ph/0605245 [hep-ph]}}.
[Erratum: Phys. Lett.B647,527(2007)].

\bibitem{Hiller:2007ur}
G.~Hiller, M.~Knecht, F.~Legger, and T.~Schietinger, ``{Photon polarization
  from helicity suppression in radiative decays of polarized $\Lambda_b$ to
  spin-3/2 baryons},''
  \href{https://dx.doi.org/10.1016/j.physletb.2007.03.056}{Phys. Lett.
  {\bfseries B649} (2007) 152--158},
\href{https://arxiv.org/abs/hep-ph/0702191}{{\ttfamily arXiv:hep-ph/0702191
  [hep-ph]}}.

\bibitem{Boer:2018vpx}
P.~Böer, M.~Bordone, E.~Graverini, P.~Owen, M.~Rotondo, and D.~Van~Dyk,
  ``{Testing lepton flavour universality in semileptonic $\Lambda_b \to
  \Lambda_c^*$ decays},'' \href{https://dx.doi.org/10.1007/JHEP06(2018)155}{JHEP
  {\bfseries 06} (2018) 155},
\href{https://arxiv.org/abs/1801.08367}{{\ttfamily arXiv:1801.08367 [hep-ph]}}.

\bibitem{Descotes-Genon:2019dbw}
S.~Descotes-Genon and M.~Novoa~Brunet, ``{Angular analysis of the rare decay
  $\Lambda_b\to \Lambda(1520)(\to NK)\ell^+\ell^-$},''
  \href{https://dx.doi.org/10.1007/JHEP06(2019)136}{JHEP {\bfseries 06} (2019)
  136},
\href{https://arxiv.org/abs/1903.00448}{{\ttfamily arXiv:1903.00448 [hep-ph]}}.

\bibitem{Das:2020cpv}
D.~Das and J.~Das, ``{The $\Lambda_b\to\Lambda^\ast(1520)(\to
  N\!\bar{K})\ell^+\ell^-$ decay at low-recoil in HQET},''
  \href{https://dx.doi.org/10.1007/JHEP07(2020)002}{JHEP {\bfseries 07} (2020)
  002}, \href{https://arxiv.org/abs/2003.08366}{{\ttfamily arXiv:2003.08366
  [hep-ph]}}.

\bibitem{Amhis:2020phx}
Y.~Amhis, S.~Descotes-Genon, C.~Marin~Benito, M.~Novoa-Brunet, and M.-H.
  Schune, ``{Prospects for New Physics searches with ${{\Lambda } ^0_{\mathrm
  {b}}} \!\rightarrow {\Lambda } (1520) {{\ell ^+} {\ell ^-}} $ decays},''
  \href{https://dx.doi.org/10.1140/epjp/s13360-021-01194-5}{Eur. Phys. J. Plus
  {\bfseries 136} no.~6, (2021) 614},
  \href{https://arxiv.org/abs/2005.09602}{{\ttfamily arXiv:2005.09602
  [hep-ph]}}.

\bibitem{Albrecht:2020azd}
J.~Albrecht, Y.~Amhis, A.~Beck, and C.~Marin~Benito, ``{Towards an amplitude
  analysis of the decay $\Lambda_b^0\to pK^-\gamma$},''
  \href{https://dx.doi.org/10.1007/JHEP06(2020)116}{JHEP {\bfseries 06} (2020)
  116}, \href{https://arxiv.org/abs/2002.02692}{{\ttfamily arXiv:2002.02692
  [hep-ph]}}.

\bibitem{Bernlochner:2021vlv}
F.~U. Bernlochner, M.~F. Sevilla, D.~J. Robinson, and G.~Wormser,
  ``{Semitauonic b-hadron decays: A lepton flavor universality laboratory},''
  \href{https://dx.doi.org/10.1103/RevModPhys.94.015003}{Rev. Mod. Phys.
  {\bfseries 94} no.~1, (2022) 015003},
  \href{https://arxiv.org/abs/2101.08326}{{\ttfamily arXiv:2101.08326
  [hep-ex]}}.

\bibitem{Roberts:1992xm}
W.~Roberts, ``{Semileptonic decays of heavy Lambda's into excited baryons},''
  \href{https://dx.doi.org/10.1016/0550-3213(93)90331-I}{Nucl. Phys. B
  {\bfseries 389} (1993) 549--562}.

\bibitem{Leibovich:1997az}
A.~K. Leibovich and I.~W. Stewart, ``{Semileptonic $\Lambda_b$ decay to excited
  $\Lambda_c$ baryons at order $\Lambda_{\rm QCD} / m_Q$},''
  \href{https://dx.doi.org/10.1103/PhysRevD.57.5620}{Phys. Rev. {\bfseries D57}
  (1998) 5620--5631},
\href{https://arxiv.org/abs/hep-ph/9711257}{{\ttfamily arXiv:hep-ph/9711257
  [hep-ph]}}.

\bibitem{Mannel:2015osa}
T.~Mannel and D.~van Dyk, ``{Zero-recoil sum rules for $\Lambda_b \to
  \Lambda_c$ form factors},''
  \href{https://dx.doi.org/10.1016/j.physletb.2015.10.016}{Phys. Lett. B
  {\bfseries 751} (2015) 48--53},
  \href{https://arxiv.org/abs/1506.08780}{{\ttfamily arXiv:1506.08780
  [hep-ph]}}.

\bibitem{Ikeno:2015xea}
N.~Ikeno and E.~Oset, ``{Semileptonic $\Lambda_c$ decay to $\nu l^+$ and
  $\Lambda(1405)$},'' \href{https://dx.doi.org/10.1103/PhysRevD.93.014021}{Phys.
  Rev. D {\bfseries 93} no.~1, (2016) 014021},
  \href{https://arxiv.org/abs/1510.02406}{{\ttfamily arXiv:1510.02406
  [hep-ph]}}.

\bibitem{Cohen:2019zev}
T.~D. Cohen, H.~Lamm, and R.~F. Lebed, ``{Precision Model-Independent Bounds
  from Global Analysis of $b \to c \ell \nu$ Form Factors},''
  \href{https://dx.doi.org/10.1103/PhysRevD.100.094503}{Phys. Rev. D {\bfseries
  100} no.~9, (2019) 094503}, \href{https://arxiv.org/abs/1909.10691}{{\ttfamily
  arXiv:1909.10691 [hep-ph]}}.

\bibitem{Nieves:2019kdh}
J.~Nieves, R.~Pavao, and S.~Sakai, ``{$\Lambda _b$ decays into $\Lambda
  _c^*\ell \bar{\nu }_\ell $ and $\Lambda _c^*\pi ^-$ $[\Lambda _c^*=\Lambda
  _c(2595)$ and $\Lambda _c(2625)]$ and heavy quark spin symmetry},''
  \href{https://dx.doi.org/10.1140/epjc/s10052-019-6929-7}{Eur. Phys. J. C
  {\bfseries 79} no.~5, (2019) 417},
  \href{https://arxiv.org/abs/1903.11911}{{\ttfamily arXiv:1903.11911
  [hep-ph]}}.

\bibitem{Nieves:2019nol}
J.~Nieves and R.~Pavao, ``{Nature of the lowest-lying odd parity charmed baryon
  $\Lambda_c(2595)$ and $\Lambda_c(2625)$ resonances},''
  \href{https://dx.doi.org/10.1103/PhysRevD.101.014018}{Phys. Rev. D {\bfseries
  101} no.~1, (2020) 014018}, \href{https://arxiv.org/abs/1907.05747}{{\ttfamily
  arXiv:1907.05747 [hep-ph]}}.

\bibitem{Bordone:2021bop}
M.~Bordone, ``{Heavy Quark Expansion of $\Lambda_b \rightarrow \Lambda^*(1520)$
  Form Factors beyond Leading Order},''
  \href{https://dx.doi.org/10.3390/sym13040531}{Symmetry {\bfseries 13} no.~4,
  (2021) 531}, \href{https://arxiv.org/abs/2101.12028}{{\ttfamily
  arXiv:2101.12028 [hep-ph]}}.

\bibitem{Papucci:2021pmj}
M.~Papucci and D.~J. Robinson, ``{Form factor counting and HQET matching for
  new physics in
  \ensuremath{\Lambda}b\textrightarrow{}\ensuremath{\Lambda}c*l\ensuremath{\nu}},''
  \href{https://dx.doi.org/10.1103/PhysRevD.105.016027}{Phys. Rev. D {\bfseries
  105} no.~1, (2022) 016027}, \href{https://arxiv.org/abs/2105.09330}{{\ttfamily
  arXiv:2105.09330 [hep-ph]}}.

\bibitem{Ablikim:2015prg}
{\bfseries BESIII} Collaboration, M.~Ablikim {\em et~al.}, ``{Measurement of
  the absolute branching fraction for $\Lambda^+_{c}\to \Lambda e^+\nu_e$},''
  \href{https://dx.doi.org/10.1103/PhysRevLett.115.221805}{Phys. Rev. Lett.
  {\bfseries 115} no.~22, (2015) 221805},
  \href{https://arxiv.org/abs/1510.02610}{{\ttfamily arXiv:1510.02610
  [hep-ex]}}.

\bibitem{Ablikim:2016vqd}
{\bfseries BESIII} Collaboration, M.~Ablikim {\em et~al.}, ``{Measurement of
  the absolute branching fraction for $\Lambda_c^+\rightarrow \Lambda
  \mu^+\nu_{\mu}$},''
  \href{https://dx.doi.org/10.1016/j.physletb.2017.01.047}{Phys. Lett. B
  {\bfseries 767} (2017) 42--47},
  \href{https://arxiv.org/abs/1611.04382}{{\ttfamily arXiv:1611.04382
  [hep-ex]}}.

\bibitem{Ablikim:2018woi}
{\bfseries BESIII} Collaboration, M.~Ablikim {\em et~al.}, ``{Measurement of
  the absolute branching fraction of the inclusive semileptonic $\Lambda_c^+$
  decay},'' \href{https://dx.doi.org/10.1103/PhysRevLett.121.251801}{Phys. Rev.
  Lett. {\bfseries 121} no.~25, (2018) 251801},
  \href{https://arxiv.org/abs/1805.09060}{{\ttfamily arXiv:1805.09060
  [hep-ex]}}.

\bibitem{Ablikim:2019hff}
{\bfseries BESIII} Collaboration, M.~Ablikim {\em et~al.}, ``{Future Physics
  Programme of BESIII},''
  \href{https://dx.doi.org/10.1088/1674-1137/44/4/040001}{Chin. Phys. C
  {\bfseries 44} no.~4, (2020) 040001},
  \href{https://arxiv.org/abs/1912.05983}{{\ttfamily arXiv:1912.05983
  [hep-ex]}}.

\bibitem{Bondar:2013cja}
{\bfseries Charm-Tau Factory} Collaboration, A.~E. Bondar {\em et~al.},
  ``{Project of a Super Charm-Tau factory at the Budker Institute of Nuclear
  Physics in Novosibirsk},''
  \href{https://dx.doi.org/10.1134/S1063778813090032}{Phys. Atom. Nucl.
  {\bfseries 76} (2013) 1072--1085}.

\bibitem{Luo:2019xqt}
Q.~Luo, W.~Gao, J.~Lan, W.~Li, and D.~Xu,
  \href{https://dx.doi.org/10.18429/JACoW-IPAC2019-MOPRB031}{``{Progress of
  Conceptual Study for the Accelerators of a 2-7GeV Super Tau Charm Facility at
  China},''} in {\em {10th International Particle Accelerator Conference}}.
\newblock 6, 2019.

\bibitem{Kou:2018nap}
{\bfseries Belle-II} Collaboration, W.~Altmannshofer {\em et~al.}, ``{The Belle
  II Physics Book},'' \href{https://dx.doi.org/10.1093/ptep/ptz106}{PTEP
  {\bfseries 2019} no.~12, (2019) 123C01},
  \href{https://arxiv.org/abs/1808.10567}{{\ttfamily arXiv:1808.10567
  [hep-ex]}}. [Erratum: PTEP 2020, 029201 (2020)].

\bibitem{Zyla:2020zbs}
{\bfseries Particle Data Group} Collaboration, P.~A. Zyla {\em et~al.},
  ``{Review of Particle Physics},''
  \href{https://dx.doi.org/10.1093/ptep/ptaa104}{PTEP {\bfseries 2020} no.~8,
  (2020) 083C01}.

\bibitem{Meinel:2021grq}
S.~Meinel and G.~Rendon, ``{Charm-baryon semileptonic decays and the strange
  $\Lambda^*$ resonances: New insights from lattice QCD},''
  \href{https://dx.doi.org/10.1103/PhysRevD.105.L051505}{Phys. Rev. D {\bfseries
  105} (2022) L051505}, \href{https://arxiv.org/abs/2107.13084}{{\ttfamily arXiv:2107.13084
  [hep-ph]}}.

\bibitem{Zwicky:2013eda}
R.~Zwicky, ``{Endpoint symmetries of helicity amplitudes},''
  \href{https://arxiv.org/abs/1309.7802}{{\ttfamily arXiv:1309.7802 [hep-ph]}}.

\bibitem{Hiller:2013cza}
G.~Hiller and R.~Zwicky, ``{(A)symmetries of weak decays at and near the
  kinematic endpoint},'' \href{https://dx.doi.org/10.1007/JHEP03(2014)042}{JHEP
  {\bfseries 03} (2014) 042}, \href{https://arxiv.org/abs/1312.1923}{{\ttfamily
  arXiv:1312.1923 [hep-ph]}}.

\bibitem{Hiller:2021zth}
G.~Hiller and R.~Zwicky, ``{Endpoint relations for baryons},''
  \href{https://dx.doi.org/10.1007/JHEP11(2021)073}{JHEP {\bfseries 11} (2021)
  073}, \href{https://arxiv.org/abs/2107.12993}{{\ttfamily arXiv:2107.12993
  [hep-ph]}}.

\bibitem{Pervin:2005ve}
M.~Pervin, W.~Roberts, and S.~Capstick, ``{Semileptonic decays of heavy
  $\Lambda$ baryons in a quark model},''
  \href{https://dx.doi.org/10.1103/PhysRevC.72.035201}{Phys. Rev. {\bfseries
  C72} (2005) 035201},
\href{https://arxiv.org/abs/nucl-th/0503030}{{\ttfamily arXiv:nucl-th/0503030
  [nucl-th]}}.

\bibitem{Mott:2011cx}
L.~Mott and W.~Roberts, ``{Rare dileptonic decays of $\Lambda_b$ in a quark
  model},'' \href{https://dx.doi.org/10.1142/S0217751X12500169}{Int. J. Mod.
  Phys. {\bfseries A27} (2012) 1250016},
\href{https://arxiv.org/abs/1108.6129}{{\ttfamily arXiv:1108.6129 [nucl-th]}}.

\bibitem{Gutsche:2017wag}
T.~Gutsche, M.~A. Ivanov, J.~G. Körner, V.~E. Lyubovitskij, V.~V. Lyubushkin,
  and P.~Santorelli, ``{Theoretical description of the decays $\Lambda_b \to
  \Lambda^{(\ast)}(\frac12^\pm,\frac32^\pm) + J/\psi$},''
  \href{https://dx.doi.org/10.1103/PhysRevD.96.013003}{Phys. Rev. D {\bfseries
  96} no.~1, (2017) 013003}, \href{https://arxiv.org/abs/1705.07299}{{\ttfamily
  arXiv:1705.07299 [hep-ph]}}.

\bibitem{Gutsche:2018nks}
T.~Gutsche, M.~A. Ivanov, J.~G. Körner, V.~E. Lyubovitskij, P.~Santorelli, and
  C.-T. Tran, ``{Analyzing lepton flavor universality in the decays
  $\Lambda_b\to\Lambda_c^{(\ast)}(\frac12^\pm,\frac32^-) +
  \ell\,\bar\nu_\ell$},''
  \href{https://dx.doi.org/10.1103/PhysRevD.98.053003}{Phys. Rev. D {\bfseries
  98} no.~5, (2018) 053003}, \href{https://arxiv.org/abs/1807.11300}{{\ttfamily
  arXiv:1807.11300 [hep-ph]}}.

\bibitem{Feldmann:2011xf}
T.~Feldmann and M.~W.~Y. Yip, ``{Form factors for $\Lambda_b \to \Lambda$
  transitions in the soft-collinear effective theory},''
  \href{https://dx.doi.org/10.1103/PhysRevD.85.014035}{Phys. Rev. D {\bfseries
  85} (2012) 014035}, \href{https://arxiv.org/abs/1111.1844}{{\ttfamily
  arXiv:1111.1844 [hep-ph]}}. [Erratum: Phys.Rev.D 86, 079901 (2012)].

\bibitem{Aoki:2010dy}
{\bfseries RBC, UKQCD} Collaboration, Y.~Aoki {\em et~al.}, ``{Continuum Limit
  Physics from 2+1 Flavor Domain Wall QCD},''
  \href{https://dx.doi.org/10.1103/PhysRevD.83.074508}{Phys. Rev. {\bfseries
  D83} (2011) 074508},
\href{https://arxiv.org/abs/1011.0892}{{\ttfamily arXiv:1011.0892 [hep-lat]}}.

\bibitem{Blum:2014tka}
{\bfseries RBC, UKQCD} Collaboration, T.~Blum {\em et~al.}, ``{Domain wall QCD
  with physical quark masses},''
  \href{https://dx.doi.org/10.1103/PhysRevD.93.074505}{Phys. Rev. {\bfseries
  D93} no.~7, (2016) 074505},
\href{https://arxiv.org/abs/1411.7017}{{\ttfamily arXiv:1411.7017 [hep-lat]}}.

\bibitem{Hashimoto:1999yp}
S.~Hashimoto, A.~X. El-Khadra, A.~S. Kronfeld, P.~B. Mackenzie, S.~M. Ryan, and
  J.~N. Simone, ``{Lattice QCD calculation of $\bar{B} \to D \ell \bar{\nu}$
  decay form-factors at zero recoil},''
  \href{https://dx.doi.org/10.1103/PhysRevD.61.014502}{Phys. Rev. {\bfseries
  D61} (1999) 014502},
\href{https://arxiv.org/abs/hep-ph/9906376}{{\ttfamily arXiv:hep-ph/9906376
  [hep-ph]}}.

\bibitem{ElKhadra:2001rv}
A.~X. El-Khadra, A.~S. Kronfeld, P.~B. Mackenzie, S.~M. Ryan, and J.~N. Simone,
  ``{The Semileptonic decays $B \to \pi \ell \nu$ and $D \to \pi \ell \nu$ from
  lattice QCD},'' \href{https://dx.doi.org/10.1103/PhysRevD.64.014502}{Phys.
  Rev. {\bfseries D64} (2001) 014502},
\href{https://arxiv.org/abs/hep-ph/0101023}{{\ttfamily arXiv:hep-ph/0101023
  [hep-ph]}}.

\bibitem{Blum:2012uh}
T.~Blum, T.~Izubuchi, and E.~Shintani, ``{New class of variance-reduction
  techniques using lattice symmetries},''
  \href{https://dx.doi.org/10.1103/PhysRevD.88.094503}{Phys. Rev. {\bfseries
  D88} no.~9, (2013) 094503},
\href{https://arxiv.org/abs/1208.4349}{{\ttfamily arXiv:1208.4349 [hep-lat]}}.

\bibitem{Shintani:2014vja}
E.~Shintani, R.~Arthur, T.~Blum, T.~Izubuchi, C.~Jung, and C.~Lehner,
  ``{Covariant approximation averaging},''
  \href{https://dx.doi.org/10.1103/PhysRevD.91.114511}{Phys. Rev. {\bfseries
  D91} no.~11, (2015) 114511},
\href{https://arxiv.org/abs/1402.0244}{{\ttfamily arXiv:1402.0244 [hep-lat]}}.

\bibitem{Aoki:2012xaa}
{\bfseries RBC, UKQCD} Collaboration, Y.~Aoki, N.~H. Christ, J.~M. Flynn,
  T.~Izubuchi, C.~Lehner, M.~Li, H.~Peng, A.~Soni, R.~S. Van~de Water, and
  O.~Witzel, ``{Nonperturbative tuning of an improved relativistic heavy-quark
  action with application to bottom spectroscopy},''
  \href{https://dx.doi.org/10.1103/PhysRevD.86.116003}{Phys. Rev. {\bfseries
  D86} (2012) 116003},
\href{https://arxiv.org/abs/1206.2554}{{\ttfamily arXiv:1206.2554 [hep-lat]}}.

\bibitem{Supplemental}
{Please see Supplemental Material at
  \url{https://arxiv.org/src/2107.13140/anc} for files
  containing the form-factor parameter values and covariance matrices.}

\bibitem{Hussain:2017lir}
M.~M. Hussain and W.~Roberts, ``{$\Lambda_c$ Semileptonic Decays in a Quark
  Model},'' \href{https://dx.doi.org/10.1103/PhysRevD.95.053005}{Phys. Rev. D
  {\bfseries 95} no.~5, (2017) 053005},
  \href{https://arxiv.org/abs/1701.03876}{{\ttfamily arXiv:1701.03876
  [nucl-th]}}. [Addendum: Phys.Rev.D 95, 099901 (2017)].

\bibitem{Lang:2015hza}
C.~B. Lang, D.~Mohler, S.~Prelovsek, and R.~M. Woloshyn, ``{Predicting positive
  parity $B_s$ mesons from lattice QCD},''
  \href{https://dx.doi.org/10.1016/j.physletb.2015.08.038}{Phys. Lett.
  {\bfseries B750} (2015) 17--21},
\href{https://arxiv.org/abs/1501.01646}{{\ttfamily arXiv:1501.01646 [hep-lat]}}.

\bibitem{Briceno:2014uqa}
R.~A. Brice\~no, M.~T. Hansen, and A.~Walker-Loud, ``{Multichannel 1
  $\rightarrow$ 2 transition amplitudes in a finite volume},''
  \href{https://dx.doi.org/10.1103/PhysRevD.91.034501}{Phys. Rev. D {\bfseries
  91} no.~3, (2015) 034501}, \href{https://arxiv.org/abs/1406.5965}{{\ttfamily
  arXiv:1406.5965 [hep-lat]}}.

\bibitem{Briceno:2015csa}
R.~A. Brice\~no and M.~T. Hansen, ``{Multichannel 0 $\to$ 2 and 1 $\to$ 2
  transition amplitudes for arbitrary spin particles in a finite volume},''
  \href{https://dx.doi.org/10.1103/PhysRevD.92.074509}{Phys. Rev. D {\bfseries
  92} no.~7, (2015) 074509}, \href{https://arxiv.org/abs/1502.04314}{{\ttfamily
  arXiv:1502.04314 [hep-lat]}}.

\bibitem{Hansen:2021ofl}
M.~T. Hansen, F.~Romero-L\'opez, and S.~R. Sharpe, ``{Decay amplitudes to three
  hadrons from finite-volume matrix elements},''
  \href{https://dx.doi.org/10.1007/JHEP04(2021)113}{JHEP {\bfseries 04} (2021)
  113}, \href{https://arxiv.org/abs/2101.10246}{{\ttfamily arXiv:2101.10246
  [hep-lat]}}.

\bibitem{Aaltonen:2008eu}
{\bfseries CDF} Collaboration, T.~Aaltonen {\em et~al.}, ``{First Measurement
  of the Ratio of Branching Fractions $B(\Lambda^0_b \to \Lambda^+_{c} \mu^{-}
  \bar{\nu}_\mu) / B(\Lambda^0_b \to \Lambda^+_{c} \pi^{-})$},''
  \href{https://dx.doi.org/10.1103/PhysRevD.79.032001}{Phys. Rev. D {\bfseries
  79} (2009) 032001}, \href{https://arxiv.org/abs/0810.3213}{{\ttfamily
  arXiv:0810.3213 [hep-ex]}}.

\bibitem{XSEDE}
J.~{Towns}, T.~{Cockerill}, M.~{Dahan}, I.~{Foster}, K.~{Gaither},
  A.~{Grimshaw}, V.~{Hazlewood}, S.~{Lathrop}, D.~{Lifka}, G.~D. {Peterson},
  R.~{Roskies}, J.~R. {Scott}, and N.~{Wilkins-Diehr}, ``{XSEDE: Accelerating
  Scientific Discovery},''
  \href{https://dx.doi.org/10.1109/MCSE.2014.80}{Computing in Science
  Engineering {\bfseries 16} no.~5, (2014) 62--74}.

\bibitem{Edwards:2004sx}
{\bfseries SciDAC, LHPC, UKQCD} Collaboration, R.~G. Edwards and B.~Joo, ``{The
  Chroma software system for lattice QCD},''
  \href{https://dx.doi.org/10.1016/j.nuclphysbps.2004.11.254}{Nucl. Phys. B
  Proc. Suppl. {\bfseries 140} (2005) 832},
  \href{https://arxiv.org/abs/hep-lat/0409003}{{\ttfamily
  arXiv:hep-lat/0409003}}.

\bibitem{Chroma}
R.~G. Edwards, B.~Joó, {\em et~al.}, ``{Chroma}.''
\newblock \url{https://github.com/JeffersonLab/chroma}.

\bibitem{JOO2015139}
B.~Joó, M.~Smelyanskiy, D.~D. Kalamkar, and K.~Vaidyanathan,
  \href{https://dx.doi.org/10.1016/B978-0-12-803819-2.00023-9}{``{Chapter 9 -
  Wilson Dslash Kernel From Lattice QCD Optimization},''} in {\em {High
  Performance Parallelism Pearls}}, pp.~139 -- 170.
\newblock Morgan Kaufmann, Boston, 2015.

\bibitem{QPhiX}
B.~Joó {\em et~al.}, ``{QPhiX Dslash and Solver Library}.''
\newblock \url{https://github.com/jeffersonlab/qphix}.

\bibitem{QLUA}
A.~Pochinsky, S.~Syritsyn, {\em et~al.}, ``{QLUA}.''
\newblock \url{https://usqcd.lns.mit.edu/w/index.php/QLUA}.

\bibitem{MDWF}
A.~Pochinsky, S.~Syritsyn, {\em et~al.}, ``{Möbius Domain Wall inverter}.''
\newblock \url{https://github.com/usqcd-software/mdwf}.

\bibitem{USQCD}
{\bfseries USQCD} Collaboration, ``{USQCD Software}.''
\newblock \url{https://usqcd-software.github.io}.

\end{thebibliography}
\end{document}